\documentstyle[preprint,aps,epsfig]{revtex}
\begin{document}
\date{\today}
\tighten
\draft
%
%
\title{ Critical properties of two--dimensional Josephson junction\\
 arrays with zero-point quantum fluctuations}
\author{Cristian Rojas and Jorge V. Jos\'{e}\\
{\it Department of Physics and Center for Interdisciplinary
Research on Complex Systems,\\ Northeastern University, Boston
Massachusetts 02115, USA}}
\date{\today}
\maketitle
%
%
\begin{abstract}
We present results from an extensive analytic and numerical study of a 
two-dimensional model of a square array of ultrasmall Josephson 
junctions.  We include the ultrasmall self and mutual capacitances of the 
junctions, for the same parameter ranges as those produced in the 
experiments.  The model Hamiltonian studied includes the Josephson, $E_J$,
as well as the charging, $E_C$, energies between superconducting islands.  
The corresponding quantum partition function is expressed in different
calculationally convenient ways within its path-integral representation.  
The phase diagram is analytically studied using a WKB renormalization 
group (WKB-RG) plus a  self-consistent harmonic approximation (SCHA) 
analysis,  together with  non-perturbative quantum  Monte Carlo simulations. 
Most of the results presented here pertain to the superconductor to normal 
(S-N) region, although some results for the insulating to normal (I-N) 
region are also included.  We find very good agreement between the WKB-RG 
and QMC results when compared to the experimental data.  To fit the data, 
we only used the experimentally determined capacitances as fitting 
parameters.  The WKB-RG analysis in the S-N region predicts a low 
temperature instability i.e. a Quantum Induced Transition (QUIT).  We carefully analyze the possible existence of the QUIT via the QMC 
simulations and carry out a finite size analysis of $T_{QUIT}$ as a function 
of the magnitude of imaginary time axis $L_\tau$.  We find that for some 
relatively large values of $\alpha=\frac{E_C}{E_J}$ 
($1\leq \alpha \leq 2.25)$, the $L_\tau\to\infty$ limit does appear to give 
a {\it non-zero} $T_{QUIT}$, while for $\alpha \ge 2.5$, $T_{QUIT}=0$.  
We use the SCHA to analytically understand the $L_\tau$ dependence of the 
QMC results with good agreement between them.  Finally, we also carried 
out a WKB-RG  analysis in the I-N region and found no evidence of a low 
temperature  QUIT, up to lowest order in ${\alpha}^{-1}$.

\end{abstract}
%
%
\section{\bf Introduction}
\label{sec:intro-jja}
The physics of Josephson junctions arrays (JJA) has been a subject of 
significant interest in the last ten years \cite{mooij-schon}. A large 
number of studies, both experimental 
\cite{mooij_van-wees_geerligs_peters_fazio_schon,van-der-zant_geerling_mooij-92,tighe-tuominen-hergenrother-tinkham,van-der-zant-thesis,delsing-chen-haviland-harada-claeson,van_der_zant-fritschy-orlando-mooij}
and theoretical
\cite{fazio-schon-91,doniach,lozovik,zwerger,fishman-stroud,jose_1984,jacobs-jose-novotny-goldman,jacobs-jose,simkin,granato-contentino,roddick-stroud,ariosa-beck,kim-choi,zaikin,simanek,kim-choi2}
have been devoted to them. Initially, part of the 
interest in JJA came from their close relation to one of the most 
extensively studied theoretical spin models, i.e. the classical 
2-D XY model for which JJA give a concrete experimental realization.  

In non-dissipative JJA the two main contributions to the energy are the 
Josephson coupling between superconducting islands due to Cooper
pair tunneling, and the electrostatic energy arising from local
deviations from charge neutrality.
In the initial experimental studies, the size of the islands was
large enough so that the charging energy contributions were very small, 
thus making the arrays' behavior effectively semi-classical.
Recent advances in submicron technology have made it possible to fabricate 
relatively large arrays of ultrasmall superconducting islands separated 
by 
insulating barriers. These islands can have areas of 
the order of a few ${\mu}{\rm m}^2$, 
with self capacitances $C_{\rm s}\approx 3\times 10^{-2}$ fF, 
and nearest neighbors' mutual capacitance $C_{\rm m}\approx 1$ fF
\cite{van-der-zant_geerling_mooij-92}. Note 
that the mutual capacitance can be at least two orders of magnitude larger 
than the self capacitance. Therefore, these arrays have charging energy
contributions, $E_c$, large enough so that quantum fluctuation 
effects are of paramount importance.

In the Delft \cite{van-der-zant_geerling_mooij-92} and Harvard 
\cite{tighe-tuominen-hergenrother-tinkham} experiments, the island sizes 
were kept constant, while varying the normal state junction resistance, 
which in turn changes the Josephson coupling energy, $E_J$. This allows 
one to obtain arrays with values of the quantum parameter 
\begin{equation}
\label{alpha_m}
                \alpha_{\rm m} = \frac{E_{C_{\rm m}}}{E_J}
\end{equation}
in the range [0.13--4.55] 
\cite{van-der-zant_geerling_mooij-92}, or
values as high as 33 \cite{tighe-tuominen-hergenrother-tinkham}. In 
this equation we have used the definition of charging energy,
\begin{equation}
\label{def-charging-energy}
                E_{C_{\rm m}} = \frac{e^2}{2C_{\rm m}}.
\end{equation}
The experimental systems can be modeled by a quantum 
generalization of the classical XY model, because the phase of the order 
parameter associated with each one of the islands is canonically 
conjugate to its excess Cooper pair number. The magnitude of $\alpha_{\rm m}$
determines the relevance of the quantum fluctuations. For small 
$\alpha_{\rm m}$ the quantum fluctuations of the phases are small and
the system is well modeled by a renormalized classical 2-D XY model.

The nature of the phase transition in the classical 2-D XY model 
is well understood, whereas in its quantum mechanical generalization 
there still are unsettled issues. One of the most notorious of these 
is the possibility of having a low temperature instability of the 
superconducting state. 
A possible reentrant transition was originally found within a mean field 
theory treatment of the self-capacitive XY model
\cite{simkin,kim-choi,zaikin,simanek}.  An explicit 
two-dimensional study of the self-capacitative XY model, within a WKB 
renormalization group (WKB-RG) analysis also found evidence of a low 
temperature reentrant instability, triggered by a quantum fluctuation 
induced proliferation of vortices \cite{jose_1984}.  

Recently, Kim and Choi  have studied the quantum induced fluctuations
in these arrays, using a variational method \cite{kim-choi2}.
They found that there is a range of values of the ratio of charging
to Josephson energy, for which there is a low temperature reentrance 
from a superconducting to a normal state. Similar results had 
been obtained by Simanek, also using a variational calculation, 
see for example Ref. \cite{simanek}.

A non-perturbative quantum Monte Carlo study of the self-capacitive model 
found a low temperature transition, but between two superconducting 
states \cite{jacobs-jose-novotny-goldman}.  
The fully frustrated version of this model was also 
studied by quantum QMC and it yielded a larger jump discontinuity in 
the superfluid density as compared to the one in zero field as well as
the critical temperature one order of magnitude 
higher \cite{jacobs-jose-novotny-goldman}.  A 
more recent analysis of the WKB-RG analysis has shown that, to lowest 
order in the quantum fluctuations,  it must have the same critical temperature 
for a quantum induced phase transition (QUIT) \cite{jose-rojas}.  
A recent QMC study of the fully frustrated 
self-capacitive model by Mikalopas et al. 
\cite{mikalopas-jarrel-pinski-chung-novotny}  
has suggested that the unusually large jump in the superfluid 
density is dominated by metastability effects due to the particular 
nature of the excitations in the frustrated model. 
This result is in agreement then with the reanalysis of the RG 
equations. However, this study
was carried out at relatively high temperatures and the question about  the
existence of a QUIT, both in the frustrated and unfrustrated cases
remains open.  We deal extensively with the later  question  here.
Other studies find within MFT that
to have reentrance it is necessary to include off-diagonal 
capacitances \cite{fishman-stroud}, while others do not agree with 
this finding
\cite{doniach,granato-contentino,roddick-stroud,ariosa-beck}.
The search for a reentrant type transition is  encouraged by 
some evidence of low temperature instabilities found  
experimentally in arrays of Josephson junctions 
\cite{mooij_van-wees_geerligs_peters_fazio_schon}, ultrathin amorphous
films \cite{belevtsev-komnik-fomin}, a multiphase high-$T_c$
system \cite{seyoum-riitano-bennett-wong}, and in granular superconductors
\cite{lin-shao-wu-hor-jin-chu-evans-bayuzick}.

Most theoretical studies have been carried out using 
the self-capacitive model and different kinds of MFT
or self consistent harmonic approximations (SCHA)\cite{ariosa-beck}.
As already mentioned, these studies do not agree among each other 
on some of the properties of the phase diagram, in particular about the 
possible existence of a low temperature instability of the 
superconducting state. 
No study has been carried out that closely represents the experimental 
systems where both the self and mutual capacitances 
are explicitly included.  The goal of this paper is to  
consider a model that is expected to represent 
the characteristics of the Delft experiments.  In particular, we 
concentrate on calculating the phase diagram using different 
theoretical tools.

One of the main results of this paper is presented in Fig. 
\ref{fig:fig01-phase-diagram} 
which shows the $\alpha_{\rm m}$ vs. $T$ phase
diagram for an array with $C_{\rm m}> C_{\rm s}$ both for the
unfrustrated ($f=0$) and fully frustrated ($f=1/2$) cases. The left hand side
of this diagram shows the superconducting to normal phase boundary (S--N)
as data points with error bars joined by a continuous line.
These data points were calculated using a QMC 
method, to be described later in the paper.
We also show (as squares) the experimental results  
taken from Refs. \cite{van-der-zant_geerling_mooij-92,van-der-zant-thesis}.
For $f=0$ at small values of $\alpha_{\rm m}$, the theoretical and experimental 
results agree quantitatively quite well with each other
and with the semiclassical WKB-RG approximation. On the other hand, 
they only qualitatively agree for the $f=1/2$ case and
on the superconducting to insulating phase boundary. The 
normal to insulating transition line is shown to the 
right of the phase diagram. The latter is just a 
tentative boundary since our numerical calculations were not
reliable enough to give the definitive location of this line,
as also happens in experiments. 
The error bars in the calculated points used to draw the N-I line
represent a crude estimation of the region where the inverse 
dielectric constant is different from zero. However, the
issue of convergence of the calculation to the path integral limit
is not resolved by these error bars. As we will 
explain in the main body of the paper we found further evidence for a low 
temperature instability of the superconducting state in our  
numerical calculations.
We found that this instability depends strongly on the magnitude of $\alpha_{\rm m}$ and
the finite size of the imaginary time axis in the QMC
calculations. The latter finding sets strict constraints on some of the 
reentrant type behavior found in previous theoretical studies. 
Other studies have found reentrance very close to the 
superconducting to insulating transition \cite{simanek}. 
This possibility is harder to study from our Monte Carlo calculations.
 
The physical content of the phase diagram 
is generally understood in terms of the interplay between the Josephson 
and charging energies. For small $\alpha_{\rm m}$ and high temperature 
the spectrum of excitations is dominated by thermally excited 
vortices, which drive the superconducting to normal transition 
as the temperature increases, 
while the charging energy contributes with weak quantum fluctuations of the 
phases. The latter produces, after averaging over the quantum fluctuations, an effective
classical action with a renormalized Josephson coupling that
lowers the critical temperature \cite{jose_1984}.

For large $\alpha_{\rm m}$ and low temperatures the charging energy
dominates. The excitations in this limit are due to the 
thermally assisted Cooper pair tunneling 
that produces charged polarized islands. 
At low temperatures, there is not enough thermal energy to overcome the 
electrostatic coulomb blockade so that the Cooper pairs are localized 
and the array is insulating.
As the temperature increases, the electric dipole excitations can unbind, 
driving an insulating to conducting transition (I--C). In the limit 
$C_{\rm m}\gg C_{\rm s}$, it was suggested that the I-C transition could 
be of the Berezinskii-Kosterlitz-Thouless (BKT) type, 
for in this case the interaction between charges is essentially logarithmic 
\cite{yaks-87,widom-badjou,mooij_van-wees_geerligs_peters_fazio_schon}. 
However, for the experimental samples it has been shown 
that rather than a true phase transition what is measured appears to be 
a crossover between an insulating to conducting phase, characterized 
by thermally activated processes
\cite{tighe-tuominen-hergenrother-tinkham,delsing-chen-haviland-harada-claeson}.
It is likely that the reason for the crossover is the short 
screening length present in the samples ($\Lambda \approx 20$ lattice 
sites). Both experimental groups
\cite{tighe-tuominen-hergenrother-tinkham,delsing-chen-haviland-harada-claeson}
find that a simple energetic argument gives an explanation for
the activation energy found in the experiments. Furthermore, the nature 
of this crossover may be linked to thermal as well as to dynamical 
effects.  As we shall see, theoretically the I-N phase is hard to study 
in detail. 

For large values of $\alpha_{\rm m}$ the model can be approximated by
a 2-D lattice Coulomb gas, where a perturbative expansion can be carried out 
using $E_J$ as a small parameter. This type of calculation
was performed in Ref. \cite{fazio-schon-91} in the limit 
$C_{\rm s}\ll C_{\rm m}$. The analysis lead to a 2-D Coulomb gas 
with a renormalized coupling constant. Here we will extend this
calculation to obtain a more accurate estimation of the
renormalized coupling constant. We do this because we are interested  
in seeing if it is possible to have a QUIT instability in 
the low temperature insulating phase. 
This possibility is suggested by the dual symmetry of the effective action 
between charges and vortices found in Ref. \cite{fazio-schon-91}, and
the fact that the $\alpha_{\rm m}$ perturbative expansion shows
a low temperature QUIT instability in the superconducting phase. 
We find that the results of a first order expansion in $\alpha_{\rm m}^{-1}$ 
does not present this type of low temperature instability.

Among the most interesting regions of the phase 
diagram is when the Josephson and charging energies are comparable.
For this nonperturbative case, using a path integral 
formulation and the Villain 
approximation \cite{jose-kadanoff-nelson-kirpatrick}, 
an effective action for logarithmically interacting 
charges and vortices was derived in Ref. \cite{fazio-schon-91} in 
the case where $C_{\rm s}\ll C_{\rm m}$. The action of the two 
Coulomb gases shows an almost dual symmetry, so that at an 
intermediate value of  $\alpha_{\rm m}$, both the S--N
and the I--C transitions converge to a single 
point as $T\rightarrow 0$. A similar picture was derived in Ref. 
\cite{granato-contentino} using a short range electrostatic interaction
and a mean field renormalization group calculation. A nonperturbative 
calculation is needed to determine the actual shape of the phase
diagram in this region.  We further discuss this point in the main body of 
the paper.

It has been argued that at $T=0$ the self-capacitive model is in the same 
universality class as the 3-D XY model \cite{doniach,zwerger}, where 
the ratio $\alpha_{\rm s}=(q^2/2C_{\rm s})/E_J$ would
play the role of temperature. This analogy would result in a transition at 
some finite value of $\alpha_{\rm s}$ from a  superconducting to an insulating
phase. Moreover, there should be a marked signature in the nature of 
the correlation functions when crossing over from
a 2-D XY model at high temperatures to a 3-D XY model as $T\rightarrow 0$.
When $\alpha_s=0$, the correlation functions decay algebraically in the 
critical region, with a temperature dependent exponent.  At T=0 and 
$\alpha_s \neq  0$ there is a single critical point at
 $\alpha_s =\alpha_c$, so 
that the correlations decay exponentially above and below $\alpha_c$, and 
algebraically at  $\alpha_s =\alpha_c$.  The question is then, how do we go 
from algebraic to exponential correlations as T changes?  
This can only happen by 
having a change of analyticity in the correlations, thus the possibility 
of having a QUIT in the self-capacitive model.

The situation is different in the mutual-capacitance dominated limit, of 
experimental interest.  When $C_s=0 $ the model is equivalent to having two 
interacting lattice Coulomb lattice gas models.  
The general critical properties 
of this case  are not fully understood at present.  A further complication 
arises when $C_s$ is small but non-zero.  The map to a higher dimensional 
known model does not work in this case, and the problem has to be studied 
on its own right.  Because of the essential differences between the 
self-capacitive and the mutual-capacitance dominated models one can not 
just take results from one case and apply them to the other.  It is the 
goal of this paper to explicitly study the mutual capacitance dominated 
model, but with nonzero $C_s$.
A brief report on some of the results of this paper has appeared 
elsewhere \cite{jose-rojas}.

The outline of the rest of the paper is the following: In Section
\ref{sec:model} we define the model studied and derive the path integral 
formulation of the partition function used in our calculations. In Section
\ref{sec:wkb-rg} we present a derivation of the path integral 
used in the semiclassical analysis, in the limit where the Josephson
energy dominates. We carry out a WKB expansion up to 
first order in $\alpha_{\rm m}$, finding an effective 
classical action where the charging energy contributions are taken 
into account as a renormalization
of the Josephson coupling. In section \ref{subsec:rg} we find general 
renormalization group (RG) equations from which we obtain the phase 
diagram for small $\alpha_{\rm m}$. 
In Section \ref{sec:insulating-normal} we study the large $\alpha_{\rm m}$ 
limit, 
in which the charging energy is dominant over the Josephson energy. There
we obtain an effective 2-D Coulomb gas model with a quantum fluctuations
renormalized coupling constant.
In Section \ref{sec:mc-operators} we discuss our  QMC calculations
and define the physical quantities calculated.  In  Section 
\ref{sec:simulation}
we present some technical details of the implementation of the 
QMC simulations. In Section \ref{sec:results-for-f=0} 
we give the QMC results for $f=0$ mainly, but also
for $f=1/2$. There we make a direct comparison   between the semiclassical 
approximation results, the QMC calculations and experiment which lead to 
the phase diagram  discussed above. In that section we also present an
$L_\tau$ dependent analysis of the apparent $T_{QUIT}$ for three 
relatively large values of $\alpha_m
= 2.0, 2.25$ and $2.5$. The 
$L_\tau\to\infty$ extrapolation of the results leads to a {\it finite}
$T_{QUIT}$ for $\alpha_m= 2.0$ and $2.25$, while for $\alpha_m=2.5$ we
get a $T_{QUIT}(L_\tau=\infty)=0$. In section \ref{sec:harmonic}
we discuss a self-consistent harmonic approximation (SCHA) analysis, 
that we use 
to analytically study the phase diagram,  and that helps us understand 
the finite  size effects of the imaginary-time lattices 
studied in the QMC calculations. At the end 
of the paper there are two appendices where we give  more technical details 
of the analysis. In Section \ref{sec:conclusions} we restate the main 
results of this paper.


\section{\bf The Model and the Path Integral Formalism}
\label{sec:model}

In this section we define the Josephson junction array model 
considered in this paper together with the path integral formulation of 
its corresponding partition function.

We assume that each superconducting island in a junction
can be characterized by a Ginzburg-Landau order parameter 
$\Psi(\vec r)=|\Psi _0(\vec r)|e^{i \phi (\vec r)}$, where $\vec r$ is a 
two-dimensional vector denoting the position of each island.
If the coherence length of the Cooper pairs is larger than the size of
the islands, we can assume that the phase of the 
order parameter is constant in each island. Moreover,
the amplitude of the order parameter is expected to have small
fluctuations about an electrically neutral island and can then be taken as constant trough 
the array. We will assume that the charge fluctuations have an effect
on the electrostatic energy but not on the Josephson contribution 
to the Hamiltonian. The gauge invariant Hamiltonian studied here is
\begin{eqnarray}
\label{hamiltonian}
          {\hat {\cal H}}=\hat H_C+\hat H_J 
          &=&
          {{q^2}\over{2}}\sum_{<\vec r_1,\vec r_2>}\hat n(\vec r_1)
          {\bf {\bf C}^{-1}}(\vec r_1,\vec r_2)\hat n(\vec r_2)+
          E_J \sum_{<\vec r_1,\vec r_2>} \Big[1-\cos\Big(\phi(\vec r_1)-
                 \phi(\vec r_2) - A_{\vec r_1,\vec r_2}\Big)\Big]          
          ,\nonumber\\
\end{eqnarray}
where $q=2e$; $\hat \phi (\vec r)$ is the quantum phase operator and 
$\hat n (\vec r)$ is
its canonically conjugate number operator, which measures the excess number
of Cooper pairs in the $\vec r$ island. These operators satisfy the 
commutation relations
$ [\hat n(\vec r_1),\hat\phi(\vec r_2)]=-i\delta_{\vec r_1,\vec r_2}
$ \cite{anderson}. Here $A_{\vec r_1,\vec r_2}$ is defined by the line 
integral that joins the sites located at $\vec r_1$ and $\vec r_2$,
$ A_{\vec r_1,\vec r_2} = \frac{2\pi}{\Phi_0} \int_{\vec r_1}^{\vec r_2}
\vec A\cdot d\vec l$, where $\vec A$ is the vector potential 
and $\Phi_0$ is the flux quantum. In Eq. (\ref{hamiltonian}) 
$\hat{H}_C$ is the charging energy due to the electrostatic
interaction between the excess Cooper pairs in the islands. The 
${\bf C}^{-1}(\vec r_1,\vec r_2)$ matrix is the electric field propagator
and its inverse, ${\bf C}(\vec r_1,\vec r_2)$, is the geometric 
capacitance matrix, which must 
be calculated by solving Poisson's equation subject to the appropriate 
boundary conditions. This is not easy to do in general 
and typically this matrix is 
approximated, both theoretically and in the experimental analysis of the 
data, by  diagonal plus nearest neighbor contributions 
\cite{fazio-schon-91}:
\begin{equation}
\label{capacitance-matrix}
{\bf C}(\vec r_1,\vec r_2) = (C_{\rm s}+zC_{\rm m})\delta_{\vec 
r_1,\vec r_2} - C_{\rm m} \sum_{\vec d} \delta_{\vec r_1,
\vec r_2+\vec d}\, .
\end{equation}
Here the vector $\vec d$ runs over nearest neighboring islands,
$z$ is the coordination number, $C_{\rm s}$ is the  
self-capacitance of each island, and $C_{\rm m}$ is the mutual capacitance 
between nearest neighbor islands. In the  experimental arrays, typically
$C_{\rm m} \sim 10^2C_{\rm s} \sim 1 {\rm fF}$ 
\cite{van-der-zant_geerling_mooij-92}.

The second term in Eq.(\ref{hamiltonian}) is the Josephson energy, which 
represents the probability of  Cooper pair tunneling between 
nearest neighboring islands. The Josephson coupling energy 
$E_J=\Phi_0 i_{c}/(2\pi)$ is assumed to be temperature independent, where 
$i_{c}$ is the junction critical current and $\Phi_0$ the flux quantum.
Here we are interested in calculating the thermodynamic properties of the 
model defined by $\hat {\cal {H}}$. The quantity of interest is the 
partition function
\begin{equation}
\label{partition_function}
Z \equiv {\rm Tr}\left\{ e^{-\beta\hat {\cal H}}\right\}\label{zeta}.
\end{equation}
The trace is taken either over the phase variables, $\hat\phi$, or 
the numbers operator, $\hat{n}$.
To evaluate the partition function we will use its path integral 
representation
\cite{schulman-book,kleinert-book}. To derive the path integral we use 
the states
\begin{equation}
\label{charge-phase}
           <n(\vec r_1)|\phi(\vec r_2)> = \delta_{\vec r_1,\vec r_2} 
           \frac{\exp\{ i n(\vec r_1)\phi(\vec r_1)\}}{\sqrt{2\pi}}.
\end{equation}
We will also use the fact that both $\{|n(\vec r)>\}$ and 
$\{|\phi(\vec r)>\}$ form complete sets.
To start we write the partition function as a trace in the phase
representation
\begin{equation}
\label{z-in-phases}
       Z = \prod_{\vec r} \int_{0}^{2\pi} d\phi(0,\vec r)
           <\{\phi(0,\vec r)\}| \exp\left\{-\beta \hat {\cal H}\right\}|
           \{\phi(0,\vec r)\}>.
\end{equation}
As usual we use the Trotter formula
\begin{eqnarray}
\label{trotter}
\exp\left\{-\beta(\hat H_C(\hat n)+\hat H_J(\hat \phi))\right\} =
& &\left[\exp\{-(\beta/L_{\tau}) \hat H_C(\hat n)\}
\exp\{-(\beta/L_{\tau})\hat H_J(\hat \phi)\}\right]^{L_{\tau}} \nonumber\\
& & + O\left(1/L_{\tau}^2\right).
\end{eqnarray}
Next we introduce $L_\tau-1$ complete sets $\{|\phi(\tau,\vec r)>\}$,
$\tau=1,2,\dots,L_\tau-1$ in Eq. (\ref{z-in-phases}) so that 
\begin{eqnarray}
\label{z-with-trotter}
      Z = \prod_{\vec r} \prod_{\tau=0}^{L_\tau-1} \int_{0}^{2\pi}
               d\phi(\tau,\vec r) 
    & & <\{\phi(0,\vec r)\}|\exp\left\{-(\beta/L_\tau){\hat{\cal H}}
               \right\}|\{\phi(1,\vec r)\}> \times \nonumber \\ 
            & &\times <\{\phi(1,\vec r)\}|\exp\left\{-(\beta/L_\tau)
              {\hat{\cal H}}\right\}|\{\phi(2,\vec r)\}> \times \nonumber\\
            & & \times\cdots \times <\{\phi(L_\tau-1,\vec r)\}|\exp\left
               \{-(\beta/L_\tau){\hat{\cal H}}\right\}|\{\phi(0,\vec r)\}> 
               \nonumber\\
            & & + O\left(1/L_{\tau}^2\right).
\end{eqnarray}
At this point we need to calculate the short time propagator, 
\begin{eqnarray}
\label{short-time-prop}
           <\{\phi(\tau,\vec r)\}|e^{-(\beta/L_\tau){\hat{\cal H}}
            }|\{\phi(\tau+1,\vec r)\}> = 
            \sum_{n(\tau,\vec r)= -\infty}^{\infty}
            & & <\{\phi(\tau,\vec r)\}|e^{-(\beta/L_\tau)
                {\hat{\cal H}} }|\{n(\tau,\vec r)\}>\times\nonumber\\
            & & \times <\{n(\tau,\vec r)\}|\{\phi(\tau+1,\vec r)\}>,
\end{eqnarray}
where we used a summation over the complete set $|\{n(\tau,\vec r)\}>$.
From Eqs. (\ref{charge-phase}) and (\ref{trotter}) this propagator can 
be written as
\begin{eqnarray}
\label{short-time-prop-2}
     <\{\phi(\tau,\vec r)\}|e^{-(\beta/L_\tau){\hat{\cal H}}
            }|\{\phi(\tau+1,\vec r)\}> = \prod_{\vec r}\frac{1}{2\pi}
            \sum_{n(\tau,\vec r)=-\infty}^{\infty} 
      & & e^{i\ n(\tau,\vec r)[\phi(\tau+1,\vec r)-\phi(\tau,\vec r)]}
          \times \nonumber\\
      & & e^{-(\beta/L_\tau){\cal H}(\{n(\tau,\vec r)\},\{\phi(\tau,
          \vec r)\}) } + \nonumber\\
      & & + O(1/L_\tau^2).
\end{eqnarray}
Inserting this equation in Eq. (\ref{z-with-trotter}) we obtain the 
following  path integral representation of the partition function
\begin{eqnarray}
\label{z-path-integral}
         Z = & & \
                 \prod_{\tau=0}^{L_{\tau}-1} \prod_{\vec r} \int_{0}^{2\pi}
                 \frac{d\phi(\tau,\vec r)}{2\pi} \sum_{\{ n(\tau,\vec r)\}=
                  -\infty}^{\infty}\exp\left[i\sum_{\tau=0}^{L_\tau-1}
                   n(\tau,\vec r)[\phi(\tau+1,\vec r)-\phi(\tau,\vec r)]
                   \right] \times \nonumber \\                   
                   & &\times \exp\left[-\frac{\beta}{L_\tau}\sum_{\tau=0}
                      ^{L_\tau-1}\bigg[H_J(\{\phi(\tau,\vec r)\}) + 
                      \sum_{\vec r_1,\vec r_2}\frac{q^2}{2} n(\tau,\vec r_1)
                      {\bf C}^{-1}(\vec r_1,\vec r_2)n(\tau,\vec r_2)\bigg]
                      \right] + \nonumber \\
                      & &  + O(1/L_\tau^2).\nonumber\\
\end{eqnarray}
together with the important  boundary condition 
$\phi(L_\tau,\vec r)=\phi(0,\vec r)$.
These equations are our starting point 
for the semiclassical approximation analysis discussed in the next section.

%
%

\section{\bf WKB and renormalization group equations}

\subsection{Semiclassical limit}

\label{sec:wkb-rg}
The semiclassical limit corresponds to taking
$q^2\rightarrow 0$, or $\alpha_{\rm m}\rightarrow 0$.
The summations over $\{n(\tau,\vec r)\}$  in Eq. (\ref{z-path-integral})
can be carried out and the result leads to
$\phi(\tau+1,\vec r)=\phi(\tau,\vec r)$, for $\tau=0,1,\dots,L_\tau-1$. In
other words, in this limit all the phase variables are constant along the 
imaginary time axis, and we recover the classical 2-D XY model 
\cite{berezinskii-kosterlitz-thouless,jose-kadanoff-nelson-kirpatrick}. 
As the charging energy increases the value of $\phi(\tau,\vec r)$ 
fluctuates along the $\tau$-axis; these fluctuations suppress the XY phase 
coherence in the model lowering its critical temperature. 
For the self-capacitive model ($C_{\rm m}=0$), at $T=0$, one can
map the model to an anisotropic three-dimensional XY model 
\cite{doniach,zwerger}.
This model should have a transition between ordered and disorder phases at a critical coupling 
$(E_{C_{\rm s}}/E_J)_c$.  Here $E_{C_{\rm s}}=e^2/2C_{\rm s}$, so we would
expect the phase boundary to go all the way down to $T=0$ for large 
enough charging energy. 

In this section we study the change of the critical temperature
as $E_{C_{\rm m}}$ increases, for small values of the ratio
$\alpha_{\rm m}=E_{C_{\rm m}}/E_J$. We start by eliminating
the $\{n's\}$  the from Eq.(\ref{z-path-integral}) using the 
Poisson summation formula
\begin{equation}
\label{poisson}
      \sum_{n=-\infty}^{\infty} f(n) = \sum_{m=-\infty}^{\infty}
      \int_{-\infty}^{\infty} f(x) e^{2\pi imx} dx,
\end{equation}
obtaining
\begin{equation}
\label{z-with-m}
Z = \prod_{\tau=0}^{L_{\tau}-1} \sqrt{{\rm det}[ C ] }
    \prod_{\vec r} \int_{0}^{2\pi}\!\!\sqrt{\frac{L_{\tau}}
    {2\pi\beta q^2}}  d\phi(\vec r,\tau)\!\!\!\!\sum_{\{ m(\vec r,\tau)\}
     = -\infty}^{\infty}\!\!\!\!\exp\bigg[-\frac{1}{\hbar}S[\{\phi\},\{m\}]
     \bigg].
\end{equation}
Here we defined the action
\begin{eqnarray}
\label{action}
\frac{1}{\hbar}S[\{\phi\},\{m\}] = & & 
     \sum_{\tau = 0 }^{L_{\tau}-1} \Bigg[\frac{ \beta}{L_{\tau}} 
     H_J(\{\phi(\tau,\vec r)\})
     +\frac{L_{\tau}}{2\beta q^2}\!\!\sum_{\vec r_1,\vec r_2}
       [\phi(\tau\!+\!1,\vec r_1) - \phi(\tau,\vec r_1) 
       + 2\pi m(\tau,\vec r_1) ] \times \nonumber \\
       & &\times{\bf C}(\vec r_1,\vec r_2)[\phi(\tau\! +\! 1,\vec r_2)
       -\phi(\tau,\vec r_2) + 2\pi m(\tau,\vec r_2)]\Bigg] + 
               \nonumber \\
               & &+ O(1/L_{\tau}^2).
\end{eqnarray}
It is convenient to write the paths in the  partition function
separated into a constant part, that  
corresponds to the classical model, plus a quantum fluctuating 
contribution, over which we will perform the integrations to find 
an effective classical action. First we eliminate the summations 
over the $\{m's\}$.  This is done at the same time that the integrals over the 
$\{\phi's\}$ are extended from $[0,2\pi)$ to $(-\infty,\infty)$. 
After a couple of standard variable changes \cite{schulman-book,kleinert-book}
we get an action where the phases and the charges are separated,
\begin{eqnarray}
\label{new-action}
       \frac{1}{\hbar}\int_{0}^{\beta\hbar} & &  d\tau L_E 
       = \frac{1}{2}
       \frac{(2\pi)^2}{\beta q^2}\sum_{\vec r_1,\vec r_2} m(\vec r_1)
        {\bf C}
       (\vec r_1,\vec r_2) m(\vec r_2) + \frac{1}{\hbar}\int_{0}^{\beta\hbar}
       d\tau \times \nonumber \\ 
       & &\times\Bigg[ \frac{\hbar^2}{2q^2} \sum_{\vec r_1,\vec r_2}
          \frac{d\psi}{d\tau}(\tau,\vec r_1){\bf C}(\vec r_1,\vec r_2)
          \frac{d\psi}{d\tau}(\tau,\vec r_2) + H_J(\{\psi(\tau,\vec r)+
          (2\pi/\beta\hbar)m(\vec r)\tau\})\Bigg].\nonumber\\
\end{eqnarray}
Here the variables $\psi(\beta\hbar,\vec r) = \psi(0,\vec r)$, and
the integers $m(\vec r)$ are called the winding numbers.
This equation shows that the winding numbers are the charge degrees of
freedom and that the coupling between phases and charges appears only
in the Josephson term.  We can also see from this equation that in the 
semi-classical limit (small charging energy) the charge fluctuations are 
exponentially suppressed. This is more so for the $m's$ because they have
a discrete excitation spectrum. Therefore, to lowest order in the 
semiclassical analysis we will set $m(\vec r)=0$, leaving integrals 
only over the phases. Next we separate the $\psi's$ into a constant 
plus a fluctuating part
\begin{equation}
\label{constant-fluctuating}
       \psi(\tau,\vec r) = \overline\phi(\vec r) + \phi_f(\tau,\vec r).
\end{equation}
At this point we use the following argument 
\cite{jose-kadanoff-nelson-kirpatrick,m-fisher}. First 
that the Lagrangian is invariant under the transformation 
$\psi(0,\vec r)\rightarrow\psi(0,\vec r)+2\pi l(\vec r)$ for all integers
$l(\vec r)$, so that we can extend the limits of 
integration over $\psi(0,\vec r)$ to $(-\infty,\infty)$, 
safe for an extra overall multiplicative constant.
Now, the limits of integration  for $\overline\phi(\vec r) \epsilon
(-\infty,\infty)$, and because of the periodicity of $\phi_f$, we can 
Fourier series expand it as
\begin{equation}
\label{fourier-series}
      \phi_f(\tau,\vec r) = (\beta\hbar)^{-1/2}\sum_{k=1}^{\infty}
      [\phi_k(\vec r)e^{i\omega_k\tau}+C.C.],
\end{equation}
where the $\omega_k=2\pi k/\beta\hbar$ are the Bose-Matsubara frequencies.
We have then the partition function \cite{kleinert-book}
\begin{eqnarray}
\label{z-fourier}
     Z = \prod_{\vec r}\sqrt{{\rm det}{[\bf C]}}\int_{-\infty}^{\infty} 
         \frac{d\overline\phi(\vec r)}{(2\pi\beta q^2)^{1/2}} 
         \prod_{k=1}^{\infty}& &\Bigg[\frac{\omega_k^2\hbar}{\pi q^2} 
            {\rm det}{[\bf C]} \int_{-\infty}^{\infty}d{\rm Re}
            \phi_k(\vec r)\int_{-\infty}^{\infty}d{\rm Im}
            \phi_k(\vec r)\Bigg] \times \nonumber \\
         & &\times \exp\left\{-\frac{1}{\hbar}S\big[\{\overline\phi\},
            \{\phi_f\}\big]\right\}.
\end{eqnarray}
Next, we expand the Josephson term in the action up to 
second order in $\phi_f$, for higher order terms are 
suppressed in the integrations. After performing the 
integrations over the Euclidean time $\tau$, and the Gaussian integrations,
the effective partition function reads
\begin{eqnarray}
\label{effective-z}
     Z_{\rm eff} = \prod_{\vec r} \sqrt{{\rm det}{[\bf C]}}\int_{-\infty}
                   ^{\infty} & &
                   \frac{d\overline\phi(\vec r)}{(2\pi\beta q^2)^{1/2}} 
                   \ \ \exp\left\{-\beta H_J(\{\overline\phi\})\right\}
                   \times \nonumber \\
                   & &\times \prod_{k=1}^{\infty} \left[ {\rm det}\left\{
                      \delta_{\vec r_1,\vec r_2} + \frac{q^2}{\hbar^2
                      \omega_k^2}\sum_{\vec r}{\bf C}^{-1}(\vec r_1,\vec r)
                      \frac{\partial^2H_J}{\partial\phi(\vec r)\partial\phi
                      (\vec r_2)}\Bigg|_{\overline\phi}\right\}\right]^{-1}.
                   \nonumber\\
\end{eqnarray}
Here we want to expand this partition function in powers of the charging
energy. This is equivalent to expanding in powers of $q^2$, so our next 
step is to expand the determinant. We use the following identities
\begin{eqnarray}
\label{det-equations}
      {\rm det}[{\bf I}+{\bf D}]&=& \exp\{{\rm Tr}[{\rm ln}({\bf I}+{\bf 
D})
                                  ]\}, \\
      {\rm ln}({\bf I}+{\bf D}) &=& -\sum_{n=1}^{\infty} \frac{(-1)^n}{n}
                                     {\bf D}^n,
\end{eqnarray}
where ${\bf I}$ is the identity matrix. To lowest order in $q^2$
and using the result 
$\sum_{k=1}^{\infty}[q^2/(\hbar\omega_k)^2] = (q\beta)^2/24$, the 
effective partition function for $\overline\phi$ is then
\begin{eqnarray}
\label{z_eff-phi}
      Z_{\rm eff} = \prod_{\vec r} \sqrt{{\rm det}{[\bf C]}}
     \int_{-\infty}^{\infty}
     & & 
     \frac{d\overline\phi(\vec r)}{(2\pi\beta q^2)^{1/2}} \exp\Bigg\{
     -\beta H_J(\{\overline\phi\})\nonumber\\
     & & 
     -\frac{(q\beta)^2}
     {24}\sum_{\vec r_1,\vec r_2}{\bf C}^{-1}(\vec r_1,\vec r_2)\frac
     {\partial H_J}{\partial\phi(\vec r_1)\partial\phi(\vec r_2)}\Bigg|
     _{\overline\phi}\Bigg\}.
\end{eqnarray}

To further advance the calculation we now use the properties of the 
Josephson energy. We start by using the fact that it is a 
local nearest neighbor interaction, so from Eq. (\ref{hamiltonian})
\begin{equation}
\label{josephson-energy}
        H_J(\{\phi\}) = \sum_{\vec r}\sum_{\vec d} f\!\left(\phi
                        (\vec r+\vec d)-\phi(\vec r)\right),
\end{equation}
with the $\vec d$ running over the nearest neighbors to 
$\vec r$ in the lattice. From this equation we can see that the 
second derivative of $H_J(\{\phi\})$ is given by
\begin{eqnarray}
\label{secon-derivative_hj}
      \frac{\partial^2 H_J}{\partial\phi(\vec r_1)\partial\phi(\vec r_2)} =
      \sum_{\vec d}\Bigg[ 
       & &
      f''\!\left(\phi(\vec r_1)-\phi(\vec r_1-\vec d)
      \right)\left(\delta_{\vec r_1,\vec r_2}-\delta_{\vec r_1,\vec r_2+
      \vec d}\right) + \nonumber \\
      & & + f''\!\left(\phi(\vec r_1+\vec d)-\phi(\vec r_1)\right)\left(
            \delta_{\vec r_1,\vec r_2}-\delta_{\vec r_1+\vec d,\vec r_2}
            \right)\Bigg],
\end{eqnarray}
where $f''\!(x)=d^2f(x)/dx^2$.

In this paper we consider a periodic array, which implies that 
the inverse capacitance matrix is invariant under translations
and rotations, that is ${\bf C}^{-1}(\vec r_1,\vec r_2)={\bf C}^{-1}
(|\vec r_1-\vec r_2|)$. In particular, this makes ${\bf C}^{-1}(\vec r,
\vec r\pm\vec d) = {\bf C}^{-1}(|\vec d|)$, independent of the direction 
of $\vec d$. Notice that here we are using $\vec d$ to denote the 
vectors that connect nearest neighboring islands, therefore in a 
periodic and symmetric array all of them have the same magnitude,
allowing us to take the terms ${\bf C}^{-1}(|\vec d|)$ out of the 
summations. From these considerations the trace gives
\begin{equation}
\label{trace}
    \sum_{\vec r_1,\vec r_2}{\bf C}^{-1}(\vec r_1,\vec r_2)\frac
     {\partial^2 H_J}{\partial\phi(\vec r_1)\partial\phi(\vec r_2)}\Bigg|
     _{\overline\phi} = 2\left[{\bf C}^{-1}(|\vec 0|)-{\bf C}^{-1}
     (|\vec d|)\right]\sum_{\vec r}\sum_{\vec d} f''\!\left(\overline
     \phi(\vec r+\vec d)-\overline\phi(\vec r)\right).
\end{equation}
Next notice that since $f''\!(x) =-f(x)$ (up to a constant), both terms in
the argument of the exponential in Eq. (\ref{z_eff-phi}) are the same 
cosine function of the classical phase variables, with only different coupling 
constants. Finally, the effective semi-classical partition function can 
be written as
\begin{equation}
\label{final-z_eff}
     Z_{\rm eff} = \prod_{\vec r} \sqrt{{\rm det}{[\bf C]}} \int_{-\infty}
                   ^{\infty}\frac{d\overline\phi(\vec r)}{(2\pi\beta q^2)
                   ^{1/2}}\exp\left\{-\beta_{\rm eff} 
H_J(\{\overline\phi\})
                    \right\},
\end{equation}
where the effective temperature is explicitly given by
\begin{equation}
\label{beta_eff}
        \beta_{\rm eff} = \beta - q^2 \frac{\beta^2}{12}\left[{\bf C}^{-1}
                          (|\vec 0|)-{\bf C}^{-1}(|\vec d|)\right].
\end{equation}
Notice that to obtain this result we have used an argument that
could be questionable, namely the extension of the $\phi(0,\vec r)$ 
to the $(-\infty,\infty)$ range in the path integrals. All 
the other approximations are consistent with the semiclassical 
approximation 
and the symmetries used are exact. 
However, to continue we will now restore the $[0,2\pi)$ range of the
phases to use the results known from the BKT theory.
As we will show later in the paper, 
the nonperturbative QMC results do agree quantitatively with the WKB 
results and experimental results to be discussed later.  A similar 
effective result was first obtained but for the self-capacitive model in 
\cite{jose_1984}.

One of the important properties of Eqs.(\ref{final-z_eff}) and 
(\ref{beta_eff}) is that up to this point we have made no 
assumptions about the structure of the capacitance matrix that go 
beyond translational invariance.  Later on we will make specific 
choices of this matrix when we make direct contact with experimental 
findings \cite{van-der-zant_geerling_mooij-92}.


\subsection{Renormalization group analysis}
\label{subsec:rg}

Now that we have expressed the quantum mechanical problem as a 
modified 2-D classical XY model we can directly apply the well known 
results for this model 
\cite{berezinskii-kosterlitz-thouless,jose-kadanoff-nelson-kirpatrick}.
The standard physical picture of the excitation spectrum in this model is 
of spin-waves plus vortex pair excitations. At low temperatures the energy to
create an isolated vortex grows logarithmically with the size of the
system, therefore excitations are created as bounded vortex-antivortex 
pairs. As the temperature increases, the vortex pair density increases
until they unbind at a critical dimensionless temperature 
$T_{\rm BKT} = 0.894(5)$
\cite{gupta-baillie,janke-nather}. The BKT scenario is best understood 
in terms of a renormalization group (RG) analysis 
\cite{berezinskii-kosterlitz-thouless,jose-kadanoff-nelson-kirpatrick}.
The RG flow diagram is obtained from a perturbative expansion in powers
of the vortex pair fugacity $y$. To lowest order in $y$, the RG 
equations corresponding to our problem are
\begin{eqnarray}
\label{rg-eq-1}
      \frac{dK_{\rm eff}}{dl} &=& -4\pi^3 K_{\rm eff}^2 y^2, \\
\label{rg-eq-2}
      \frac{dy}{dl} &=& [2-\pi K_{\rm eff}]y.
\end{eqnarray}
Here we have used the following definitions:
\begin{eqnarray}
\label{rg-definition-k_eff}
	K_{\rm eff} &=& K - xK^2, \\
\label{rg-definition-x}
        x &=& \frac{q^2}{12E_J}\left[{\bf C}^{-1}(|\vec 0|)-{\bf C}^{-1}
              (|\vec d|)\right],\\
\label{rg-definition-k}
        K &=& \beta E_J.
\end{eqnarray}
Then the equations for the coupling constants $K$ and $y$ are
\begin{eqnarray}
\label{rg-effective-1}
	\frac{dK}{dl} &=& 4\pi^3 K^2 y^2 \frac{(1-xK)^2}{(2Kx-1)}, \\
\label{rg-effective-2}
        \frac{dy}{dl} &=& \left[2-\pi K(1-xK)\right]y.
\end{eqnarray}
To find the critical temperature, we use the initial conditions
from the temperature and the bare vortex pair fugacity 
\begin{eqnarray}
\label{K-initial}
       K_{\rm eff}(l=0) &=& \beta_{\rm eff} E_J\left[1+\frac{1}
                            {2\beta_{\rm eff}E_J}
                            +O\left\{\frac{1}{(\beta_{\rm eff}E_J)^2}
                             \right\}\right]^{-1}, \\
\label{y-initial}
       y(l=0) &=& \exp\left\{-\frac{\pi^2}{2} K_{\rm eff}(l=0)\right\}.
\end{eqnarray}
The RG equations have two nontrivial fixed points (for $x<\pi/8$). One
corresponds to the effective BKT thermal fluctuations driven 
transition, and the other to a quantum fluctuations induced transition
(QUIT) \cite{jose_1984,jacobs-jose-novotny-goldman,jacobs-jose}. 

One way to analyze the structure of the RG flow in the $(y,K)$ 
phase space is to use a conserved quantity associated with 
Eqs. (\ref{rg-eq-1}) and (\ref{rg-eq-2})
\begin{equation}
\label{conserved-eff}
       A = -\pi {\rm ln}K_{\rm eff}-\frac{2}{K_{\rm eff}}+2\pi^3 y^2.
\end{equation}
Using Eq. (\ref{rg-definition-x}) and expanding up to first order in
$x$ we find
\begin{equation}
\label{conserved}
      A = \pi x K - \pi {\rm ln}K-\frac{2}{K}+2\pi^3 y^2.
\end{equation}

Figure \ref{fig:fig02-rg-flow} shows the RG flows obtained from
numerically solving the RG equations for different values of $A$, 
where 
the arrows indicate the direction of increasing $l$. We have also plotted 
the set of initial conditions from Eqs.(\ref{K-initial}) 
and (\ref{y-initial}) as a discontinuous line. One important flow line is the 
separatrix between the lines for which $y(l\!\rightarrow\!\infty)
\rightarrow\infty$ and those for $y(l\!\rightarrow\!\infty)
\rightarrow 0$. This line has $A_c=-\pi[1+{\rm ln}(2/\pi)]$, which is
determined by the condition that it must pass through the point $(y=0,
K_{\rm eff} = 2/\pi)$.  The critical temperature is obtained from the 
intersection of the separatrix with the initial 
conditions given in Eqs. (\ref{K-initial}) and (\ref{y-initial}). 
This intersection exists only if $x$ is less than the critical value
$x_c<\!\pi/8$, to be estimated below. Fortunately, we do not need to 
find this intersection explicitly since we already
know the critical value of the effective coupling $K_{\rm eff}$,
which is the usual critical coupling of the classical XY model,
$K_{\rm eff}^{(c)}=K_{c}^{XY}\approx 1.1186$ \cite{gupta-baillie}. 
Therefore, the values of the two critical couplings are
\begin{eqnarray}
\label{eq-critical-temps}
           K(1-xK) &=& K_{c}^{XY},\\
\label{both-critical-temps}
           K_{\pm} &=&\frac{1}{2x}\left[1\pm \sqrt{1-4xK_{c}^{XY}}\right],
\end{eqnarray}
which leads to $x_c=1/(4K_{c}^{XY})\approx 0.2235$.
   
We note from Fig. 2 that the $K^{-1}$ axis 
can be divided into three different regions. If we set $K_{+}=K_{\rm QUIT}$,
and $K_{-}=K_{\rm BKT}$, then in the region  
$[K_{\rm QUIT}^{-1},K_{\rm BKT}^{-1}]$, as $l$ increases, 
the fugacity of the vortex-antivortex pairs decreases.  
In the limit $l\!\rightarrow
\!\infty$ the energy to create a macroscopic vortex pair becomes 
infinite. Therefore,
the system is superconducting for temperatures in this interval. For
temperatures $T>E_J K_{BKT}^{-1}$ the renormalized fugacity $y(l)$ 
increases 
and the low $y$ approximation breaks down. For these temperatures, the
array is normal. For $T<E_J K_{QUIT}^{-1}$, 
the vortex pair density increases due to the quantum fluctuations, 
leading us to think that there may be  
a low temperature transition driven by the quantum fluctuations (QUIT).

To obtain the results described above we used a high temperature
perturbative calculation. Therefore the QUIT results are in principle 
outside the regimen of validity of the WKB-RG calculation. 
We need, then, other calculations and approaches valid at low 
temperatures to prove or
disprove the existence of the QUIT. Expanding Eq. (\ref{both-critical-temps}),
up to first order in $x$ and using Eq. (\ref{rg-definition-k}) we find 
the critical temperatures,
\begin{eqnarray}
\label{T_BKT}
            T_{BKT} &\approx& T_{BKT}^{(0)} - \frac{E_J}{k_{\rm B}} x + 
                              O(x^2), \\
\label{T_QUIT}
            T_{QUIT} &\approx& \frac{E_J}{k_{\rm B}} x + O(x^2).
\end{eqnarray}

Note that these equations are applicable not only in 2-D, for if
the system described by Eq. (\ref{final-z_eff}) has a transition point at
some $K_{\rm eff}^{c}$ then the equation $K_{\rm eff}^{c} = K-xK^2$ has 
two solutions for $K$. In this argument we should notice that the 
existence 
of the second solution for $K$ depends on the higher order terms in the 
$x$ expansion. Note that the change of $T_{BKT}$ for small $x$ is 
correctly given by the small $x$ result. The existence of a low 
temperature quantum phase is of a nonperturbative nature, however.  
That is one of the reasons why we resort to using the nonperturbative 
quantum QMC approach latter in the paper.

Another interesting property of Eqs. (\ref{T_BKT}) and (\ref{T_QUIT})
is that the first order correction does not depend on the specific 
value of $T_{BKT}^{(0)}$. In particular, if we 
add a magnetic field to the Hamiltonian in Eq. (\ref{hamiltonian}),
all the calculations leading to Eqs. (\ref{final-z_eff}) and 
(\ref{beta_eff}) would be unchanged. Therefore, if $T_{c}^{(0)}(B)$ is the 
superconducting to normal transition temperature for the array in
a finite magnetic field $B$ at $x=0$, then to first order in $x$ we must have
\begin{equation}
\label{tc-with-mag-field}
           T_{c}(B)\approx T_{c}^{(0)}(B)-(E_J/k_{\rm B})x+O(x^2).
\end{equation}
This equation is in agreement with the results obtained in 
Ref. \cite{jacobs-jose-novotny-goldman}.
Furthermore, we notice that to lowest order in $x$, the $T_{QUIT}$ 
must be the same with then without a magnetic field.  This result 
was not noted before and it can be used as a test of the QMC calculations,
in particular those of Mikalopas et al. 
\cite{mikalopas-jarrel-pinski-chung-novotny}.

To make comparisons with experiment, we need to specify the 
capacitance matrix.  In particular, if we use Eq. 
(\ref{capacitance-matrix}) and the specific geometry of the array, 
we find that $x$ is given by
\begin{eqnarray}
\label{x}
          x &=& \frac{q^2}{12 z E_J C_{\rm m}}\left[1-
                C_{\rm s}{\bf C}^{-1}(|\vec 0|)\right],\nonumber\\
            &=& \frac{q^2}{12 z E_J C_{\rm s}} g(C_{\rm m}/C_{\rm s}).
\end{eqnarray} 
The function $g(w)$ can be written as an elliptic integral for a 
two-dimensional square lattice \cite{kleinert-book-2}. For a general lattice
geometry, we find the following limiting behavior
\begin{equation}
\label{limits-g}
      g(w) \approx \left\{\begin{array}{ll}
                          z-z(1+z)w,   & \mbox{if $w \ll 1$,} \\ 
                          w^{-1}\left\{1-(4\pi w)^{-1}\ln w\right\},
                                   & \mbox{if $w \gg 1$.}
                          \end{array}
                   \right. 
\end{equation}
Using Eqs. (\ref{T_BKT}) and (\ref{limits-g}) we get
\begin{equation}
\label{limits-T_BKT}
      \frac{k_{\rm B} T_{BKT}}{E_J} \approx \frac{k_{\rm B} T_{BKT}^{(0)}}
      {E_J} - \left\{ \begin{array}{ll}
                          (2/3)\alpha_{\rm s} + O(\alpha_{\rm s}^2),
                               & \mbox{if $C_{\rm s}\gg C_{\rm m}$,} \\
                               & \\
                          (2/3z)\alpha_{\rm m} + O(\alpha_{\rm m}^2),
                               & \mbox{if $C_{\rm s}\ll C_{\rm m}$.}
                      \end{array}
               \right.
\end{equation}
This result is in agreement with the Monte Carlo calculation of the
superconducting to normal transition temperature carried out in 
Ref. \cite{jacobs-jose-novotny-goldman} for the self-capacitive model.
As we will show later in this paper, it is also in good agreement with 
our QMC calculations for the model dominated by the mutual capacitances.
%


\section{\bf Insulating to Normal Cross Over.}
\label{sec:insulating-normal}
So far we have studied the normal to superconducting transition
in the limit where the Josephson energy dominates over the charging energy.
In the opposite limit, when the relevant excitations are charge
fluctuations, the transition is expected to be from a normal conducting state, where the
charges are free to move, to an insulating state where the charges are 
bound into neutral dipole pairs (see Fig. \ref{fig:fig01-phase-diagram}). It 
has been suggested that in the limit $(C_{s}/C_{\rm m}) \rightarrow 0$ 
this I-N transition would be of a BKT type   
\cite{yaks-87,widom-badjou,mooij_van-wees_geerligs_peters_fazio_schon,fazio-schon-91}.
Experimental results have shown, however, that the behavior of the 
fabricated samples is better explained by a crossover from a normal 
to an insulating phase
\cite{tighe-tuominen-hergenrother-tinkham,van-der-zant-thesis,delsing-chen-haviland-harada-claeson}.
In finite systems, like the experimental ones, we would expect 
a rounding of the transition. Furthermore, in the experiments, the 
screening length is shorter than the sample size
$\Lambda\sim\sqrt{C_{\rm m}/C_{\rm s}}\approx 18$ lattice spacings
\cite{van-der-zant_geerling_mooij-92}.
Minnhagen et al. have argued that for any finite screening length, the
transition is washed out even for an infinite array 
\cite{minnhagen-olsson-xu}. 
This is not difficult to understand since in the BKT scenario the 
superconducting to normal transition depends on the unscreened nature of 
the vortex logarithmic interaction 
\cite{berezinskii-kosterlitz-thouless,jose-kadanoff-nelson-kirpatrick}.

In this section we present results from a perturbative calculation 
of the effect of the Josephson energy on the expected I-N crossover 
temperature. We start with Eqs. (\ref{new-action})
and (\ref{constant-fluctuating}) leading to
\begin{eqnarray}
\label{z-constant-fluctuating}
           Z = \sqrt{{\rm det}{[\bf C]}} \int_{0}^{2\pi} 
               \prod_{\vec r}
               \sqrt{\frac{L_{\tau}}{2\pi \beta q^2}} d\overline\phi(\vec r)
               \!\sum_{m(\vec r )=-\infty }^{\infty} \int_{-\infty}^
               {\infty}                
               & &\prod_{\tau = 1}^{L_{\tau}-1}\!\sqrt{{\rm det}
               {[\bf C]}}
               \prod_{\vec r} \sqrt{\frac{L_{\tau}}{2\pi \beta q^2}}
               d\phi_f(\tau,\vec r) \times \nonumber \\
               & &\times\exp\bigg[-\frac{1}{\hbar}\int_{0}^{\beta\hbar} 
               d\tau L_E \bigg],
\end{eqnarray}
where the action is now given by
\begin{eqnarray}
\label{action-m-over_phi-phi_f}
       \frac{1}{\hbar}\int_{0}^{\beta\hbar} d\tau L_E = \frac{1}{2}
       \frac{(2\pi)^2}{\beta q^2}\sum_{\vec r_1,\vec r_2} 
        & &
       m(\vec r_1){\bf C}
       (\vec r_1,\vec r_2) m(\vec r_2) + \frac{1}{\hbar}\int_{0}^{\beta\hbar}
       d\tau \times \nonumber \\
       & & \times \Bigg[ \frac{\hbar^2}{2q^2} \sum_{\vec r_1,\vec r_2}
          \frac{d\phi_f}{d\tau}(\tau,\vec r_1){\bf C}(\vec r_1,\vec r_2)
          \frac{d\phi_f}{d\tau}(\tau,\vec r_2) + \nonumber \\
       & & \ \ \ \ \ \ + H_J(\{\overline\phi(\vec r)+\phi_f(\tau,\vec r)+
                         (2\pi/\beta\hbar)m(\vec r)\tau\})\Bigg],
       \nonumber\\
\end{eqnarray}
with the boundary condition
\begin{equation}
\label{boundary-for-phi_f}
           \phi_f(0,\vec r) = \phi_f(\beta\hbar,\vec r) = 0.
\end{equation}

Since we are interested in the charge degrees of freedom, our task here
is to integrate out the phases. This limit has been studied before,
in particular in Ref. \cite{fazio-schon-91}. Here we are not only 
interested in the crossover temperature, but we mostly want 
to ascertain if there 
is an equivalent QUIT in the insulating phase at low temperatures.

Since we are at the limit $E_J\ll E_C$, the Josephson energy can 
be treated as perturbation.  We expand the exponential
\begin{equation}
\label{h_j-expantion}
      \exp\left[-\frac{1}{\hbar}\int_{0}^{\beta\hbar}d\tau H_J(\tau)\right]
      \approx 1-\frac{1}{\hbar}\int_{0}^{\beta\hbar}d\tau H_J(\tau)+
      \frac{1}{2\hbar^2}\int_{0}^{\beta\hbar}\int_{0}^{\beta\hbar}
      d\tau d\tau' H_J(\tau)H_J(\tau')+\dots 
\end{equation}
We note that Eq. (\ref{z-constant-fluctuating}) can be
written as
\begin{equation}
\label{z-with-z_eff}
       Z = Z_{\phi} \prod_{\vec r} \sum_{m(\vec r)=-\infty}^{\infty}
           \exp\Bigg[ -\frac{(2\pi)^2}{2\beta q^2}\sum_{\vec r_1,\vec r_2}
                       m(\vec r_1){\bf C}(\vec r_1,\vec r_2) m(\vec r_2)
               \Bigg] Z_{\rm eff}(\{m\}),
\end{equation}
where $Z_{\phi}$ contains only phase degrees of freedom and can formally 
be written as
\begin{eqnarray}
\label{z_phi}
      Z_{\phi} &=& \prod_{\vec r} \int_{-\infty}^{\infty} {\cal D}
                   \phi_f(\vec r)
	           \exp\Bigg[-\frac{1}{\hbar}S_f[\phi_f]\Bigg], \\
\label{s_f}
         S_f[\phi_f] &=& -\frac{\hbar^2}{2q^2}\int_{0}^{\beta\hbar}
                            d\tau \sum_{\vec r_1,\vec r_2} \frac{d\phi_f}
                            {d\tau}(\tau,\vec r_1){\bf C}(\vec r_1,\vec r_2)
                            \frac{d\phi_f}{d\tau}(\tau,\vec r_2).
\end{eqnarray}
Here we have used the following short hand notation for the measure
\begin{equation}
\label{path-integral-measure}
          {\cal D} 
	  \phi_f(\vec r) = \lim_{L_\tau\rightarrow\infty}
          \prod_{\tau = 1}^{L_{\tau}-1}
          \!\!\sqrt{{\rm det}{[\bf C]}} \prod_{\vec r} 
          \sqrt{\frac{L_{\tau}}{2\pi \beta q^2}} d\phi_f(\tau,\vec r),
\end{equation}
noting that strictly speaking the integrals over a finite number of
$L_\tau$'s have to be calculated before the limit $L_\tau\rightarrow\infty$
is taken \cite{feynman-stat}.

All the interactions between phases and charges are contained in 
$Z_{\rm eff}(\{m\})$, the effective partition function for the 
charges. The details of the explicit evaluations 
of $Z_{\rm eff}$ are given in Appendix A. The result for
Eq. (\ref{z-with-z_eff}) can then be written, up to second order in
$E_J$, as
\begin{eqnarray}
\label{z-with-z_eff-2}
     Z = Z_{\phi} \prod_{\vec r}
          & &
          \sum_{m(\vec r)=-\infty}^{\infty}           
           \exp\!\Bigg[ -\frac{1}{2\tilde K}\sum_{<\vec r_1,\vec r_2>}
                       (m(\vec r_1)-m(\vec r_2))^2 - \frac{1}{2\tilde K}\left(
                       \frac{C_{\rm s}}{C_{\rm m}}\right)\sum_{\vec r} 
                        m(\vec r)^2 +  \nonumber \\
          & & + \frac{\tilde K^2}{2} \left(\frac{(2\pi)^2 E_J 
                              C_{\rm m}}{q^2}\right)^2\!\!\!
                \sum_{<\vec r_1,\vec r_2>} 
                {\cal I}\Big(m(\vec r_1)-m(\vec r_2),\tilde K
                [1-C_{\rm s}{\bf C}^{-1} (|\vec 0|)]\Big)\Bigg] + 
\nonumber\\
          & & + O(E_j^4).
\end{eqnarray}
Here we have defined 
\begin{eqnarray}
\label{def-k}
       \tilde K &=& \frac{\beta q^2}{(2\pi)^2 C_{\rm m}}, \\
\label{def-i} 
       {\cal I}(m,\tilde K) &=& \int_{0}^{1/2}\!\! dx_1\ \cos(2\pi m x_1)\ 
                       \exp\!\Big[-\big\{2(2\pi)^2/z\big\}\tilde K \ 
                       x_1(1-x_1)\Big].
\end{eqnarray} 
The function ${\cal I}(m,\tilde K)$ is an even function of $m$, so it 
can be expanded in a Taylor series in $m^2$.  One way to do it is to take
$\cos(x_1)\approx1+(1/2)x_1^2-(1/24)x_1^4+\dots$.  This is a good 
approximation if the coefficient in the exponential is large. Since 
we are interested in discrete values of $m$, and considering that 
values of $m$ greater than one are suppressed even near the transition 
point, we can use the following approximation
\begin{equation}
\label{approx-i}
        {\cal I}(m,\tilde K)={\cal I}(0,\tilde K)-\Big[{\cal I}(0,\tilde K)
                             -{\cal I}(1,\tilde K)\Big]m^2.
\end{equation}
With this approximation we can write Eq. (\ref{z-with-z_eff-2}) as
\begin{equation}
\label{z-with-z_eff-3}
     Z = Z_{\phi} \prod_{\vec r} \sum_{m(\vec r)=-\infty}^{\infty}
           \exp\!\Bigg[ -\frac{1}{2\tilde K_{\rm eff}}
                        \sum_{<\vec r_1,\vec r_2>}
                       (m(\vec r_1)-m(\vec r_2))^2 - \frac{1}{2\tilde K}\left(
                       \frac{C_{\rm s}}{C_{\rm m}}\right)\sum_{\vec r} 
                        m(\vec r)^2 \Bigg] + O(E_j^4).
\end{equation}
The effective coupling constant is given by
\begin{eqnarray}
\label{k-eff}
     \tilde K_{\rm eff} &=& \tilde K \Bigg[1 + \left(\frac{(2\pi)^2 E_J 
                            C_{\rm m}} {q^2}\right)^2 h\Big(\tilde K
                            [1-C_{\rm s}{\bf C}^{-1}(|\vec 0|)]\Big)
                            \Bigg]^{-1}, \\
\label{def-h}
         h(w) &=& w^3 \int_{0}^{1/2} dx [1-\cos(2\pi x)] \exp\!\Big[
                  -\big\{2(2\pi)^2/z\big\}w \ x(1-x)\Big].
\end{eqnarray}
The function $h(w)$ has the following limiting asymptotic behavior
\begin{equation}
\label{limits-h}
    h(w) =\left\{ \begin{array}{ll}
                     (1/2)w^3\Big[1-w\ (1/z)(12+(2\pi)^2/3)\Big]+O(w^5),
                             & \mbox{if $w\ll 1$,} \\
                             &                     \\
                     \!\frac{(z/2)^3}{(2\pi)^4}\Big[1+\frac{12}{(2\pi)^2}
                        (z/2)w^{-1}-\frac{1}{(2\pi)^2}(z/2)^2\ w^{-2}\Big] 
                         + O(w^{-3})	,
                           & \mbox{if $w\gg 1$.}
                      \end{array}
                   \right.
\end{equation}
                   
We now use the fact that for the experimental systems
$C_{\rm s}\ll C_{\rm m}$,
so that in the limit $(C_{\rm s}/C_{\rm m})\rightarrow 0$ we can 
use
\begin{equation}
\label{limit-cs_cm}
      \lim_{(C_{\rm s}/C_{\rm m})\rightarrow 0} \Big[1-C_{\rm s}{\bf C}^{-1}
      (|\vec 0|)\Big] = \lim_{(C_{\rm s}/C_{\rm m})\rightarrow 0} \Bigg[1+
      \frac{1}{4\pi}\left(\frac{C_{\rm s}}{C_{\rm m}}\right) \ln\left(
      \frac{C_{\rm s}}{C_{\rm m}}\right) \Bigg] = 1.
\end{equation}
The end result is a discrete Gaussian model with an effective coupling
constant given by Eq. (\ref{k-eff}). This effective model can 
be transformed into a Villain model 
\cite{janke-nather,itzykson-drouffe-book}.
The critical points  of this model are given by the equation 
$\tilde K_{\rm eff} = \tilde K_{c}^{\rm V}$,
where $\tilde K_c^{\rm V}$ is the critical coupling for the Villain model,
$\tilde K_{c}^{\rm V} \approx 0.752(5)$ for a square array 
\cite{janke-nather}. In other words we have to solve the equations,
\begin{eqnarray}
\label{K-crit}
   \tilde K_c &=& \tilde K_c^{\rm V} + \Omega\ h(\tilde K_c), \\
\label{omega}
      \Omega &=& \tilde K_c^{\rm V}\ (\pi^4/4)\ (E_J/E_{C_{\rm m}})^2.
\end{eqnarray}
From these equations we find the first order correction to the crossover 
temperature for a square array
\begin{equation}
\label{tc-first-correc}
      \frac{T_c}{T_c^{(0)}} \approx 1 - 0.259\ \left(\frac{E_J}{E_{C_{\rm 
                            m}}}
                                   \right)^2 + O\Big((E_J/E_{C_{\rm }})^4
                                                \Big).
\end{equation}
Here we have used $h(\tilde K_c^{\rm V})\approx 0.0106$.

An important property of Eq. (\ref{K-crit}) is that we can show that 
it has only one solution, since the function $h(\tilde K)$ is 
concave for small $\tilde K$ and it has an inflection point at 
$\tilde K_{\rm infl}$, 
\begin{equation}
\label{inflection}
      \tilde K_{\rm infl}\approx z\ \frac{6.2}{2(2\pi)^2}.
\end{equation}
A sufficient condition for Eq. (\ref{K-crit}) to have only one solution is
$\tilde K_c^{\rm V}>\tilde K_{\rm infl}$. This condition is satisfied 
for  square as well as triangular arrays.  This result shows that there 
is no insulating QUIT phase and it is in clear 
contrast to the existence of the QUIT found using the WKB-RG approximation 
in the superconducting phase.
%

\section{\bf Quantum Monte Carlo Results}
%
%
\subsection{Definition of Physical Quantities Calculated}
\label{sec:mc-operators}
The two important physical parameters in our analysis are the temperature 
and $\alpha=\frac{E_c}{E_j}$.  Since in the experiments the 
self-capacitance is much smaller than the mutual 
capacitance, the relevant quantum parameter here is 
\begin{equation}
\label{alpha-m}
      \alpha_m = \frac{E_{C_{\rm m}}}{E_J} = \frac{e^2}{2C_{\rm m}E_J}.
\end{equation}
In the region where $\alpha_m$ is small, the phases dominate and we 
expect a superconducting to normal transition. 
The quantity we will use to characterize the coherent superconducting 
phase is the helicity modulus \cite{fisher-barber-jasnow,ohta-jasnow}
defined as  
\begin{equation}
\label{y-def}
         \Upsilon = \frac{\partial^2 F}{\partial A_{\vec r,\vec r
                    +\hat x}^2}\Bigg|_{A=0}.
\end{equation}
Here $\hat x$ is the unitary vector in the $x$ direction.
The superfluid density per unit mass, $\rho_s$, is proportional $\Upsilon$,
with $\rho_s(T) = \frac{1}{V} \left(\frac{ma}{\hbar}\right)^2 \Upsilon(T)$,
 where $a$ is the distance between superconducting islands,
 $m$ is the mass of the Cooper pairs, and V is the volume. 
From Eqs. (\ref{z-path-integral}) and (\ref{y-def}) we get 
\begin{eqnarray}
\label{order-parameter-sc-nor}
       \frac{1}{E_J L_x L_y}\Upsilon(T) = & &
       \frac{1}{L_x L_y L_\tau}\Bigg[
       \left< \sum_{\tau=0}^{L_\tau-1}\sum_{\vec r} 
       \cos\Big(\phi(\tau,\vec r)-\phi(\tau,\vec r+\hat x)-A_
       {\vec r,\vec r+\hat x}\Big) \right> -
       \nonumber \\
       & & -\frac{E_J \beta}{L_\tau}\Bigg\{\left< \Bigg[
       \sum_{\tau=0}^{L_\tau-1}\sum_{\vec r}
       \sin\Big(\phi(\tau,\vec r)-\phi(\tau,\vec r +\hat x)-A_{\vec r,
       \vec r+\hat x}\Big)\Bigg]^2\right>- \nonumber \\
       & &\ \ \ \ \ \ \ \ \ \ \ -\left< \sum_{\tau=0}^{L_\tau-1}
       \sum_{\tau,\vec r}\sin\Big(\phi(\vec r)-\phi
       (\tau,\vec r +\hat x)-A_{\vec r,\vec r+\hat x}\Big)
       \right>^2\Bigg\}\Bigg]. 
       \nonumber\\
\end{eqnarray}
The quantity we shall use to probe the possible charge coherence in the
 array is the inverse dielectric constant of the
gas of Cooper pairs, defined as \cite{minnhagen-warren,grest}, 
\begin{equation}
\label{ep}
           \frac{1}{\varepsilon} = \lim_{\vec k\rightarrow 0} \left[
           1-\frac{q^2}{k_BT}\frac{1}{{\bf C}(\vec k)} <n(\vec k)n(-\vec k)>
           \right].
\end{equation}
We can obtain the Fourier transform ${\bf C}(\vec k)$ from 
Eq. (\ref{capacitance-matrix}) for the capacitance matrix to get, 
\begin{equation}
\label{c-k}
       {\bf C}(\vec k) = C_s + 2C_m[1-\cos(k_x)]+2C_m[1-\cos(k_y)].
\end{equation}
The Fourier transform of the charge number is defined by
\begin{equation}
\label{n-k}
       n(\vec k) = \frac{1}{\sqrt{L_x L_y}}\sum_{\vec r} n(\vec r) 
                    \exp\left[i\vec k \cdot\vec r\right].
\end{equation}
Using this equation we can obtain a path integral representation for 
this correlation function, given by
\begin{eqnarray}
\label{op-for-nn}
        <n(\vec r_1)n(\vec r_2)> = 
        & & 
        -\frac{1}{Z}\prod_{\tau=0}^{L_\tau-1}
        \sqrt{{\rm det}{[\bf C]}} \prod_{\vec r} \int_{0}^{2\pi}
        \sqrt{\frac{L_\tau}{2\pi \beta q^2}} d\phi(\tau,\vec r)
        \sum_{\{ m(\tau,\vec r)\}=-\infty}^{\infty} \times\nonumber\\
        & & \times\frac{\partial^2}{\partial\phi(L_\tau,\vec r_1)\partial\phi
            (L_\tau,\vec r_2)}\Bigg\{\exp\bigg[-\frac{1}{\hbar}
            S[\{\phi\},\{m\}]\bigg]\Bigg\}\Bigg|_{\phi(L_\tau,\vec r)=
            \phi(0,\vec r)}.
\end{eqnarray}
The action is given in Eq. (\ref{action}). This equation becomes
\begin{eqnarray}
\label{nxny}
       <n(\vec r_1)n(\vec r_2)> = 
         & &\lim_{L_\tau\rightarrow\infty}\Bigg\{
       \frac{L_\tau}{\beta q^2}{\bf C}(\vec r_1,\vec r_2) - \frac{1}{Z} 
       \prod_{\tau=0}^{L_{\tau}-1} \sqrt{{\rm det} {[\bf C]} } \prod_{\vec r} 
       \int_{0}^{2\pi}\sqrt{\frac{L_{\tau}} {2\pi\beta q^2} } 
       d\phi(\vec r,\tau) \times\nonumber\\
       & &\times\sum_{\{ m(\vec r,\tau)\}= -\infty}^{\infty}
          \Bigg(\frac{1}{\hbar}\frac{\partial S}{\partial
          \phi(L_\tau,\vec r_1)}\ \frac{1}{\hbar}\frac{\partial S}
          {\partial\phi(L_\tau,\vec r_2)} \Bigg)
          \exp\bigg[-\frac{1}{\hbar}S[\{\phi\},\{m\}]\bigg]\Bigg\},
         \nonumber\\
\end{eqnarray}
with
\begin{equation}
\label{ds_dphi}
            \frac{1}{\hbar}\frac{\partial S}{\partial\phi(L_\tau,\vec r_1)} =
            \frac{L_\tau}{\beta q^2} \sum_{\vec r}{\bf C}(\vec r_1,\vec r)
            \left[\phi(L_\tau,\vec r)-\phi(L_\tau-1,\vec r)+2\pi 
            m(L_\tau-1,\vec r)\right].
\end{equation}
Notice, that this is not a well behaved operator since in the limit 
$L_\tau\rightarrow\infty$ we would have to subtract two large numbers
and the path integral in the second term in Eq. (\ref{nxny}) would diverge. 
This divergence is canceled out by the first term in Eq. (\ref{nxny}).
This can be seen explicitly by doing the calculation of $\epsilon^{-1}$
seting $E_J=0$, which leads to
\begin{equation}
\label{nxny-with-ms}
      <n(\vec r_1)n(\vec r_2)> = \frac{1}{\beta q^2}\ {\bf C}(\vec r_1,
      \vec r_2) + \left(\frac{2\pi}{\beta L_\tau}\right)^2 
        \sum_{\vec r_3,\vec r_4}{\bf C}(\vec r_1,\vec r_3) 
        {\bf C}(\vec r_2,\vec r_4)  <m(\vec r_3)m(\vec r_4)>.
\end{equation}
This result can be put into Eq. (\ref{ep}) to obtain a finite inverse 
dielectric constant,
\begin{equation}
\label{ep-2}
            \frac{1}{\varepsilon} = \lim_{\vec k\rightarrow 0} \left[
            \frac{(2\pi)^2}{\beta q^2}\ {\bf C}(\vec k) <|m(\vec k)|^2>
            \right].
\end{equation}
Here we have used the Fourier transform defined in Eq. (\ref{n-k})
and the $m(\vec r)$ defined as 
$m(\vec r) = \sum_{\tau=0}^{L_\tau-1} m(\tau,\vec r)$.
Note that in general this operator will not exactly be the
inverse dielectric constant of a gas of Cooper pairs, since it will
depend on $L_\tau$. But we expect that it does contain 
most of the relevant information of the inverse dielectric constant
of our charged system. In our Monte Carlo calculations we have used
the general result Eq. (\ref{nxny}) valid for $E_J\neq 0$ and finite
$L_\tau$.

\subsection{\bf The Simulation Approach}
\label{sec:simulation}
Up to now we have seen that the partition function defined by the
Hamiltonian in Eq. (\ref{hamiltonian}) can be expressed in different 
convenient representations for analytic analyses.
To carry out our QMC calculations, we have used what is, in principle, 
the most straightforward representation of $Z$ given by Eqs. 
(\ref{z-with-m}) and (\ref{action});  
it involves the phases and the charge integer as statistical variables. 
This representation is general enough to be used over all the whole
parameter range covered in the phase diagram.

In this case we have a set of angles 
$\phi(\tau,\vec r)\in [0,2\pi)$, located at the nodes of a three-dimensional 
lattice, with two space dimensions, $L_x$ and $L_y$,  and one 
imaginary time dimension, $L_\tau$. The periodic boundary condition, 
comes from the trace condition in Eq.
(\ref{partition_function}), and we also have chosen to use periodic 
boundary conditions in both space directions. The link variables 
$m(\tau,\vec r)$ are defined in the bonds between two nodes
in the $\tau$ direction and they can take any integer value.

We have basically used the standard Metropolis algorithm to move about in 
phase space \cite{metropolis}. As the phases are updated we restrict their 
values to the interval $[0,2\pi)$. Moreover, the shifts along a $\tau$-column 
and the individual phase moves are adjusted to keep the 
acceptance rates in the range $[0.2,0.3]$.

If $\alpha_m$ is small, the system is in the semiclassical limit.  In 
this case the fluctuations of the phases along the imaginary time axis as 
well as the fluctuations in the $m$'s are suppressed by 
the second term in Eq. (\ref{action}). Attempts to change a phase variable
will have a very small success rate. Therefore we implemented
two kinds of Monte Carlo moves in the phase degrees of freedom. In one 
sweep of the array we update the $L_x\times L_y$ imaginary time columns,
by shifting all the phases along a given column by the same angle. This 
move does not change the second term in Eq. (\ref{action}), and thus it 
probes  only the Josephson energy \cite{jacobs-jose-novotny-goldman}. 
To account for phase fluctuations
along the imaginary time axis, which become more likely as  
$(\alpha_m/T)$ increases,  we also make local updates of the phases
along the planes.

Another aspect of the implementation of the QMC algorithm is the order in 
which we visit the array. This is relevant for the optimization of the 
computer code in different computer architectures. In a scalar machine 
we have used an algorithm that updates column by column in the array. 
For a vector machine we have used the fact that for local updates, like 
the ones we use, the lattice can be separated into four sublattices in 
a checkerboard-like pattern. This separation
is done in such a way that each of the sublattices can be updated 
using a long vector loop without problems of data dependency. Using this
last visiting scheme, the cpu time grows sublinearly with the size of the 
array. One of the problems that this type of visiting scheme has in a 
vector machine, like the Cray C90, is that the array's dimensions 
have to be even, and this produces memory conflicts. We have not made 
attempts to optimize this part of the code. We have not used parallel 
machines in our calculations but the same type 
of checkerboard visiting scheme would lead to a fast algorithm.

We followed Ref. \cite{jacobs-jose-novotny-goldman} and replaced the 
U(1) symmetry of the problem by a discrete ${\bf Z}_N$ subgroup. We took
$N=5000$.  This allows us to use integer arithmetic for the values 
of the phase variables, and to store lookup tables for the Josephson 
cosine part of the Boltzmann factors. This simplification can not be used for the 
charging energy part of the Boltzmann factors, except in the
$C_{\rm m}=0$ case, where the $m$'s can be summed up in a virtually 
exact form. In the latter case we can also store lookup
tables using the following definition of an effective potential $V_{\rm 
eff}$,
\begin{equation}
\label{charging-boltzman}
         \exp\left[-\left(\frac{L_\tau C_{\rm s}}{q^2\beta}\right)
         V_{\rm eff}(\phi)\right] = \sum_{m=-\infty}^{\infty} 
         \exp\left[-\frac{1}{2}\left(\frac{L_\tau C_{\rm s}}
                   {q^2\beta}\right)(\phi+2\pi m)^2\right].
\end{equation}
We notice that this summation can be evaluated numerically to any 
desired accuracy.

We calculated the thermodynamic averages after we had made $N$ 
visits to the array updating the phases and $M$ visits updating
the $m$'s. Typically, if $\alpha_{\rm m}$ is small we used $N=4$
and $M=1$. In the opposite limit we used $N=1$ and
$M=8,10,...$.  This is so because our local updating algorithms
for the $m$'s have serious decorrelation time problems, due to the 
long range interaction among the charges. We typically found that
in order to get reasonably small statistical errors, we needed 
to perform, in most cases, about $N_{\rm meas}=2^{12}=4096$ measurements
of the thermodynamical quantities, other times we took up to 
$N_{\rm meas}=2^{13}=8192$ measurements.

Once we have a long stationary string of values for the measured 
operators we calculated their mean values and uncertainties.  
We also  have used the algorithm proposed in 
Ref. \cite{flyvbjerg-petersen} for the 
efficient calculation of the helicity modulus.  
This method has a bias problem due to the last term 
in Eq. (\ref{order-parameter-sc-nor}). However, in the zero
magnetic field case this problem is not present, since this term is 
identically zero.


\subsection{\bf Results for f=0}
\label{sec:results-for-f=0}
In this subsection we present the bulk of our Monte Carlo results.
We have mostly calculated the helicity modulus in the 
small $\alpha_{\rm m}$ region and the inverse dielectric 
constant in the large $\alpha_{\rm m}$
regime, and both quantities in the intermediate region. 

Most of the calculations we performed were for parameter values close to 
or at the experimental ones. In particular, the ratio 
between the self and mutual capacitances was kept fixed between the values 
$C_{\rm s}/C_{\rm m}\approx 0.01$ and 0.03, with the bulk of the 
calculations carried out for 0.01. We found that for the 
helicity modulus both values gave essentially the same results.
Almost all of the calculations were done by lowering the temperature,
in order to reduce the possibility for the system to be trapped
in metastable states \cite{mikalopas-jarrel-pinski-chung-novotny}.

We have a clear physical understanding of the behavior of the system 
in the very small $\alpha_{\rm m}$ limit, since this limit 
is close to the classical 2-D XY model. Moreover, we have the 
semiclassical calculation results,  mentioned before, 
up to first order in $\alpha_{\rm m}$, 
which, as we shall see, agree very well with the Monte Carlo results. 
In this limit the results are solid because the discrete
imaginary time path integral calculations converge very rapidly to the 
infinite $L_\tau$ limit. Therefore in this section we will discuss our 
numerical results for increasing values of $\alpha_{\rm m}$. This 
will allow us to go from a well understood physical
and calculational picture to the nonperturbative region of parameter 
space which is less understood.  Here is where we will explore the 
limits of our numerical calculational schemes. The end result will be 
that a significant portion of the phase diagram can be understood. However, 
some of the most interesting intermediate regimes of the phase diagram are 
still very difficult to fully understand with our present calculational 
techniques.

In Fig. \ref{fig:fig03-alpha-0p5} we show a typical curve for the helicity 
modulus as a function of temperature, in the small $\alpha_{\rm m}$ limit. 
As $\alpha_m$ increases $\Upsilon$ flattens in the superconducting region.
In order to calculate the transition 
temperature we used the fact that the critical temperature and the helicity 
modulus still satisfy the universal relation,
\begin{equation}
\label{tc-hel}
              \Upsilon(T_c) = \frac{2}{\pi}\ T_c.
\end{equation}
Based on the first order results from the semiclassical approximation 
analysis we know that this universal result is
independent of $\alpha_{\rm m}$. In other words, we can determine the
critical temperature by the intercept of $\Upsilon(T)$
with the line $(2/\pi)T$, as shown in Fig. \ref{fig:fig03-alpha-0p5}.

As can be seen from Fig. \ref{fig:fig03-alpha-0p5}, at high temperatures and 
small $\alpha_{\rm m}$ the asymptotic limit 
$L_\tau\rightarrow\infty$ is already reached for small 
$L_\tau$.  From Eq. (\ref{action}), we can see that the parameter that 
determines this rate of convergence is
\begin{equation}
\label{conv-parameter}
            P = \frac{L_\tau }{(\beta E_J)\alpha_{\rm m}}.
\end{equation}
The deep quantum limit is reached for a relatively large $P\gg 1$. 
This progression is shown in Fig. \ref{fig:fig04-alpha-1p25} 
where we plot the helicity modulus as a 
function of temperature for a relatively large 
$\alpha_{\rm m}=1.25$,  $L_x=L_y=20$, and 
three values of $L_\tau$. It can be seen that convergence is reached 
for $P>5$, as found before in the self capacitive model in 
Ref. \cite{jacobs-jose-novotny-goldman}.

As shown in Fig. \ref{fig5-alpha-1p25}, for a larger $\alpha_{\rm m}$ 
the behavior changes and the departure from the 
$L_\tau\rightarrow\infty$ limit is manifested as a small dip in $\Upsilon$ at 
low temperature \cite{jacobs-jose}. As the temperature is lowered 
$\Upsilon$ shows an upward behavior. This is also seen in Fig. 
 \ref{fig:fig06-alpha-2p25} 
from more extensive calculation for a still larger $\alpha_m$'s. This finite 
$L_\tau$ behavior can be understood in terms of a  plane decoupling along the 
imaginary time direction. A way to see this is to notice that if we take 
both contributions to the action given in Eq.(\ref{action}) as 
independent, both of them would yield a low temperature transition.  
If all the $L_\tau$ planes are considered decoupled, then the Josephson 
coupling would be
$\beta/L_\tau$. Therefore in this case the N--S transition 
would happen at $T_{\rm N-S}\approx 1/L_\tau$. On the other hand, if 
we only consider the second term in Eq. (\ref{action}) we see that  
plane decoupling would take place at 
$T_{\rm decl}\approx (\alpha_{\rm m}/L_\tau)$. Now, if 
$T_{\rm N-S}>T_{\rm decl}$ which implies $\alpha_{\rm m} < 1$, then only 
the first transition would take place.
If however $\alpha_{\rm m}$ is large enough, we could have
$T_{\rm N-S}<T_{\rm decl}$, producing the observed dip in the helicity 
modulus. This is seen in Fig. \ref{fig:fig08-Y-1_e-alpha-2p25} where the 
helicity 
modulus is shown together with the inverse dielectric constant.  
There it can be seen that the dip starts at about the same temperature 
where $\epsilon^{-1}$ becomes finite, signaling that the 
nonzero winding numbers have become relevant. In this region the 
fluctuations between planes have a small energy cost in the action
given by Eq. (\ref{action}).

As we increase $\alpha_{\rm m}$ further we arrive at a point where at low 
temperatures, and fixed $L_\tau$, $\Upsilon$ goes to zero as shown 
in Fig. \ref{fig:fig09-alpha-2p5}. To understand the nature of the low 
temperature phase we have computed the equal space imaginary time 
correlation function
\begin{equation}
\label{time-corr-func}
             C_\tau(\tau) = \Big<\cos\left(\phi(\vec r,\tau)-
                             \phi(\vec r,0)\right)\Big>.
\end{equation}
We evaluated this function at three different temperatures,
for $\alpha_{\rm m}=1.75$ and $L_\tau=32$. The results are shown in
Fig. \ref{fig:fig07-corr-t-alpha-1p75} where we see that the appropriate 
value for $L_\tau$ needs to be increased as T is lowered. 
More extreme are the two temperature results
for $\alpha_{\rm m}=2.5$ and $L_\tau=64$, shown in Fig. 
\ref{fig:fig10-corr-t-alpha-2p5}. 
The upper curve corresponds to $k_BT/E_J=0.36$.  As seen in Fig. 
\ref{fig:fig09-alpha-2p5} at this temperature a value of $L_\tau=64$ is enough
to reliably calculate the helicity modulus, for in this case the 
planes are correlated with a short decorrelation time. The lower 
curve has $k_BT/E_J=0.1$. At this temperature the helicity modulus 
is zero and the correlation function has a very short decorrelation 
time. These results show that the low temperature discontinuity is 
related to a decoupling of the planes along the imaginary time axis.  

This plane decoupling does not show any dependence on  $(L_x,L_y)$ for 
the cases considered, and the curves shown in Fig. \ref{fig:fig09-alpha-2p5} 
are reproducible within the statistical 
errors for other values of $L_x$ and $L_y$. 
Again we point out that upon increasing $L_\tau$ the decoupling 
temperature moves closer to zero temperature.  We should note that 
from the WKB analysis there is a critical value for $\alpha$ above 
which the superconducting state is no longer stable.  So as we consider 
larger values of $\alpha_{\rm m}$, the superconducting state will 
become less and less stable.   As we mentioned  
the simulations were performed while lowering the temperature.  
In contrast, in Fig.\ref{fig:fig11-alpha-2p75} we show results 
from lowering the temperature for 
$\alpha_m=2.75$.  In this case $\Upsilon$ reaches a zero for $T\leq 0.2$. 
We reversed the process increasing the temperature. Up to the last 
temperature calculated the results are consistent with having zero 
$\Upsilon$ for $\alpha_{\rm m}=2.75$. 
The low temperature state arises from a decoupling transition between the 
imaginary time planes which leads to an ensemble of decorrelated
planes.  Maybe the planes could get recoupled at higher temperatures but,
as already mentioned, our local algorithm would take too long 
to realign these planes so as to produce a coherent state. 

As shown in Fig. \ref{fig:fig09-alpha-2p5} the temperature where $\Upsilon$ 
has a sharp drop changes with the size of the system. To see if in 
the limit $L_\tau\rightarrow\infty$ we still have a finite low temperature 
transition,  we tried to extract it from the data for three 
different $\alpha_m$ values by plotting them against 
$1/L_\tau$ as shown in Figs. \ref{fig:fig12(a)-finite-size-alpha-2},
\ref{fig:fig12(b)-finite-size-alpha-2p25} and 
\ref{fig:fig12(c)-finite-size-alpha-2p5}.  From these 
figures it appears that for $\alpha_{\rm m}=2.0 $ and $2.25$ there is 
a nonzero transition temperature in the $L_\tau\rightarrow\infty$ limit. 
We used a jackknife 
calculation to estimate the infinite $L_\tau$ temperature and we found 
$$T_{\rm decl}(\alpha_{\rm m}=2,L_\tau\rightarrow\infty)=(0.0183
\pm 0.009)(E_J/k_B),$$ 
and 
$$T_{\rm decl}(\alpha_{\rm m}=
2.25, L_\tau\rightarrow\infty)=(0.0067\pm 0.0025) (E_J/k_B).$$  
The same type of calculation was done for $\alpha_{\rm m}=2.5$.  
The results are shown in Fig. \ref{fig:fig12(c)-finite-size-alpha-2p5}. 
Here we found that 
$$T_{\rm decl}(\alpha_{\rm m}=2.5, L_\tau\rightarrow\infty)=
(-0.013\pm 0.005)(E_J/k_B).$$ 
Therefore from these  estimates we  
surmise that the critical value for $\alpha_m\sim 2.5$, which  is
larger than the one estimated from the WKB-RG analysis. 
However, our QMC calculations, in particular at low temperatures,
 are not precise enough to make a definitive determination of the
critical $\alpha_m$.

The calculations of the S--N transition
line seem to indicate that the superconductor to insulator zero temperature
transition occurs at $\alpha_{\rm m}\approx 3$, which is quantitatively 
different from the T=0 estimate \cite{fazio-schon-91}.

The evaluation of the inverse dielectric constant is considerably more
complicated since the insulating region we have 
$\alpha_{\rm m} > 3$, needing larger values 
of $L_\tau$ at low temperatures.  Moreover there are serious critical 
slowing down problems due to the long range charge interactions
which worsen as the size of the system increases. 

We should point out that, in comparing with the purely classical case 
\cite{lee-teitel}, the quantum 2-D Coulomb gas 
studied here has the extra complication of the 
$n-\phi$ coupling, as seen in Eq. (\ref{z-path-integral}). 
This introduces an imaginary component to the action. 
Therefore we are forced to integrate the $n$'s introducing the 
new variables $\{m\}$ leading to the action given in Eq. (\ref{action}). 
This is the action that we used to perform the Monte Carlo 
calculations.

We use the expression given in Eq. (\ref{ep-2}), which for finite $\vec k$
and $L_x=L_y$ gives $|\vec k|=2\pi/L_x$, as an upper bound for 
the inverse dielectric constant \cite{olsson}. 
The technical problems mentioned above made the calculation 
of the dielectric constant 
less reliable than that of $\Upsilon$.
The results obtained from the Monte Carlo runs were too noisy to give us 
a quantitative estimate of the conductor to insulator transition 
temperature if there was one. Our quoted results give  
only tentative values for the 
transition line. 

The results are shown in Fig. \ref{fig:fig01-phase-diagram}, where we 
also plotted the results of the Monte Carlo calculation of the 
normal to superconductor transition temperature as well as the 
experimental results from the Delft group 
\cite{van-der-zant-thesis}. We have fitted a straight 
line to the first seven points in this line and used a jackknife 
calculation of $T_c(\alpha_{\rm m})$ for small 
$\alpha_{\rm m}$.  We obtained
\begin{equation}
\label{tc-first-order}
         \frac{k_B T_c}{E_J} = (0.9430\pm0.0042)-(0.1800\pm0.0040)
                               \alpha_{\rm m} 
                               + O\left(\alpha_{\rm m}^2\right).
\end{equation}
The value of the slope is in good agreement with the semiclassical
approximation result given in Eq. (\ref{limits-T_BKT}).
The dashed line  gives $\alpha_{\rm m}=2.8$ at T=0 and joins  the 
last QMC point to $T=0$. The line is only a guide 
to the eye.  We have not performed 
detailed calculations around 
$\alpha_{\rm m}=3$ since the required values of $L_\tau$ makes reliable
calculations too computationally intensive to be carried out with 
current algorithms and computer capabilities. 


\subsection{\bf Results for f=1/2.}
\label{sec:results-for-f=1/2}
We also have performed a few calculations of the helicity modulus for the 
fully frustrated case f=1/2. The results of these calculations are shown 
in Fig. \ref{fig:fig01-phase-diagram}. The experimental results of the Delft
group show a transition temperature for $\alpha_{\rm m}=0$ at
$(k_B T/E_J)\approx 0.3$ \cite{van-der-zant-thesis}, while the 
classical fully frustrated case has a critical temperature close to 
$(k_B T/E_J)\approx 0.5$. Taking this into account we have 
rescaled the Monte Carlo results so that the calculated value
for the transition temperature for $\alpha_{\rm m}=0$, f=1/2; coincides 
with the experimental result.

Taking the experimental value of the critical temperature for the $f=0$ case, we
performed a least square fit for the five smallest $\alpha_{\rm m}$'s 
considered and found, using a jackknife calculation,
\begin{equation}
\label{exp-fit-for-f=0}
        \left(\frac{k_B T}{E_J}\right)_{f=0} \approx (0.9787\pm 0.0070) -
        (0.256\pm 0.017) \alpha_{\rm m} + O(\alpha_{\rm m}^2).
\end{equation}
In the same way, but now for the $f=1/2$ case, we found 
\begin{equation}
\label{exp-fit-for-f=1/2}
        \left(\frac{k_B T}{E_J}\right)_{f=1/2} \approx (0.3188\pm 0.0015) -
        (0.2929\pm 0.0066) \alpha_{\rm m} + O(\alpha_{\rm m}^2).
\end{equation}
These rough calculations show that the slopes of both curves are very 
close. This result can be compared 
with Eq. (\ref{tc-with-mag-field}),
which confirms that the first order correction in $\alpha_{\rm m}$ to the 
critical temperature should not depend on the value of the magnetic
field. A similar result for the equality of slopes was obtained
in Ref. \cite{jacobs-jose-novotny-goldman}, using a Monte Carlo 
calculation for the self-capacitive model.
%
%
\section{\bf Self-consistent Harmonic Approximation}
\label{sec:harmonic}
We need an alternative analytic approach, in principle exact for fixed 
and finite $L_\tau$, to further understand  the QMC results at 
low temperatures. This is important because of the strong $L_\tau$
dependence in the study of the QUIT temperature, and
the analytic WKB results are only strictly 
valid at high  temperatures and $L_\tau=\infty$. 
In this section we use a variational principle to evaluate the
free energy for the JJA within a self-consistent harmonic approximation 
(SCHA). This approximation gives increasingly better results 
as the temperature is lowered. Previous SCHA
calculations \cite{doniach,lozovik,ariosa-beck} did not explicitly include 
the charge degrees of freedom, which are of significant
importance in the analysis
presented here. As a bonus, we note that the SCHA developed here could 
be used as the basis for developing an alternative QMC algorithm  
to study the model at low temperatures and intermediate $\alpha$ values, 
were both the WKB and the standard QMC analyses have problems.

We start with the following decomposition of the Hamiltonian given in Eq.
(\ref{hamiltonian})
\begin{equation}
\label{more-split-action}
         H=H_0 + \{H_J-H_H\}. 
\end{equation}
where $H_0= \left\{H+(H_H-H_J)\right\}$, and
\begin{eqnarray}
\label{def-of-H_j}
      H_J &=& \frac{E_J}{L_\tau}\sum_{\tau=1}^{L_\tau-1}
            \sum_{<\vec r_1,\vec r_2>} \left[1-\cos\left(\phi(\tau,\vec r_1)-
            \phi(\tau,\vec r_2)\right)\right], \\
\label{def-of-H_h}
      H_H &=& \frac{E_J\Gamma}{2L_\tau}\sum_{\tau=1}^{L_\tau-1}
            \sum_{<\vec r_1,\vec r_2>} \left[\phi(\tau,\vec r_1)-
            \phi(\tau,\vec r_2)\right]^2.
\end{eqnarray}
In other words, we replace the Josephson Hamiltonian by a spin wave 
term and introduce its stiffness $\Gamma$ as the variational parameter.
$\Gamma$ is of course the helicity modulus, but to 
$\underline {emphasize}$ that
it is evaluated within the SCHA we us the $\Gamma$ notation instead of
$\Upsilon$.
The spin wave approximation does not have a phase transition at any
temperature so the variational calculation is going to give a 
vanishing $\Gamma$ as the signature of the JJA transition.
The variational free energy is given by
$F_V = F_H + \left< H_J - H_H \right>_H$, where
$\beta F_H = -\ln Z_H$, and $Z_H = {\rm Tr}\left\{\exp\{-\beta H_0\}\right\}$,
with the average $<,>$ defined as usual by $<A>_H =  
{\rm Tr}\left\{A\exp\{-\beta H_0\}\right\}/Z_H$.
We use the change of variables 
\begin{eqnarray}
\label{extra-change-of-variables}
      \psi(\tau,\vec r) &=& \psi(0,\vec r) + \sqrt{\frac{2}{L_\tau}}\
                          \sum_{n=1}^{L_\tau-1} \chi_n(\vec r) \sin
                          \left(\frac{\pi n}{L_\tau} \tau\right), \\
\label{tauss}
       \tau &=& 1,2,\dots,L_\tau-1,
\end{eqnarray}
and perform the integrals over the 
variables $\chi_n(\vec r)$.  The result is
\begin{eqnarray}
\label{z_h-mid-way}
      Z_H = 
           & &\sqrt{{\rm det}{[\bf C]}} \prod_{\vec r}\sum_{m(\vec r)=-\infty}^
            {\infty} \exp\left\{-\frac{1}{\hbar} \tilde S_m \right\}
            \left[\prod_{n=1}^{L_\tau-1}\left[1+\frac{(\beta^2 q^2 
            E_J\Gamma/C_{\rm m})}{2L_\tau^2\left[1-
            \cos\left(\frac{\pi n}{L_\tau}
            \right)\right]}\right]\right]^{-L_x L_y/2} \times\nonumber\\
            & &\times\prod_{\vec r}\int_{0}^{2\pi} \frac{\phi(0,\vec r)}
               {\sqrt{2\pi\beta q^2}} \exp\left\{-\frac{1}{2}\sum_{\vec r_1,
               \vec r_2} \phi(0,\vec r_1){\bf N}(\vec r_1,\vec r_2)
               \phi(0,\vec r_2)+\sum_{\vec r} j(\vec r)
               \phi(0,\vec r)\right\}.
            \nonumber\\
\end{eqnarray}
where
\begin{eqnarray}
\label{def-s-tilde}
      \frac{1}{\hbar}\tilde S_m &=& \frac{(2\pi)^2}{2}\left[\frac{C_{\rm m}}
      {\beta q^2}+\frac{\beta E_J\Gamma}{6}\left(1-\frac{1}{L_\tau}
      \right)\left(2-\frac{1}{L_\tau}\right)-(\beta E_J\Gamma) 
      g\!\left(\beta q\sqrt{E_J\Gamma/C_{\rm m}}\right)\right] 
      \times\nonumber\\
      & &\times\sum_{\vec r_1,\vec r_2} m(\vec r_1){\bf O}(\vec r_1,\vec r_2)
      m(\vec r_2).
\end{eqnarray}
To obtain this equation we have taken $C_{\rm s}=0$, while the general case
can also be treated as well, but since $C_{\rm s}\ll C_{\rm m}$ in the 
experiment this assumption simplifies the calculations. We 
also have used the following definitions to write the previous equations,
\begin{eqnarray}
\label{defs-for-o}
      \sum_{<\vec r_1,\vec r_2>}\left[\phi(\vec r_1)-\phi(\vec r_2)\right]^2
      &=& \sum_{\vec r_1,\vec r_2} \phi(\vec r_1) {\bf O}(\vec r_1,
          \vec r_2) \phi(\vec r_2), \\
\label{def-for-c}
      {\bf C} &=& C_{\rm m} {\bf O},\\
\label{def-for-n}
      {\bf N} &=& (\beta E_J\Gamma)\left[1-h\left(\beta q 
                  \sqrt{E_J\Gamma/C_{\rm m}}\right)\right]{\bf O},
\end{eqnarray}
with ${\bf O}$ the lattice Laplacian operator, and $g(*)$ and $f(*)$ are
functions defined in  terms of a Matsubara sum and given in Appendix B. 

The details of the variational calculation of the free energy are
presented in Appendix B. Here we discuss the main conclusions from 
the calculation. The results for $\alpha_{\rm m}=0$, and 1 are shown 
in Fig. \ref{fig:fig14-ys}. There we see that the $\alpha_{\rm m}=0$ case 
has the right low temperature linear $T$ dependence. The result for 
$\alpha_{\rm m}=1.0$ has an essentially flat low temperature
behavior for $\Gamma$, which is due to quantum phase
slips tunneling processes.
In Appendix B it is shown that the helicity modulus can be expressed as
\begin{equation}
\label{equation-for-gamma}
     \Gamma = \exp\left\{ -\frac{1}{2}\left< (\Delta\phi)^2
\right>\right\}.
\end{equation}
with the fluctuations in the phase given by,
\begin{eqnarray}
\label{Dphi-for-infty}
       < (\Delta \phi)^2 >_H &=& \frac{1}{2}\sqrt{\frac{2\alpha_{\rm m}}
                        {\Gamma}} \left(\frac{\sinh\left(\beta E_J
                        \sqrt{ 8\alpha_{\rm m}\Gamma}\right)}
                                 {\cosh\left(\beta E_J \sqrt{
                                 8\alpha_{\rm
m}\Gamma}\right)-1}\right).
\end{eqnarray}

The classical limit corresponds to setting
$\alpha_{\rm m}\rightarrow 0$, which gives
$< (\Delta\phi)^2 >_{H(cl)} = \frac{1}{2\beta E_J \Gamma}$.
Using this result and Eq. (\ref{equation-for-gamma}), at low 
temperatures  $\Gamma$ is given by
$\Gamma \approx 1 - {k_B T}/{4E_J} + O(T^2)$.
This is precisely the same low temperature behavior obtained from a 
spin wave analysis in two dimensions \cite{ohta-jasnow}. 
From $< (\Delta\phi)^2 >_{H(cl)}$  we find the 
transition temperature within this 
approximation, $\Gamma_c = \frac{1}{4} \left(\frac{k_B T_c}{E_J}\right)$,
$\Gamma_c = 1/e$, and ${k_B T_c^{(0)}}/{E_J} = {4}/{e} 
\approx 1.472$,
which is an overestimate, (since the dimensionless 2-D critical 
temperature is $T_c^{XY}\approx 0.9$).
The problem with this approximation is extending the integration intervals
from $[-\pi,\pi]$ to $(-\infty,\infty)$. As we have discussed
this is a good approximation at low temperatures but it breaks
down near the transition point. 
Our conclusions from this analysis are
that for $\alpha_m=0$the classical SCHA gives  
good results for low temperatures while it overestimates the critical 
temperature at higher ones.

In the $\alpha_m\neq   0$ case the quantization of the 
spin wave excitations leads to
a non-vanishing result for $< (\Delta\phi)^2 >_H$, given in Eq. 
(\ref{Dphi-for-infty}). For $(k_B T/E_J)\ll 1$ we get
\begin{eqnarray}
\label{quantum-Dphi-for-low-T}
   & & < (\Delta\phi)^2 >_H \approx \frac{1}{2}\sqrt{\frac{2\alpha_{\rm m}}
                         {\Gamma}} \left( 1 + 2\exp\left\{-\beta E_J 
                         \sqrt{8\alpha_{\rm m} \Gamma}\right\} \right),\\
    & & \nonumber\\
\label{quamtum-Y-for-low-T}
    & & \Gamma \approx \Gamma_0 
\left(1-\frac{1}{2}\sqrt{\frac{2\alpha_{\rm m}}
                       {\Gamma_0}} \exp\left\{-\beta E_J\sqrt{8\alpha_{\rm 
m}
                         \Gamma_0}\right\} \right),
\end{eqnarray}
where $\Gamma_0$ is the helicity modulus at zero temperature and it 
is the self-consistent solution to the equation 
 $\Gamma_0 = \exp\left\{-\sqrt{\frac{\alpha_{\rm m}}
  {8\Gamma_0}}\right\}$,
  with $\Gamma_0 \approx 1-\sqrt{\frac{\alpha_{\rm m}}{8}}$ for
 $\alpha_{\rm m}\ll 1$.
The solution to this equation is shown as a function of 
$\alpha_{\rm m}$ in Fig. \ref{fig:fig15-ys-T=0}, where we also show some 
Monte Carlo simulation results.  The $\Gamma_0$ result  
also presents a transition to a zero $\Gamma$ state at 
$\alpha_{\rm m}^{c}(T=0)=32/e^2\approx 4.33$, with a jump from 
$\Gamma_0^{c} = 1/e^2$ to zero. 
 Again the result of the 
SCHA overestimates the stability 
of the superconducting state. Both the extension of the integration 
intervals and ignoring the $m$'s in these calculations are probably 
responsible for the deviations at large $\alpha_{\rm m}$. 
 An interesting 
observation is that the result for $\Gamma_c$  is 
exact up to first order in $\alpha_{\rm m}$. This is equal to the result
we obtained from the WKB-RG analysis.  Also surprising is
that the first order correction to the critical temperature agrees 
with Eq. (\ref{limits-T_BKT}) 
\begin{equation}
\label{firt-correction-tc}
     \left(\frac{k_B T_c}{E_J}\right) = \left(\frac{k_B T_c^{(0)}}{E_J}\right)
               - \left(\frac{2}{3z}\right) \alpha_{\rm m} 
               + O(\alpha_{\rm m}^2),
\end{equation}
where $z$ is the coordination number of the lattice. 
In Fig. \ref{fig:fig17-ys-lt}
we show $\Gamma$ for $\alpha_{\rm m}=1.0$ as a function of the
temperature for increasing values of $L_\tau$,  which should be 
compared with Fig. \ref{fig:fig04-alpha-1p25}. This result strongly suggests 
that the upward tendency of 
the helicity modulus at low temperatures seen in the QMC results 
may be an artifact of the finite $L_\tau$ nature of the 
calculations. The origin of this 
increase is in the low temperature result for the finite $L_\tau$ 
calculation of Eq. (\ref{def-of-Dphi2})
\begin{equation}
\label{low-temp-dphi-finite-lt}
        < (\Delta \phi)^2 >_H \approx \frac{L_\tau}{\beta E_J \Gamma}
        \left(1-\frac{(L_\tau+2)L_\tau}{16(\beta E_J)^2\alpha_{\rm 
m}\Gamma}
        + O(T^3)\right),
\end{equation}
so that for finite $L_\tau$ and at low temperatures 
the helicity modulus is given by
\begin{equation}
\label{Y-low-temp-finite-lt}
      \Gamma \approx 1 - \frac{L_\tau}{2}\left(\frac{k_B T}{E_J}\right) +
                      O(T^2).
\end{equation}

We have been able to explain the shape of the helicity modulus
curves for low temperatures, but Figs. \ref{fig:fig04-alpha-1p25} and 
\ref{fig5-alpha-1p25} show 
that if $\alpha_{\rm m}>\alpha_{\rm m}^{*}$, where $\alpha_{\rm m}^{*}\in 
(1.25,1.75)$, then the helicity modulus has a dip before it may go to one
at low temperatures. So far we have ignored the contribution of the
Discrete Gaussian Model (DGM) to the variational free energy which
is a good approximation only for small $\alpha_{\rm m}$. 
We calculated the helicity modulus for the model ignoring the DGM 
and then including it as a continuous Gaussian model for finite $L_\tau$, 
which is a good approximation for the effective
coupling $J_{\rm eff}\ll 1$. 
For large $\alpha_{\rm m}$ the crossover point, which is when $J{\rm eff}$
becomes soft,
is seen as a finite dip in the helicity modulus.
Unfortunately, the $T^{*}$ found in the Monte Carlo calculations
is much larger than the one given by Eq. (\ref{def-of-Jeff}). 
It is apparent that the effective coupling 
$J_{\rm eff}$ does not contain all the contributions to the 
renormalization of the DGM due to the 
integration over the phases. 
To illustrate the nature of this
crossover we performed several calculations of the helicity modulus
for $L_\tau=10, 20$, and 40 with $\alpha_{\rm m}=1$.  The results are 
shown in Fig. \ref{fig:fig19-cross-over}. There we took
$\frac{L_\tau}{\beta E_J\alpha_{\rm m}}=6$ for the crossover temperature. 
These results can be compared with those of Fig. \ref{fig5-alpha-1p25}.

The Monte Carlo calculations show that this dip occurs simultaneously with the
rise in the inverse dielectric constant. This is a signal that the non-zero 
effective constant for the winding numbers becomes soft, making
their contribution to the helicity modulus non vanishing.  We note 
that the effect of a finite lattice, necessary for the Monte Carlo 
calculation, increases the softening temperature of $J_{\rm eff}$.
On the other hand, the variational calculation for $C_{\rm s}=0$ does
not depend on the size of the lattice, therefore it can not capture 
these finite space size effects.

%
\section{\bf Conclusions}
\label{sec:conclusions}
We have presented a thorough study of the $\alpha_{\rm m}$ vs. 
$T$ phase diagram for an array of ultrasmall Josephson junctions using a 
series of theoretical tools. 
One of our main goals was to perform these calculations for 
these arrays using experimentally realistic
parameters. The model we used for the JJA is defined by a Hamiltonian 
that has two contributions, a Josephson coupling and an electrostatic 
interaction between the superconducting islands. The ratio of these 
two contributions was defined as $\alpha_{\rm m}$ 
= (charging energy)/(Josephson energy). This was the important 
quantum parameter in our analysis.

For convenience of calculation
we derived different path integral formulations of the quantum 
partition function of the JJA. In the  small $\alpha_{\rm m}$ limit 
we used a WKB-RG approximation to find the first order correction in 
$\alpha_{\rm m}$ to 
the classical partition function. The result of this calculation was an
effective classical partition function of a 2-D XY model type, where the 
coupling constant is modified by the quantum fluctuations. We used the 
modified renormalization group equations for the 2-D model to find the 
superconducting to normal phase boundary. We also found that up to 
first order in $\alpha_{\rm m}$, the correction
to the transition temperature was $\underline{independent}$ of  magnetic field.
One interesting finding from 
this calculation was the possible existence of a low 
temperature instability  QUIT of the superconducting state.
We found evidence for the QUIT, but the evidence is at the border of validity
of the calculational approaches used. To have a definite theoretical proof
of the existence of the QUIT one needs to have better algorithms
and/or improvements in computer power. The results presented here are,
however, rather encouraging. Of course the ultimate test will be furnished
by experiment, and there too incipient indications of a low temperature
instability have also been reported in \cite{van-der-zant-thesis}.

In the large $\alpha_{\rm m}$ limit we used a perturbative expansion
in $1/\alpha_{\rm m}$ to find an effective partition function for a 
quantum a 2-D Coulomb gas. This model shows a renormalized 
conducting to insulating 
transition as the temperature is lowered. We did not find a low 
temperature QUIT instability in the large $\alpha_{\rm m}$ insulating phase.

We also performed extensive nonperturbative quantum Monte Carlo 
calculations of the JJA model. We concentrated our analysis on the 
helicity modulus $\Upsilon$. This quantity is directly related to 
the superfluid density in the array. Using $\Upsilon$ we determined the
superconducting to normal transition boundary. We found good agreement
between the critical temperatures obtained by QMC calculation and the 
semiclassical approximation. We also carried out a low temperature
$1/L_\tau$ extrapolation analysis of the $T_{QUIT}$  and found 
evidence for $T_{QUIT}\neq 0$ for relatively large values of 
$\alpha_m=2.0$ and $2.25$ but $T_{QUIT}= 0$ for $\alpha_m=2.5$. 
These calculations have, however, a strong $L_\tau$ dependence
and our QMC algorithm is not precise enough
to completely ascertain the nature of the low temperature phase.
Nonetheless, the results found here together with the scant
emerging experimental evidence for a low temperature instability
yields further support for the possible existence of a QUIT.

We also presented some QMC calculations of the inverse dielectric 
constant of the 2-D quantum Coulomb gas, in order to find the conducting to 
insulating phase boundary. We found that the present 
Monte Carlo path integral implementation of our model, that includes
phase and charge degrees of freedom,  
does not allow us to make reliable calculations
of this quantity. Our results for this transition are only qualitative.
Further technical improvements are needed in order to make solid 
quantitative statements about the $N-I$ phase boundary.

To use a QMC calculation to prove or disprove the presence of 
a low temperature instability in the superconducting state is not an easy 
task. However, these type of calculations give us upper temperature limits for 
the instability region. As far as our calculations could determine, the 
results for the superfluid density as a function of temperature are in
rather good agreement with experimental findings in JJA 
\cite{van-der-zant-thesis}, except for the incipient data 
on the reentrant transition in 
the nonperturbative region of $\alpha_{\rm m}$.

To further understand the QMC results at low temperatures and 
as a function of $L_\tau$, we have also implemented a self-consistent 
harmonic approximation analysis of the model, that includes phases and
charge freedoms.  We were able to make 
successful qualitative, and in some instances even quantitative 
comparisons between both calculational approaches.
One of the conclusions from these calculations is that the general
trend of the QMC results for $\Upsilon$ can 
be traced to the discretization of the imaginary time 
axis. For small $\alpha_{\rm m}$, the decrease of the helicity modulus
at low temperature is clearly due to this effect. 

Among the most significant aspects of the results presented in this 
paper is the quantitative agreement between our different calculational 
approaches and the corresponding experimental results 
in the superconducting-normal phase boundary with essentially 
only the measured capacitances as adjustable parameters. The
existence of the QUIT is also a significant result of this paper,
for it had not been studied in a model including realistic capacitances
in the model.

%
\acknowledgments
We thank J. Houlrick and M. Novotny for contributions and helpful
discussions in the early stages of the Monte Carlo study presented
here. This work has been partially supported  by  
NSF grants DMR-95-21845, and PHY-94-07194, at UC Santa Barbara.
\vskip 1.0cm
%
%
\appendix
\section*{A}
In this appendix we present the derivation of the renormalization group
equations in the insulating to normal region of the phase diagram up
to first order in ${\alpha_m}^{-1}$. We begin by writing the effective
partition function from Eqs. (\ref{z-with-z_eff}), (\ref{z_phi}), and
(\ref{s_f})
\begin{equation}
\label{z_eff}
       Z_{\rm eff}(\{m\}) = \left< \exp\Bigg[ -\frac{1}{\hbar} 
                                               \int_{0}^{\beta\hbar} d\tau 
                                               H_J(\{\overline\phi(\vec r)+
                                                     \phi_f(\tau,\vec r)+
                                                     (2\pi/\beta\hbar)
                                                     m(\vec r)\tau
                                                  \})
                                       \Bigg]
                            \right>_{\phi}.
\end{equation}
The phase average $<>_{\phi}$ is defined by
\begin{eqnarray}
\label{ave_phi}
      <A>_\phi = \frac{1}{Z_{\phi}}\prod_{\vec r} & &\int_{0}^{2\pi}
                 \frac{d\overline\phi(\vec r)}{2\pi} \int_{-\infty}
                 ^{\infty} {\cal D}\phi_f(\vec r) \ A\Big(
                  \{\overline\phi(\vec r)+
                  \phi_f(\tau,\vec r)\}\Big) \exp\Bigg[-\frac{1}{\hbar} 
                                                      S_f[\phi_f] \Bigg]. 
              \nonumber\\
\end{eqnarray}
From Eq.(\ref{hamiltonian}) the Josephson energy is
\begin{equation}
\label{h_j}
        H_J(\{\phi\}) = -E_J \sum_{<\vec r_1,\vec r_2>} \cos\Big(
                             \phi(\vec r_1) - \phi(\vec r_2)\Big).
\end{equation}
Inserting this equation into Eq. (\ref{ave_phi}), and using the expansion 
given in Eq. (\ref{h_j-expantion}), we first see that the integrations over 
the $\{\overline\phi\}$'s eliminate all the odd powers 
in $E_J$. Moreover up to the third term in 
Eq. (\ref{h_j-expantion}) we have integrals of the form
\begin{equation}
\label{integrals}
      \prod_{\vec r} \int_{0}^{2\pi} \frac{d\overline\phi(\vec r)}{2\pi}
      \exp\!\Big[\overline\phi(\vec r_1)-\overline\phi(\vec r_2)+
      \overline\phi(\vec r_3)-\overline\phi(\vec r_4)
      \Big] = \delta_{\vec r_1,\vec r_4}\ \delta_{\vec r_2,\vec r_3},
\end{equation}
where $\vec r_1\ne\vec r_2$ and $\vec r_3\ne\vec r_4$, because they are 
pairs of nearest neighbors.

Using this equation and the fact that Eq. (\ref{s_f}) is an even function
of the $\phi_f$'s we end up with 
\begin{eqnarray}
\label{approx-z_eff}
      Z_{\rm eff}(\{m\}) \approx 1 + 
       & &
      \left(\frac{E_J}{2\hbar}\right)^2
      \int_{0}^{\beta\hbar}\!\! d\tau \int_{0}^{\beta\hbar}\!\! d\tau'
         \sum_{<\vec r_1,\vec r_2>} \exp\Big[-G(\vec r_1,\vec r_2;\tau,\tau')
         \Big] \times \nonumber \\
      & & \times\cos\Big((2\pi/\beta\hbar)\big[m(\vec r_1)-m(\vec r_2)\big]
          (\tau-\tau')\Big)\ +\ O(E_J^4).
\end{eqnarray}
Here we have defined the correlation function
\begin{eqnarray}
\label{corr-func}
       G(\vec r_1,\vec r_2;\tau_1,\tau_2) &=& -\ln\Bigg\{\frac{1}{Z_\phi}
       \prod_{\vec r}\int_{-\infty}^{\infty}\!\! {\cal D}\phi_f(\vec r)
       e^{-\frac{1}{\hbar}S_f[\phi_f]}\ \exp\Bigg[\int_{0}
       ^{\beta\hbar} \sum_{\vec r} j(\tau,\vec r)\phi_f(\tau,\vec r)\Bigg]
       \Bigg\}, \nonumber\\
       & & \\
\label{j-def}
       j(\tau,\vec r) &=& i\Big[\delta(\tau-\tau_1)-\delta(\tau-\tau_2)\Big]
                         \Big[\delta_{\vec r,\vec r_1}-\delta_{\vec r,
                              \vec r_2}\Big].
\end{eqnarray}
The easiest way to perform this integral is to go to Fourier space, and
using Eq. (\ref{boundary-for-phi_f}) we can write

\begin{eqnarray}
\label{fourier-expantion}
       \phi_f(\tau,\vec r) &=& 
\sqrt{\frac{2}{\beta\hbar}}\sum_{n=1}^{\infty}
                               \psi_n(\vec r) \sin(\nu_n \tau), \\
\label{nu_n}
        \nu_n &=& \left(\frac{\pi}{\beta\hbar}\right)n.
\end{eqnarray}
From this equation, Eq. (\ref{corr-func}) can be written as
\begin{eqnarray}
\label{corr-func-in-fourier}
       G(\vec r_1,\vec r_2;\tau_1,\tau_2) &=& -\ln\Bigg\{\frac{1}{Z_\phi}
       \prod_{n=1}^{\infty} \prod_{\vec r}\int_{-\infty}^{\infty}\!\!
        d\psi_n(\vec r)
       e^{-\frac{1}{\hbar}S_f[\psi_n]}\ \exp\Bigg[\sum_{n=1}^{\infty}
         \sum_{\vec r} j_n(\vec r)\psi_n(\vec r)\Bigg]
       \Bigg\}, \nonumber\\
       & & \\
\label{s_fn}
       S_f[\psi_n] &=& \frac{1}{2}\sum_{n=1}^{\infty}\sum_{\vec r_3,\vec r_4}
                     \psi_n(\vec r_3)\left\{(\hbar\nu_n/q)^2 {\bf C}
                     (\vec r_3,\vec r_4)\right\}\psi_n(\vec r_4), \\
\label{j_n}
       j_n(\vec r) &=& \sqrt{\frac{2}{\beta\hbar}}\int_{0}^{\beta\hbar}\!\!
                        d\tau \sin(\nu_n\tau) j(\tau,\vec r).
\end{eqnarray}
Therefore the correlation function is
\begin{eqnarray}
\label{corretion-function_1}
      G(\vec r_1,\vec r_2;\tau_1,\tau_2) &=& -\frac{q^2}{2\hbar}\sum_{n=1}
      ^{\infty} \sum_{\vec r_3,\vec r_4}\frac{1}{\nu_n^2}\ \ j_n(\vec r_3)
      {\bf C}^{-1}(\vec r_3,\vec r_4) j_n(\vec r_4), \\
\label{correlation-function_2}
      &=& -q^2\beta\sum_{\vec r_3,\vec r_4}
      \int_{0}^{\beta\hbar}\!\! d\tau \int_{0}^{\beta\hbar}\!\! d\tau'
      j(\tau,\vec r_3)\left\{{\bf g}(\tau,\tau'){\bf C}^{-1}(\vec r_3,
      \vec r_4)\right\} j(\tau',\vec r_4),\nonumber\\
\end{eqnarray}
where we have used the result $(\tau<\tau')$
\begin{equation}
\label{def-g}
    {\bf g}(\tau,\tau') = \frac{1}{(\beta\hbar)^2}\sum_{n=1}^{\infty}
                           \frac{\sin(\nu_n\tau) \sin(\nu_n\tau')}{\nu_n^2}
                         = 
\left(1-\frac{\tau'}{\beta\hbar}\right)\frac{\tau}
                           {\beta\hbar}.
\end{equation}
Using Eq. (\ref{j-def}) we finally find
\begin{equation}
\label{correlation-function_3}
      G(\vec r_1,\vec r_2;\tau_1,\tau_2) = 2q^2 \beta 
\frac{|\tau_1-\tau_2|}
      {\beta\hbar}\left(1-\frac{|\tau_1-\tau_2|}{\beta\hbar}\right)\Big[
      {\bf C}^{-1}(|\vec 0|)-{\bf C}^{-1}(|\vec r_1-\vec r_2|)\Big].
\end{equation}
Inserting this equation in Eq. (\ref{approx-z_eff}), and noting
that here $|\vec r_1-\vec r_2|=|\vec d|$ we have
\begin{eqnarray}
\label{z_eff-final}
    Z_{\rm eff}(\{m\}) \approx 1 + 
     & &
    \left(\frac{E_J}{2\hbar}\right)^2\!\!
    \int_{0}^{\beta\hbar}\!\! d\tau\!\! \int_{0}^{\beta\hbar}\!\! d\tau'\!\!
    \times\nonumber\\
    & & \times
    \sum_{<\vec r_1,\vec r_2>}\!\!\exp\!\Bigg\{-\frac{2q^2\beta}{zC_{\rm m}}
    \frac{|\tau-\tau'|}{\beta\hbar}\left(1-\frac{|\tau-\tau'|}{\beta\hbar}
    \right)\Big[1-C_{\rm s}{\bf C}^{-1}(|\vec 0|)\Big] \Bigg\} \times 
    \nonumber\\
    & & \times\cos\Big((2\pi/\beta\hbar)\big[m(\vec r_1)-m(\vec r_2)\big]
          (\tau-\tau')\Big)\ +\ O(E_J^4).
\end{eqnarray}
At this point we can perform the following symmetrical change of variables
\begin{eqnarray}
\label{change-of-var}
          x_1 = \frac{1}{\sqrt{2}}\frac{(\tau-\tau')}{\beta\hbar}, \\
          x_2 = \frac{1}{\sqrt{2}}\frac{(\tau+\tau')}{\beta\hbar}.
\end{eqnarray}
Now we see that the integral over $x_2$ can be calculated explicitly
leaving only $x_1$ to be integrated.  We find then
\begin{eqnarray}
\label{zeff-final}
      Z_{\rm eff}(\{m\})\approx 1 + \frac{(\beta E_J)^2}{2}
         \sum_{<\vec r_1,\vec r_2>} 
       & &\int_{0}^{1/2}\!\! dx_1\ \exp\!
        \Bigg\{-\frac{2q^2\beta}{z C_{\rm m}} 
      \Big[1-C_{\rm s}{\bf C}^{-1}(|\vec 0|)\Big]\ x_1(1-x_1) \Bigg\}
      \times\nonumber\\
       & &\ \ \ \ \times\cos\Big(2\pi\big[m(\vec r_1)-m(\vec r_2)\big]x_1\Big) 
          +\ O(E_J^4).
\end{eqnarray}
Which is the result used in Eq. (\ref{z-with-z_eff-2}).
%
%
\section*{B}
Here we give the details of the derivation of the effective
free energy in the variational calculation, from which we obtain the
expressions for $<{\Delta\phi}^2>_{H}$ which was used to obtain the
helicity modulus in Section \ref{sec:harmonic}.

We start by defining the following effective quantities for the $m$'s and
the $\phi(0,\vec r)$ interaction terms
\begin{eqnarray}
\label{def-for-j}
      j(\vec r) &=& -\pi\beta E_J\Gamma \left[\left(1-\frac{1}{L_\tau}\right)-
                    h\left(\beta q\sqrt{E_J\Gamma/C_{\rm m}}\right)\right]
                    \sum_{\vec r_1} m(\vec r_1) {\bf O}(\vec r_1,\vec r),\\
\label{def-s-tilde-appendix}
      \frac{1}{\hbar}\tilde S_m &=& \frac{(2\pi)^2}{2}\left[\frac{C_{\rm m}}
      {\beta q^2}+\frac{\beta E_J\Gamma}{6}\left(1-\frac{1}{L_\tau}
      \right)\left(2-\frac{1}{L_\tau}\right)-(\beta E_J\Gamma) 
      g\!\left(\beta q\sqrt{E_J\Gamma/C_{\rm m}}\right)\right] 
      \times\nonumber\\
      & &\times\sum_{\vec r_1,\vec r_2} m(\vec r_1){\bf O}(\vec r_1,\vec r_2)
      m(\vec r_2).
\end{eqnarray}
The functions $h(\Lambda)$ and $g(\Lambda)$ are given in terms of  
Matsubara frequency summations
\begin{eqnarray}
\label{def-for-h}
      h(\Lambda) &=& \frac{1}{2}\left(\frac{\Lambda}{L_\tau}\right)^2\
                     \sum_{n=1}^{L_\tau-1} \frac{\sin^2(\pi n/L_\tau)}
                     {\left[\Lambda^2+2L_\tau^2(1-\cos(\pi n/L_\tau))\right]}
                     \left[\frac{1-\cos(\pi n)}{1-\cos(\pi n/L_\tau)}
                     \right]^2 \nonumber\\
                               \nonumber\\
                 &=& 1-\frac{1}{L_\tau} + \frac{2}{L_\tau}\left(
                     \frac{L_\tau}{\Lambda}\right)^2 \left[
                     \frac{\sinh( (L_\tau-1)\lambda)+\sinh(\lambda)}
                          {\sinh(L_\tau\lambda)} - 1 \right],\\
                               \nonumber\\
\label{definition-of-g}
      g(\Lambda) &=& \frac{1}{2}\left(\frac{\Lambda}{L_\tau}\right)^2
                     \sum_{n=1}^{L_\tau-1} \frac{\sin^2(\pi n/L_\tau)}
                     {\left[\Lambda^2+2L_\tau^2(1-\cos(\pi n/L_\tau))\right]} 
                     \frac{1}{\left[1-\cos(\pi n/L_\tau)\right]^2}\nonumber\\
                     \nonumber\\
                 &=& \frac{1}{L_\tau}\left(\frac{L_\tau}{\Lambda}\right)^2
                     \left[\frac{\sinh((L_\tau-1)\lambda)}{\sinh(L_\tau 
                      \lambda)}-\left(\frac{L_\tau-1}{L_\tau}\right)
                      \left\{1-\frac{1}{6}\left(\frac{\Lambda}{L_\tau}
                      \right)^2 (2L_\tau-1)\right\}\right], \nonumber\\
                 & & \\
\label{def-lambda}
      \lambda &=& \ln\left\{1+\frac{1}{2}\left(\frac{\Lambda}{L_\tau}\right)
                  \left[\left(\frac{\Lambda}{L_\tau}\right) + \sqrt
                  {4+\left(\frac{\Lambda}{L_\tau}\right)^2}\right]\right\}.
\end{eqnarray}

To complete the calculation of $Z_H$ we need to make one more approximation 
because the integral can not be explicitly calculated due to the 
finite limits of integration. Therefore we  extend the 
limits of integration to the entire real axis. We expect 
that this approximation is good for 
large $\Gamma$ or small $T$, which is our region of interest.
For small $\Gamma$ it would give an overestimation of the transition 
temperature. After extending the limits of integration we find
that Eq. \ref{z_h-mid-way} becomes
\begin{eqnarray}
\label{z_h-final}
      Z_H =  & & 
            \left(\frac{C_{\rm m}}{\beta^2 q^2 E_J\Gamma}\right)^{L_xL_y/2}
            \left[1-h\left(\beta q\sqrt{E_J\Gamma/C_{\rm m}}\right)\right]^
            {-L_xL_y/2} \times\nonumber\\
            & & \times \left[\prod_{n=1}^{L_\tau}\left[1+\frac{(\beta^2 q^2
                E_J\Gamma/C_{\rm m})}{2L_\tau^2[1-\cos(\pi n/L_\tau)]}
                \right]\right]^{-L_xL_y/2} \!\! \prod_{\vec r} 
                \sum_{m(\vec r)=-\infty}^{\infty}
                \exp\left\{-\frac{1}{\hbar}\tilde{\tilde S}_m\right\},
\end{eqnarray}
where the effective action for the $m$'s is given by
\begin{eqnarray}
\label{eff-action-for-ms}
    & & \frac{1}{\hbar}\tilde{\tilde S}_m = \frac{1}{2} J_{\rm eff}
      \sum_{\vec r_1,\vec r_2} m(\vec r_1){\bf O}(\vec r_1,\vec r_2)
      m(\vec r_2),
       \nonumber\\
\label{def-of-Jeff}
       & &J_{\rm eff} = (2\pi)^2\Bigg[\frac{C_{\rm m}}{\beta q^2}+
                       \frac{\beta E_J\Gamma}{6}
                       \left(1-\frac{1}{L_\tau}\right)
                       \left(2-\frac{1}{L_\tau}\right)
                       - (\beta E_J\Gamma) g\!\left(\beta q
                         \sqrt{E_J\Gamma/C_{\rm m}}\right) - \nonumber\\
                         \nonumber\\
       & &\ \ \ \ \ \ \ \ \ \ \ \ \ \ \ \ 
                      -\frac{\beta E_J\Gamma \left\{(1-1/L_\tau)-h
                       \left(\beta q\sqrt{E_J\Gamma/C_{\rm m}}\right)
                       \right\}^2}{4\left(1-h\left(\beta q
                       \sqrt{E_J\Gamma/C_{\rm m}}\right)\right)}\Bigg].
\end{eqnarray}
and $\bf {O}$ is the lattice  Laplacian operator.
The free energy in the harmonic approximation is therefore
\begin{eqnarray}
\label{def-of-f_h}
  \frac{\beta F_H}{L_xL_y} = \frac{\beta F_{DG}(J_{\rm eff})}{L_x L_y}+
                   \frac{1}{2}\Bigg\{& &\ln\left(\frac{\beta^2 q^2 E_J\Gamma}
                   {C_{\rm m}}\right) + 
                   \ln\left[1-h\left(\beta q\sqrt{E_J\Gamma/C_{\rm m}}
                   \right)\right]+ \nonumber\\
                   & & +\sum_{n=1}^{L_\tau-1} \ln\left[1+\frac{(\beta^2 q^2
                   E_J\Gamma/C_{\rm m})}{2L_\tau^2[1-\cos(\pi n/L_\tau)]}
                   \right]\Bigg\}. 
\end{eqnarray}
The function $F_{DG}$ is related to the partition function of the
Discrete Gaussian Model (DGM) defined by  
\begin{eqnarray}
\label{def-of-dg-free-energy}
       & &\beta F_{DG}(J) = -\ln\left[\prod_{\vec r}
       \sum_{m(\vec r)=-\infty}^{\infty} \exp\left\{-\frac{J}{2} 
       \sum_{<\vec r_1,\vec r_2>} 
       [m(\vec r_1)-m(\vec r_2)]^2 \right\}\right].
\end{eqnarray}

We will ignore this term for the moment,
since the discrete nature of its excitations makes it very small. 
We  need to calculate the average of the harmonic part of the 
Hamiltonian, i.e.
\begin{eqnarray}
\label{average-over-h_h}
      <H_H>_H &=& E_J\Gamma L_x L_y \Big<[\phi(\tau,\vec r+\vec d)-
                \phi(\tau,\vec r)]^2 \Big>_H = -\frac{\Gamma}{\beta}
                \frac{\partial \ln Z_H}{\partial\Gamma}
                 = \Gamma \frac{\partial F_H}{\partial\Gamma}\nonumber\\
               &=& \frac{L_x L_y}{2\beta}\left\{1-\frac{\beta q}{2}
                   \sqrt{\frac{E_J\Gamma}{C_{\rm m}}} 
                   \frac{h'\left(\beta q\sqrt{E_J\Gamma/C_{\rm m}}
                   \right)}{\left[1-h\left(\beta q\sqrt{E_J\Gamma/
                   C_{\rm m}}\right)\right]} + f\left(\beta q
                   \sqrt{E_J\Gamma/C_{\rm m}}\right)\right\},\nonumber\\
%
%
\end{eqnarray}
where $\vec d$ joins nearest neighbors, and $f(\Lambda)$ is 
given by a summation over Matsubara frequencies resulting in,
\begin{equation}
\label{function-f}
        f(\Lambda) =\frac{1}{2}\left\{ (L_\tau-1) - \frac{L_\tau}
                    {\left[1+(\Lambda/L_\tau)^2/2 \right]} 
                    \frac{\sinh\left((L_\tau-1)\lambda\right)}
                    {\sinh\left(L_\tau \lambda\right)}\right\}.
\end{equation}
 
On the other hand we know that if the Hamiltonian is Gaussian, then
the calculation of the average over the Josephson Hamiltonian is

\begin{equation}
\label{ave-over-Hj}
      < H_J >_H = 2E_J L_x L_y \left[ 1 - \exp\left\{
                  -\frac{1}{2} \left< [\phi(\tau,\vec r+\vec d)-\phi(\tau,
                  \vec r)]^2 \right> \right\} \right]. 
\end{equation}

Therefore, the relevant term in the variational free energy is given
by the average of the square of the lattice derivative over the harmonic 
Hamiltonian. Finally, we can write the variational free energy 

\begin{eqnarray}
\label{variational-free-energy}
       & &\frac{\beta F_V}{L_x L_y} = \frac{\beta F_H}{L_x L_y} +
       2\beta E_J \left[ 1 - \exp\left\{-\frac{1}{2} \left< (\Delta\phi)^2 
       \right> \right\} \right] - \beta E_J\Gamma \left< (\Delta\phi)^2
       \right>, \nonumber\\
       & & \\
\label{def-of-Dphi}
       & &<(\Delta\phi)^2>_H = <[\phi(\tau,\vec r+\vec d)-\phi(\tau,\vec r)
                                ]^2>_H.
\end{eqnarray}
with $<(\Delta\phi)^2>_H$ given in Eq. (\ref{average-over-h_h}).

The variational condition requires minimizing this function with
respect to $\Gamma$. Taking the derivative of this equation,
and using Eq. (\ref{average-over-h_h}), we find that 

\begin{equation}
\label{equation-for-gamma2}
     \Gamma = \exp\left\{ -\frac{1}{2}\left< (\Delta\phi)^2 
\right>\right\}.
\end{equation}
$\Gamma$ is then the normalized coupling constant of the spin wave
Hamiltonian, i.e. the normalized helicity modulus. 
Eq. (\ref{equation-for-gamma}) gives the condition to identify the phase 
transition as the point at which  $\Gamma\ne 0$ becomes a solution to the 
variational equations. The complete problem is to solve the combination of 
Eqs. (\ref{def-of-Dphi}) and (\ref{equation-for-gamma}). We can write 
Eqs. (\ref{def-of-Dphi}) as a function of $\Gamma$
\begin{equation}
\label{def-of-Dphi2}
       < (\Delta \phi)^2 >_H = \frac{1}{2\beta E_J \Gamma}
                   \left\{1-\beta E_J\sqrt{ 2\alpha_{\rm m}\Gamma }
                   \frac{h'\left(\beta E_J \sqrt{ 8\alpha_{\rm m}\Gamma }
                           \right)}
                        {\left[1-h\left(\beta E_J \sqrt{ 8\alpha_{\rm m}
                            \Gamma}\right)
                         \right]} + f\left(\beta E_J \sqrt{ 8\alpha_{\rm m}
                         \Gamma}\right)\right\}.
\end{equation}      
      
We now study the behavior of the helicity modulus as a function
of $\alpha_{\rm m}$ in the continuum limit, that is in the
$L_\tau\rightarrow\infty$ limit. In this limit we get
\begin{eqnarray}
\label{h-for-infty}
        h(\Lambda) &=& 1 - \frac{2}{\Lambda} \left( \frac{\cosh(\Lambda)-1}
                          {\sinh(\Lambda)}\right),\\
\label{g-for-infty}
        g(\Lambda) &=& \frac{1}{\Lambda^2}\left[1+\frac{\Lambda^2}{3}-
                       \Lambda\frac{\cosh(\Lambda)}{\sinh(\Lambda)}\right],\\
\label{f-for-infty}
        f(\Lambda) &=& \frac{\Lambda}{2} \frac{\cosh(\Lambda)}
                       {\sinh(\Lambda)} - \frac{1}{2},\\
\label{Dphi-for-infty2}
       < (\Delta \phi)^2 >_H &=& \frac{1}{2}\sqrt{\frac{2\alpha_{\rm m}}
                                 {\Gamma}} \left(\frac{\sinh\left(\beta E_J 
                                 \sqrt{ 8\alpha_{\rm m}\Gamma}\right)}
                                 {\cosh\left(\beta E_J \sqrt{ 
                                 8\alpha_{\rm m}\Gamma}\right)-1}\right).
\end{eqnarray}
We used Eqs. (\ref{equation-for-gamma2}) and (\ref{Dphi-for-infty2}) to find
the helicity modulus for $\alpha_{\rm m} = 0$ and 1.  The results of 
these calculations are shown in Fig. \ref{fig:fig14-ys}. This figure shows 
several interesting features.  First the helicity modulus for 
$\alpha_{\rm m}=0$ has the right shape, i. e., a linear $T$ behavior 
near $T=0$, and a jump at the transition temperature.  The transition 
temperature is, however, overestimated. Note that the $\alpha_{\rm m}=1$ 
helicity modulus has a flat behavior at low temperatures, which must be 
contrasted with the $\alpha_{\rm m}=0$ result.
%
%
%
%
%

%
%
\newpage

%
%
%
\begin{figure}
%
\caption{
  The phase diagram for unfrustrated (f=0) and 
  fully frustrated (f=1/2) arrays. Squares denote the experimental results 
  from Ref. [5]. The other symbols give the QMC results including 
  their statistical errors.  The dotted lines for f=0 is a naive extrapolation 
  of the QMC results to zero temperature, while the dot-dashed curve is the
  result of the WKB-RG calculation.
  In the f=1/2 case we have rescaled the MC results so that
  the MC curve would coincide with the experimental result
  at $\alpha_{\rm m}(f=1/2)=0$.i
}
\label{fig:fig01-phase-diagram}
\end{figure}
%
%
\begin{figure}
%
\caption{
         Renormalization group flow diagram. The discontinuous line 
         indicates the vortex pair density as a function of temperature. See 
         text for a discussion of the analysis of this RG flow diagram.
        }
\label{fig:fig02-rg-flow}
\end{figure}
%
%
\begin{figure}
%
\caption{
         The helicity modulus $\Upsilon$ vs. temperature T for 
         $\alpha_{\rm m}=0.5$, $L_x = L_y = 32$, and two different 
         values of $L_\tau = 6$, and  8. This figure shows that for 
         this value of $\alpha_{\rm m}$ the results have converged 
         to the infinite $L_\tau$ limit already for $L_\tau = 6$. The 
         line is $(2k_B T/E_J\pi)$. The intersection of this line with 
         $\Upsilon(T)$ is used to determine the normal to superconducting 
         transition temperature.
        }
\label{fig:fig03-alpha-0p5}
\end{figure}
%
%
\begin{figure}
%
\caption{
         The helicity modulus vs. temperature for $\alpha_{\rm m} = 1.25$, 
         $L_x = L_y = 20$, and three different values of $L_\tau = 8$, 16, 
         and 32. This figure shows that a convenient convergence criteria 
         for the $L_\tau\rightarrow\infty$ limit is 
         $P=L_\tau/(\beta E_J \alpha_{\rm m}) > 5$.
        }
\label{fig:fig04-alpha-1p25}
\end{figure}
%
%
\begin{figure}
%
\caption{
         The same as Fig. \ref{fig:fig04-alpha-1p25} with 
         $\alpha_{\rm m}=1.75$, $L_x = L_y = 20$, and $L_\tau = 16$, 
         32 and 64. Again the criteria for convergence $P>5$ works in 
         this figure. Note that the departure from the infinite $L_\tau$ 
         limit is in the form of a small reentrant type dip in $\Upsilon$ as 
         the temperature is lowered.
        }
\label{fig5-alpha-1p25}
\end{figure}
%
%
%
%
\begin{figure}
%
\caption{
          The helicity modulus vs. temperature for $\alpha_m = 2.25$,
          $L_x = L_y = 20$, and $L_\tau = 32, 48, 64$, and 96. 
          As in Fig. 5, $\Upsilon$ has a dip as the temperature is
          lowered and shows a reentrant--like behavior at
          lower temperatures.
        }
\label{fig:fig06-alpha-2p25}
\end{figure}

\begin{figure}
%
\caption{
         $\Upsilon$ and inverse dielectric constant $\epsilon^{-1}$ vs 
         temperature for $\alpha_{\rm m} = 2.25$, $L_x = L_y = 20$, 
         and $L_\tau = 32$. As the temperature is lowered, the drop 
         in the helicity modulus is correlated with the rise in the 
         inverse dielectric constant, signaling the decoupling
         between imaginary time planes.
        }
\label{fig:fig08-Y-1_e-alpha-2p25}
\end{figure}
%
%
\begin{figure}
%
\caption{
         $\Upsilon$ vs temperature for $\alpha_{\rm m} = 2.5$, 
         $L_x=L_y=20$, and $L_\tau=32$, 48, 64, 80, 96, and 128. 
         At this value of $\alpha_{\rm m}$ the drop in the helicity 
         modulus at low temperatures yields essentially $\Upsilon=0$.  
         This abrupt drop is probably due to having a finite $L_\tau$. 
         Note that the decrease of the $\Upsilon$ as $T$ is lowered 
         happens even before the plane decoupling temperature.
       }
\label{fig:fig09-alpha-2p5}
\end{figure}
%
%
\begin{figure}
%
\caption{
         The imaginary time correlation function $C(\tau)$ vs. $\tau$, for 
         $\alpha_{\rm m}= 1.75$, $L_x = L_y = 20$, and $L_\tau = 32$ for 
         $T=0.35$, 0.30, and 0.18. As the temperature is lowered, the 
         correlation time decreases until at $(k_BT/E_J)\approx 0.3$
         the correlation time is shorter than one lattice spacing in the time
         direction. At this point the value of $L_\tau$ is too small to produce
         a convergent calculation of the helicity modulus.
        }
\label{fig:fig07-corr-t-alpha-1p75}
\end{figure}
%
%
\begin{figure}
%
\caption{
          The imaginary time correlation function $C(\tau)$ vs $\tau$ for 
          $\alpha_{\rm m}=2.5$, $L_x = L_y = 20$, and $L_\tau = 64$, at 
          $(k_bT/E_J)=0.36$ and 0.10. As in Fig. 
          \ref{fig:fig07-corr-t-alpha-1p75} this correlation function for 
          temperatures below the decoupling temperature is consistent with 
          having zero decorrelation time.
        }
\label{fig:fig10-corr-t-alpha-2p5}
\end{figure}
%
%
\begin{figure}
%
\caption{
          The helicity modulus vs. temperature for $\alpha_{\rm m} = 2.75$, 
          $L_x = L_y = 20$, and $L_\tau = 64$. The upper curve was obtained 
          while lowering the temperature.  The lower curve corresponds to 
          increasing it. The disparity in the results is most likely due to 
          the difficulty in establishing phase coherence in the 64 equal time 
          planes as $T$ is increased, above the plane decoupling temperature.
        }
\label{fig:fig11-alpha-2p75}
\end{figure}
%
%
%
\begin{figure}
%
\caption{
         Estimate of the plane decoupling transition vs $1/L_\tau$ 
         extracted from results similar to those of Fig. 
	\ref{fig:fig06-alpha-2p25} for 
         $\alpha_{\rm m}=2.0$, $L_x =L_y = 20$ and $L_\tau=
	48, 64, 80, 96, 128$.  The line from a least
         squares fit suggests a non zero decoupling 
         temperature in the limit $1/L_\tau=0$. This figure is 
         also consistent with the condition $P\approx 5$.
        }
\label{fig:fig12(a)-finite-size-alpha-2}
\end{figure}
%

%
\begin{figure}
%
\caption{ The same as Fig. \ref{fig:fig12(a)-finite-size-alpha-2}. 
         Data extracted from Fig. \ref{fig:fig06-alpha-2p25} for 
         $\alpha_{\rm m}=2.25$ and $L_x =L_y = 20$.  The line from a least
         squares fit also seems to suggest a non zero decoupling 
         temperature in the limit $L_\tau\to\infty$. This figure is 
         also consistent with the $P\approx 5$ condition.
        }
\label{fig:fig12(b)-finite-size-alpha-2p25}
\end{figure}
%
%
%
%
\begin{figure}
%
\caption{
           Same as Fig. \ref{fig:fig12(a)-finite-size-alpha-2}. 
           In this case, extracted from Fig. \ref{fig:fig09-alpha-2p5}, 
           for $\alpha_{\rm m} = 2.5$ and $L_x = L_y = 20$. 
           The least squares straight line fit seems to suggest a 
	   negative decoupling temperature in the limit $1/L_\tau=0$.
}
\label{fig:fig12(c)-finite-size-alpha-2p5}
\end{figure}
%
%
%
\begin{figure}
%
\caption{ 
         The spin wave stiffness vs $T$ (related to the
         normalized helicity modulus)  obtained from the
	 self-consistent-harmonic approximation for $\alpha_{\rm m}=0$ and 
         $\alpha_{\rm m}=1$.
        }
\label{fig:fig14-ys}
\end{figure}
%
%
\begin{figure}
%
\caption{
         The $T=0$ helicity modulus at as a function of $\alpha_{\rm m}$, 
         obtained from the SCHA. Extrapolated $T=0$ $\Upsilon$ results 
         from the Monte Carlo simulations are shown as diamonds joined 
         by a line. From this figure it is clear that the SCHA gives
         reasonable results at low temperatures and for small values
         of $\alpha_{\rm m}$.
       } 
\label{fig:fig15-ys-T=0}
\end{figure}
%
%
%
\begin{figure}
%
\caption{
         The normalized helicity modulus $\Gamma$ for $\alpha_{\rm m}=1.0$ 
         as a function of $T$ for increasing values of $L_\tau=$ 5,10,20,  
         and 100. The extrapolated plot for $L_\tau=\infty$ is shown as a 
         solid line. 
        }
\label{fig:fig17-ys-lt}
\end{figure}
%
%
%
%
\begin{figure}
%
\caption{
         The normalized helicity modulus $\Gamma$ for $\alpha_{\rm m}=1.0$ 
         and $L_\tau=10$, 20, and 40. The crossover temperature $T^*$ is 
         calculated using $(L_\tau/(\beta E_J)\alpha_{\rm m})=6$, so that 
         as $L_\tau$ increases the crossover temperature decreases. 
         Here the solid line correspond to $L_\tau=10$, the dashed line
         to $L_\tau=20$ and the dot-dashed line to $L_\tau=40$.
         See text for discussion of these results.
        }
\label{fig:fig19-cross-over}
\end{figure}
%
%
%
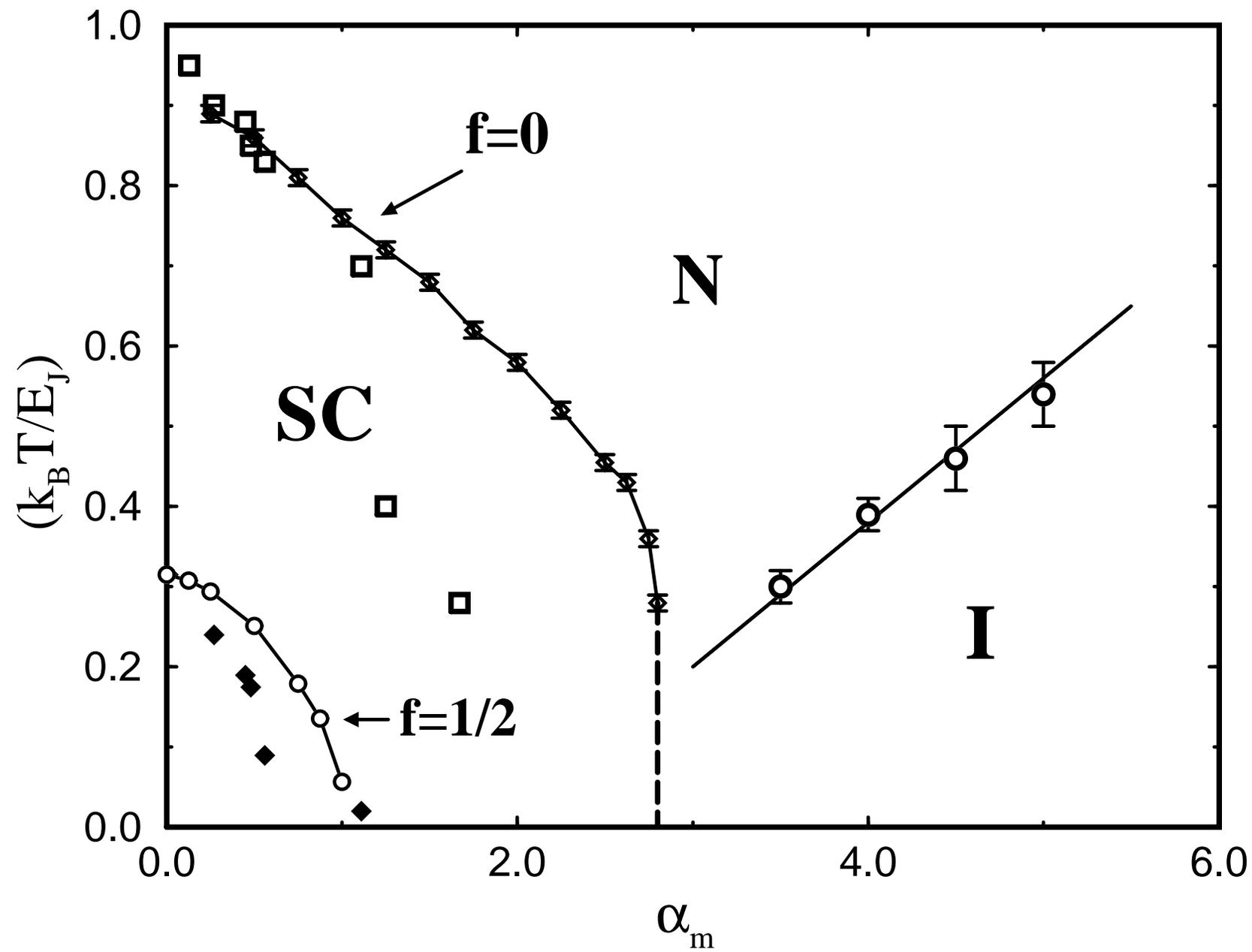
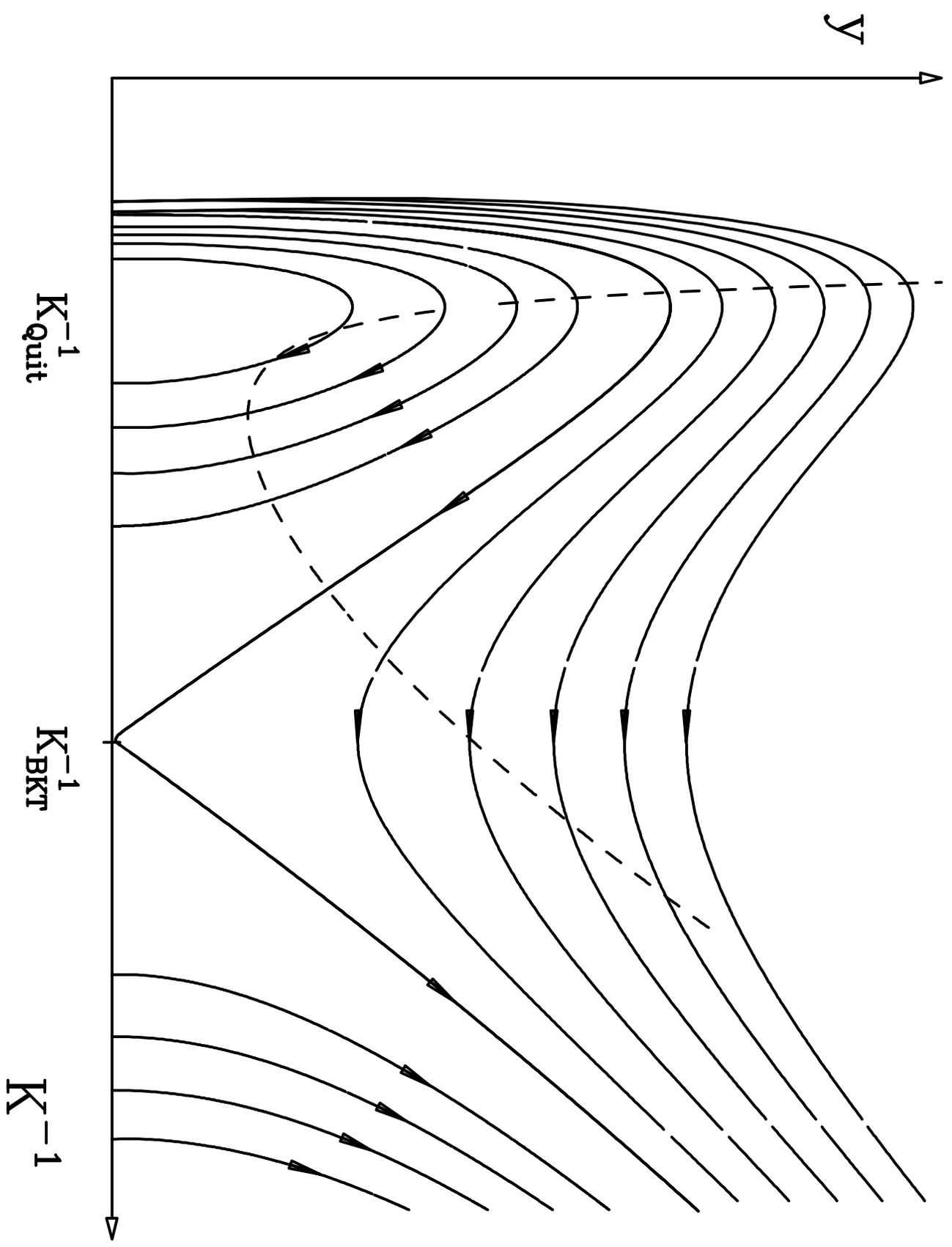
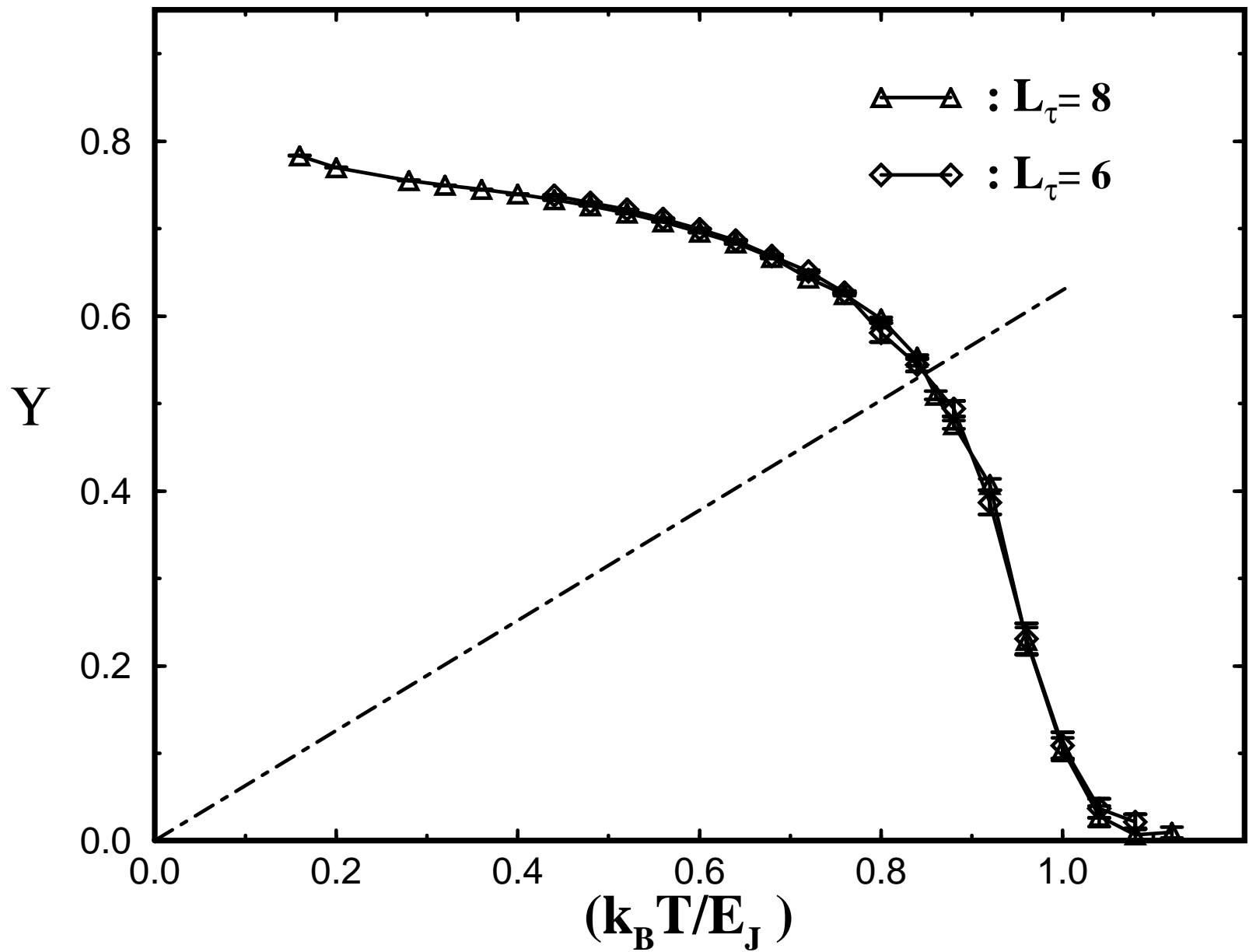
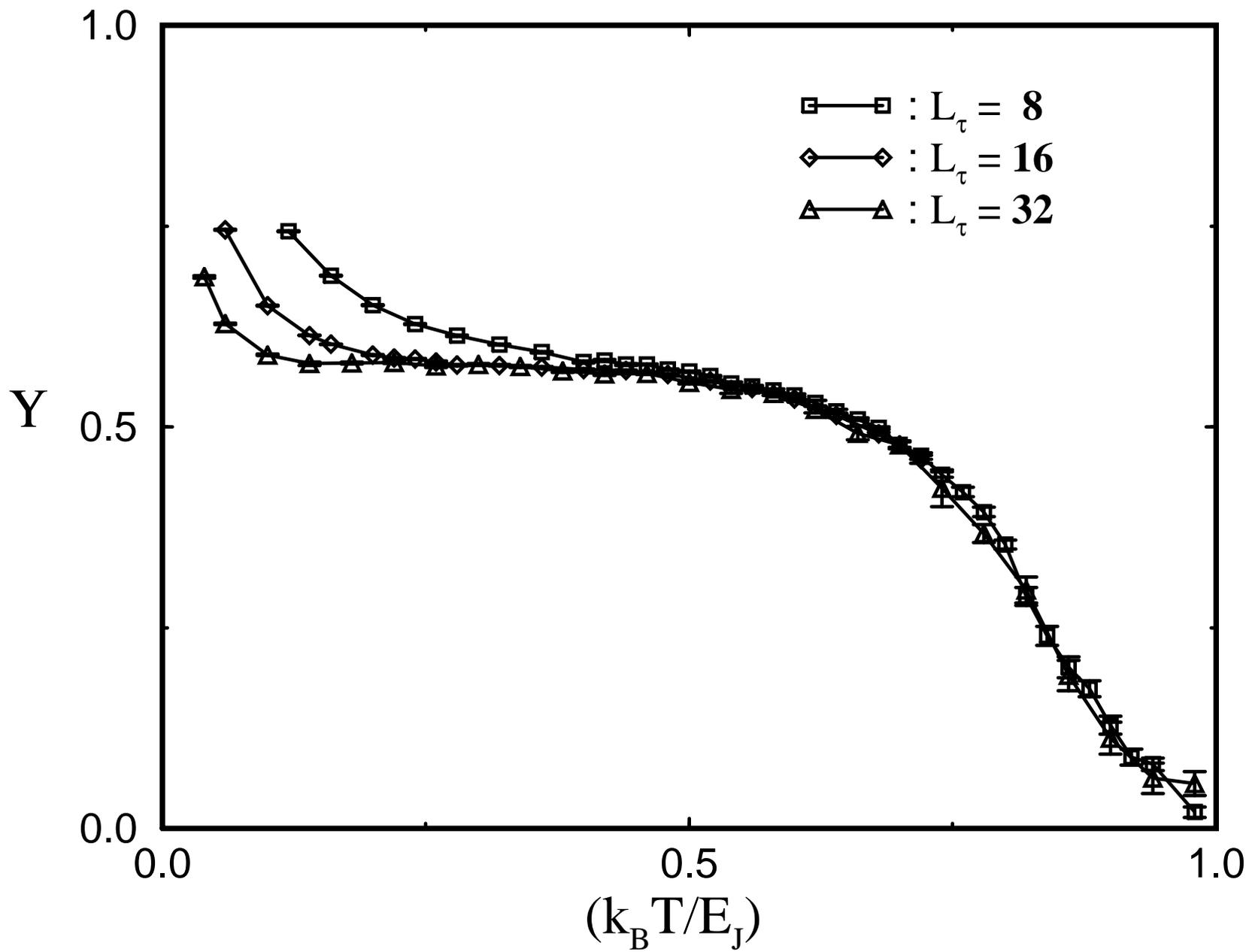
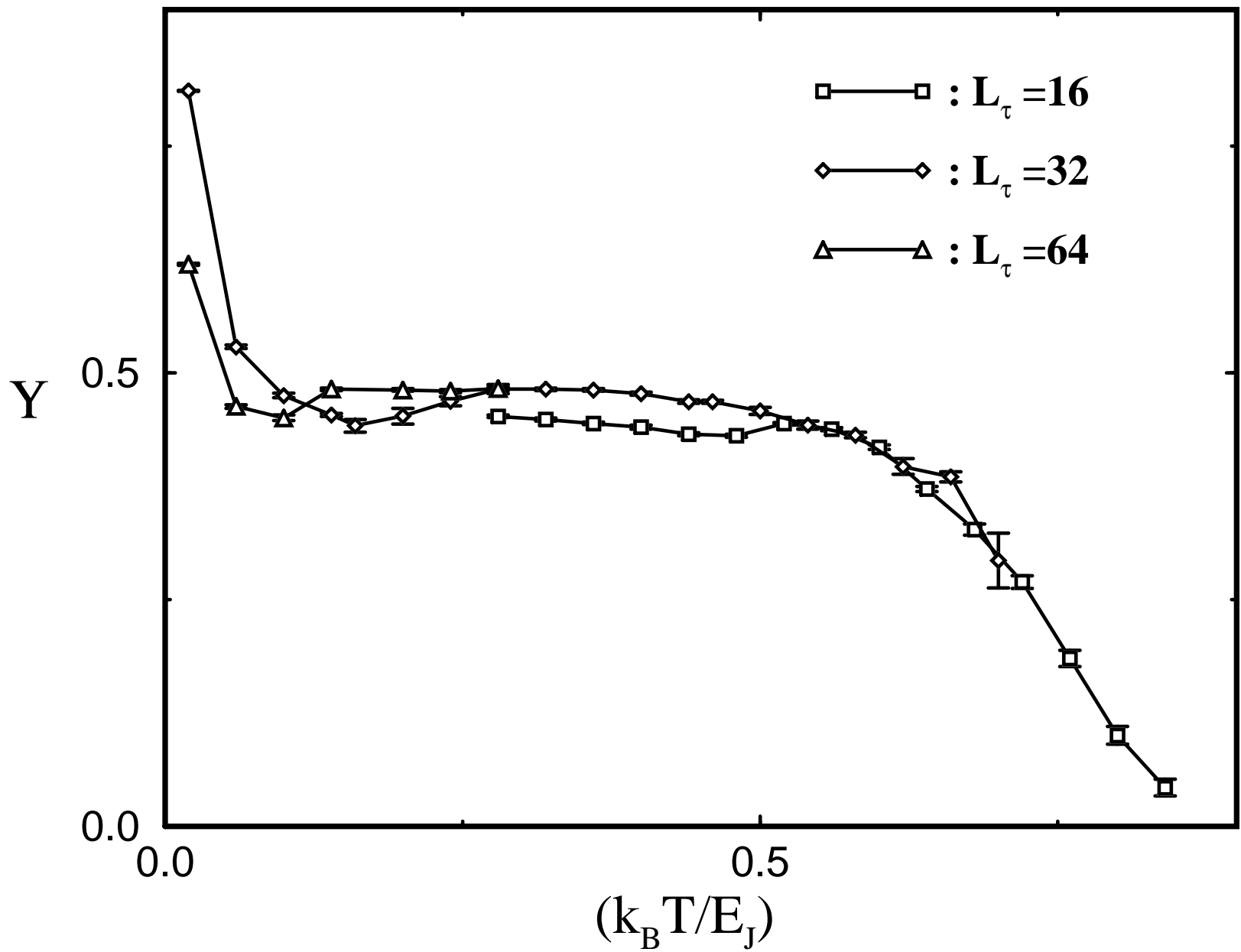
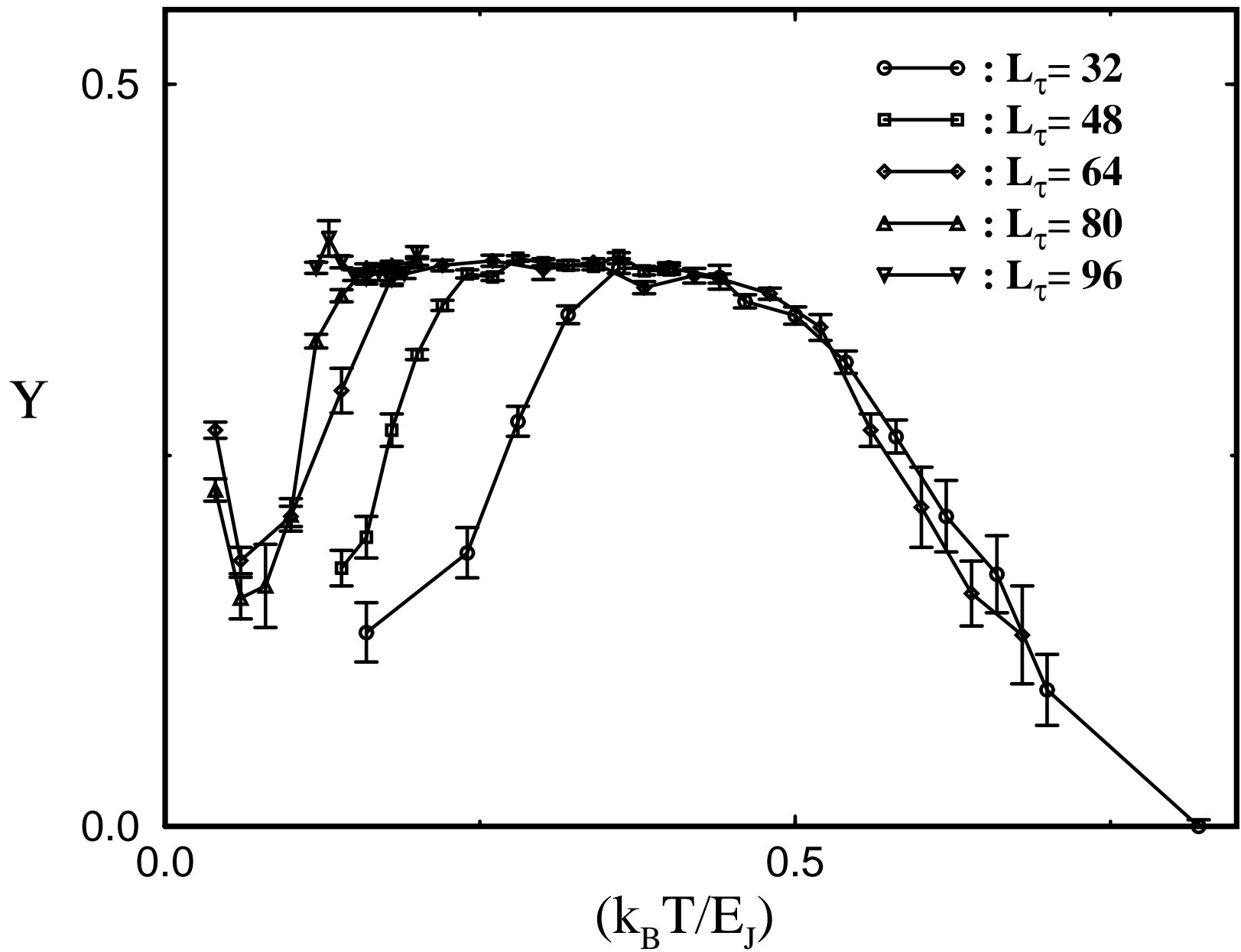
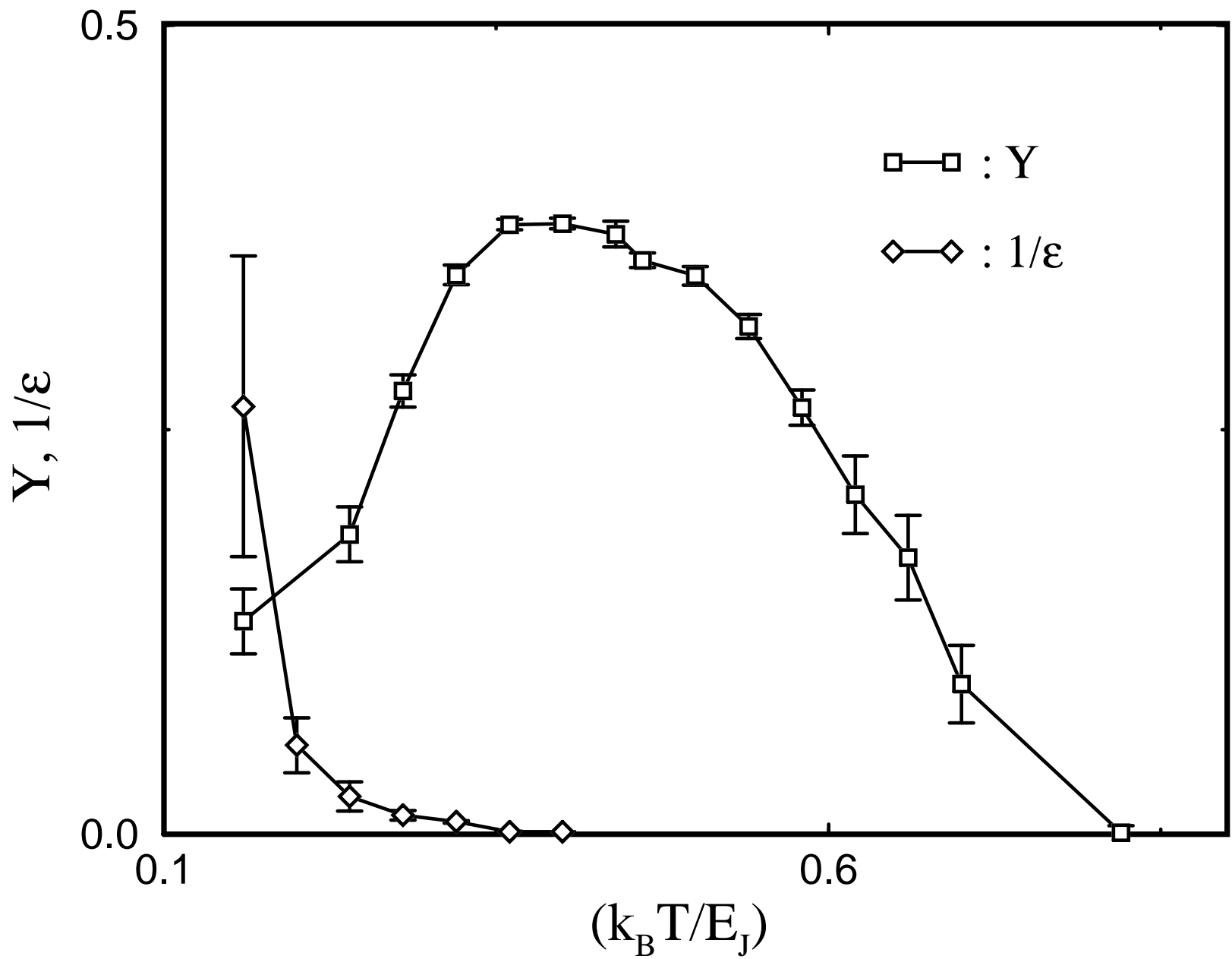
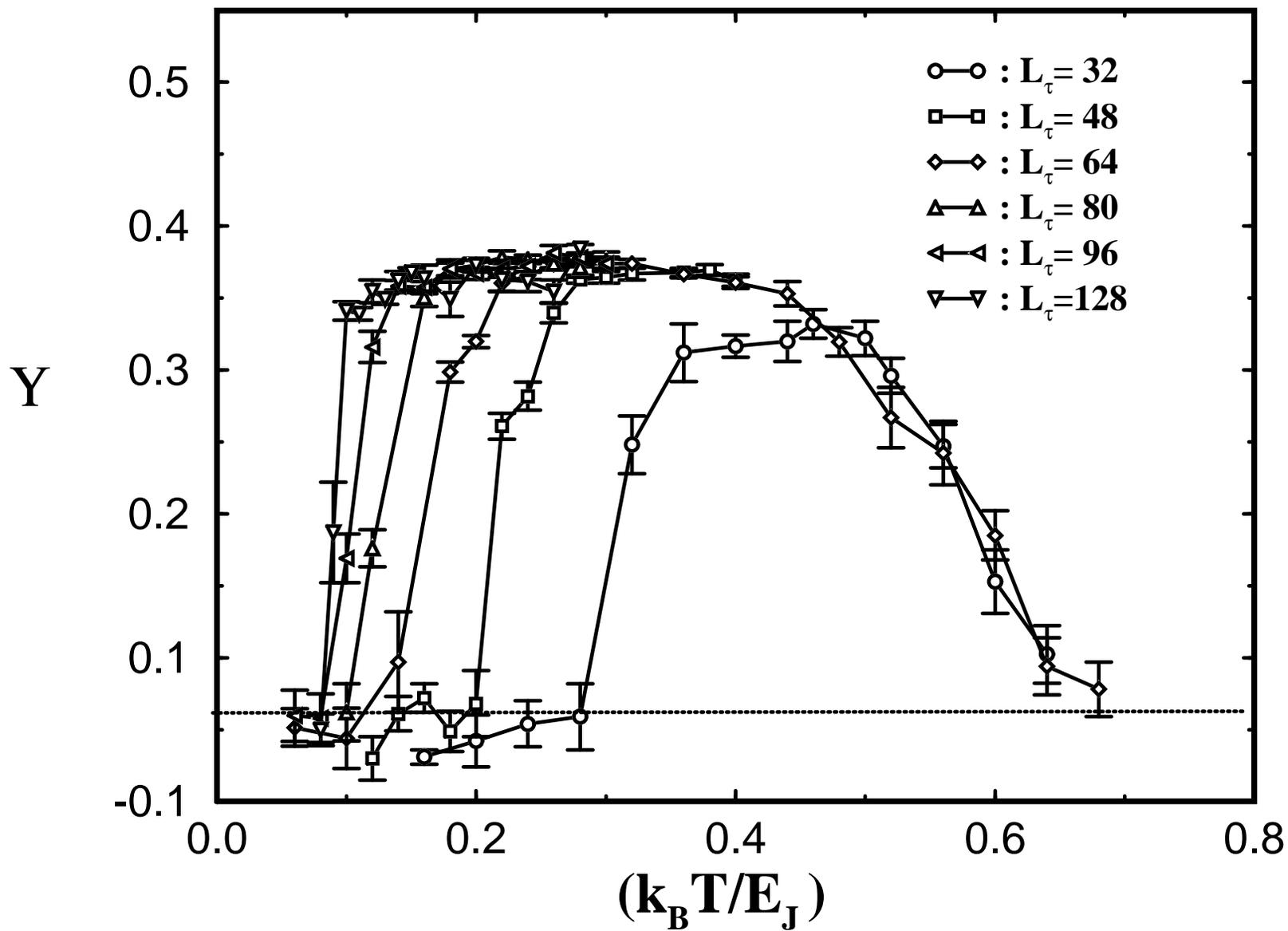
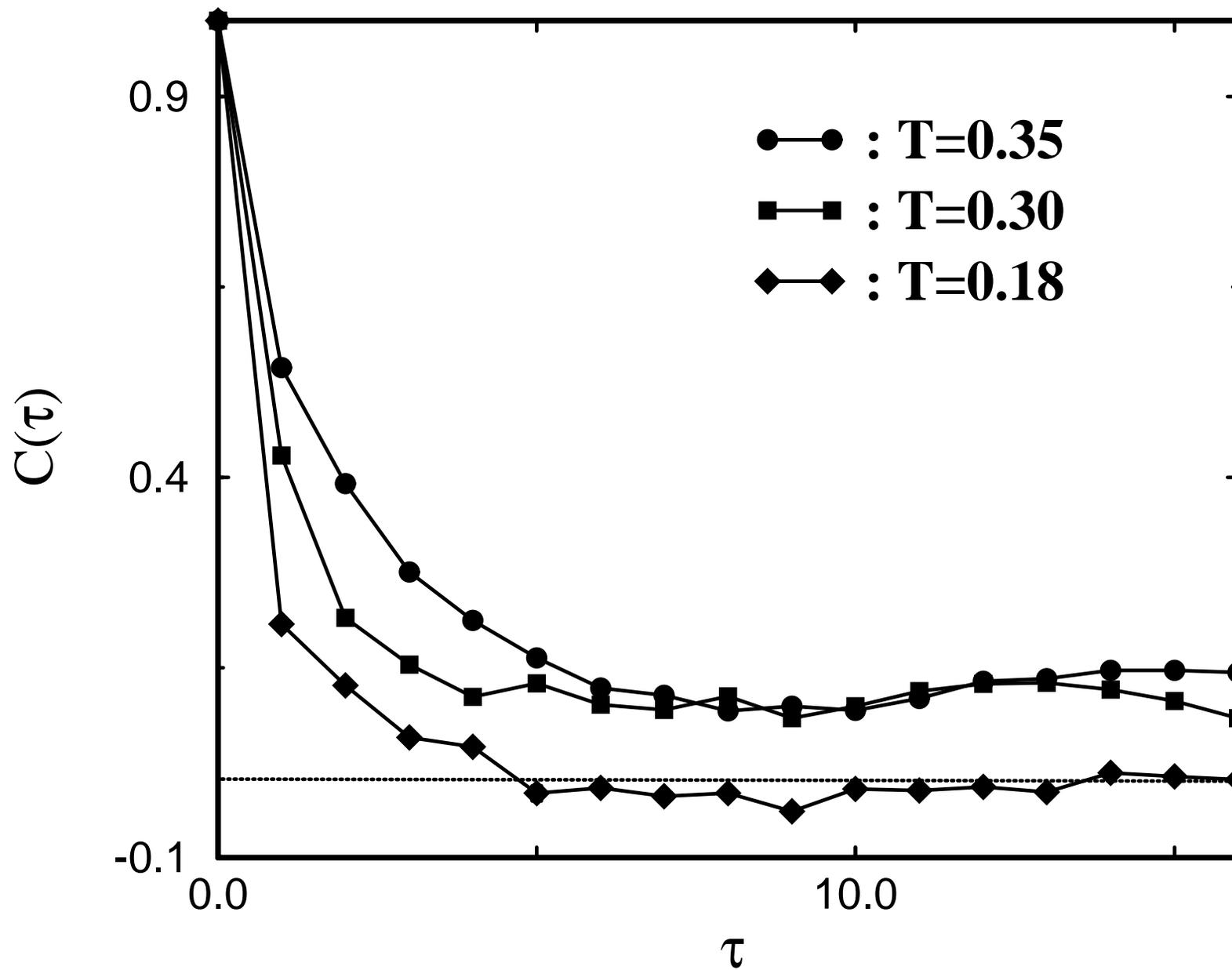
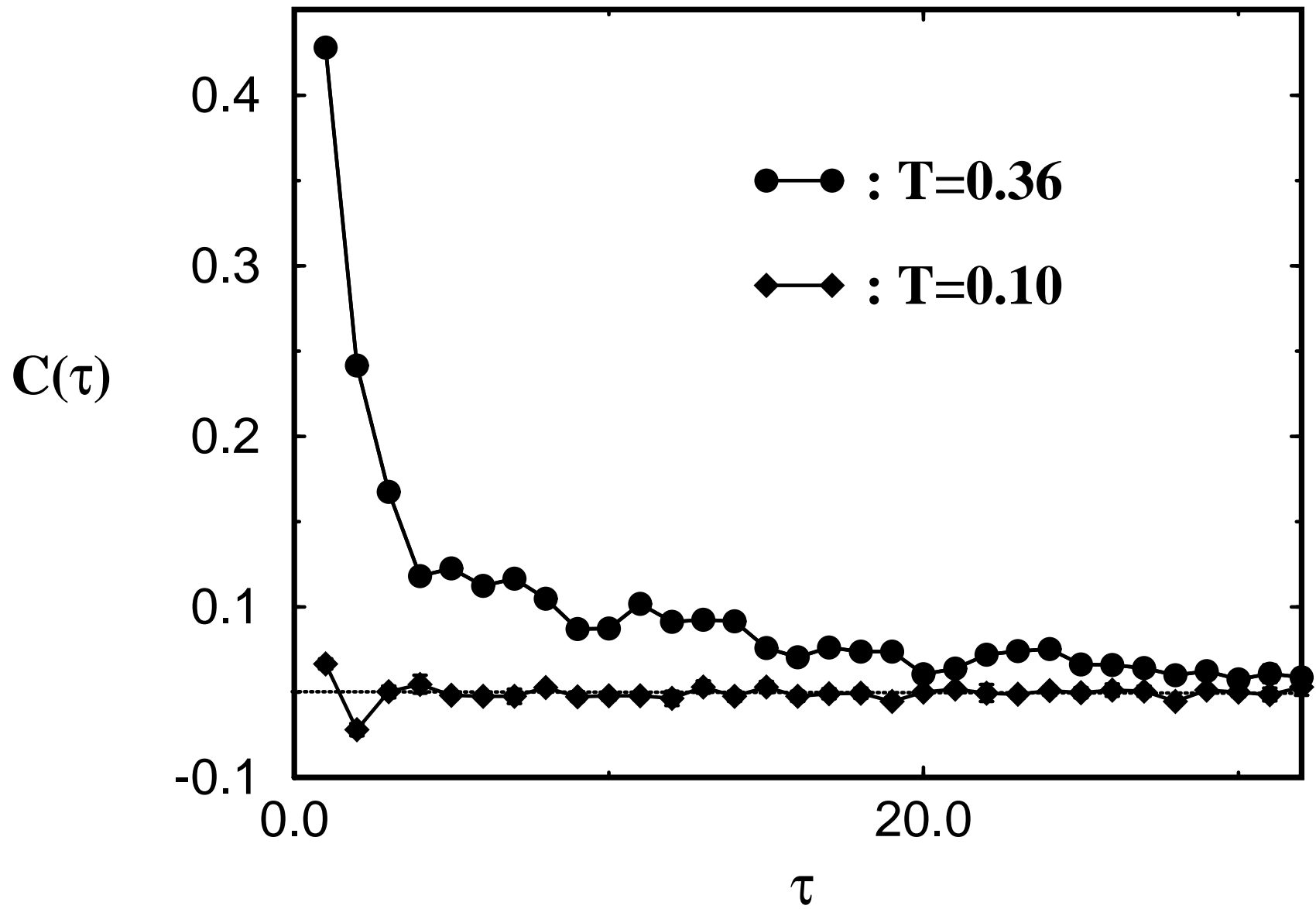
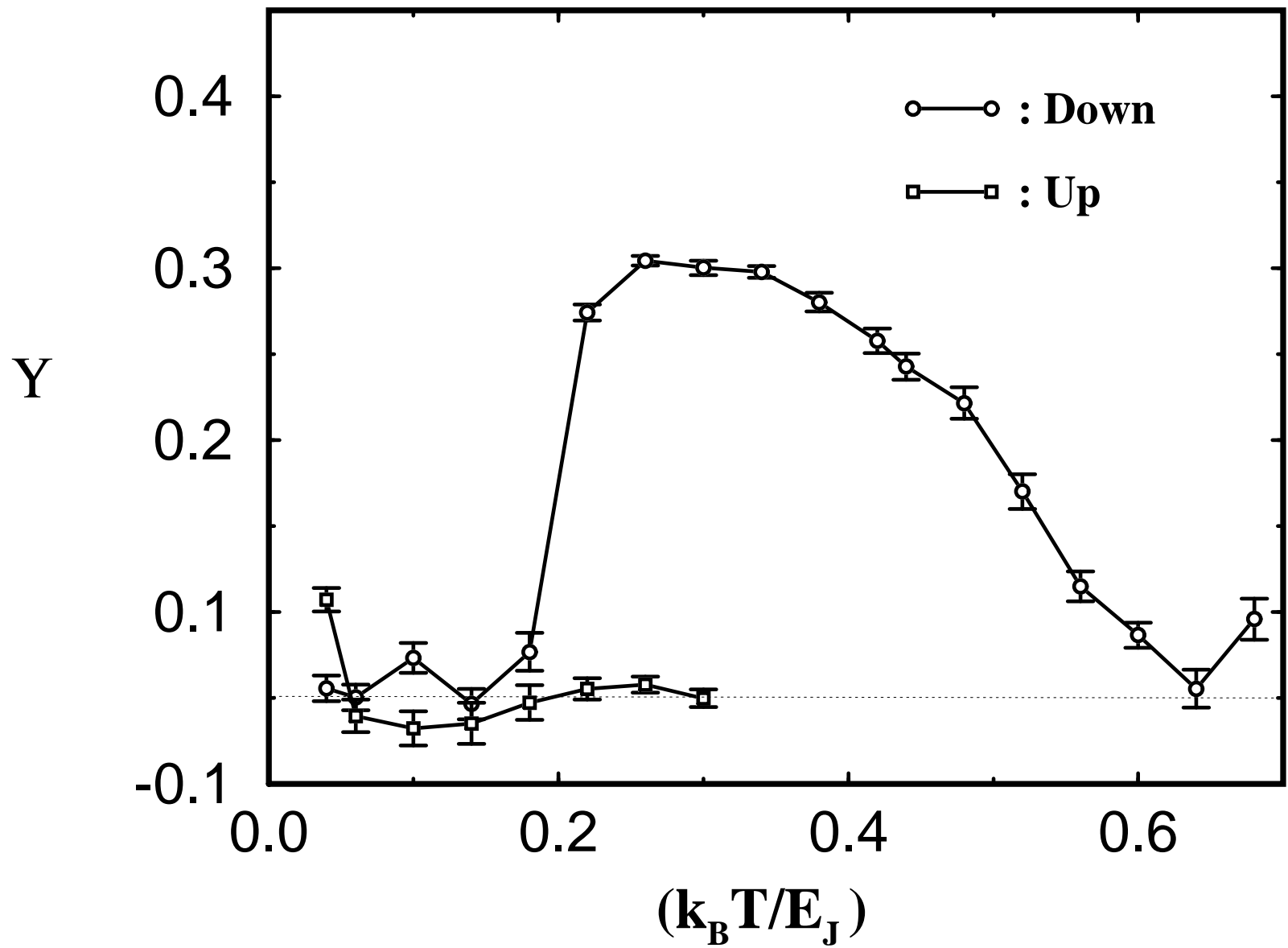
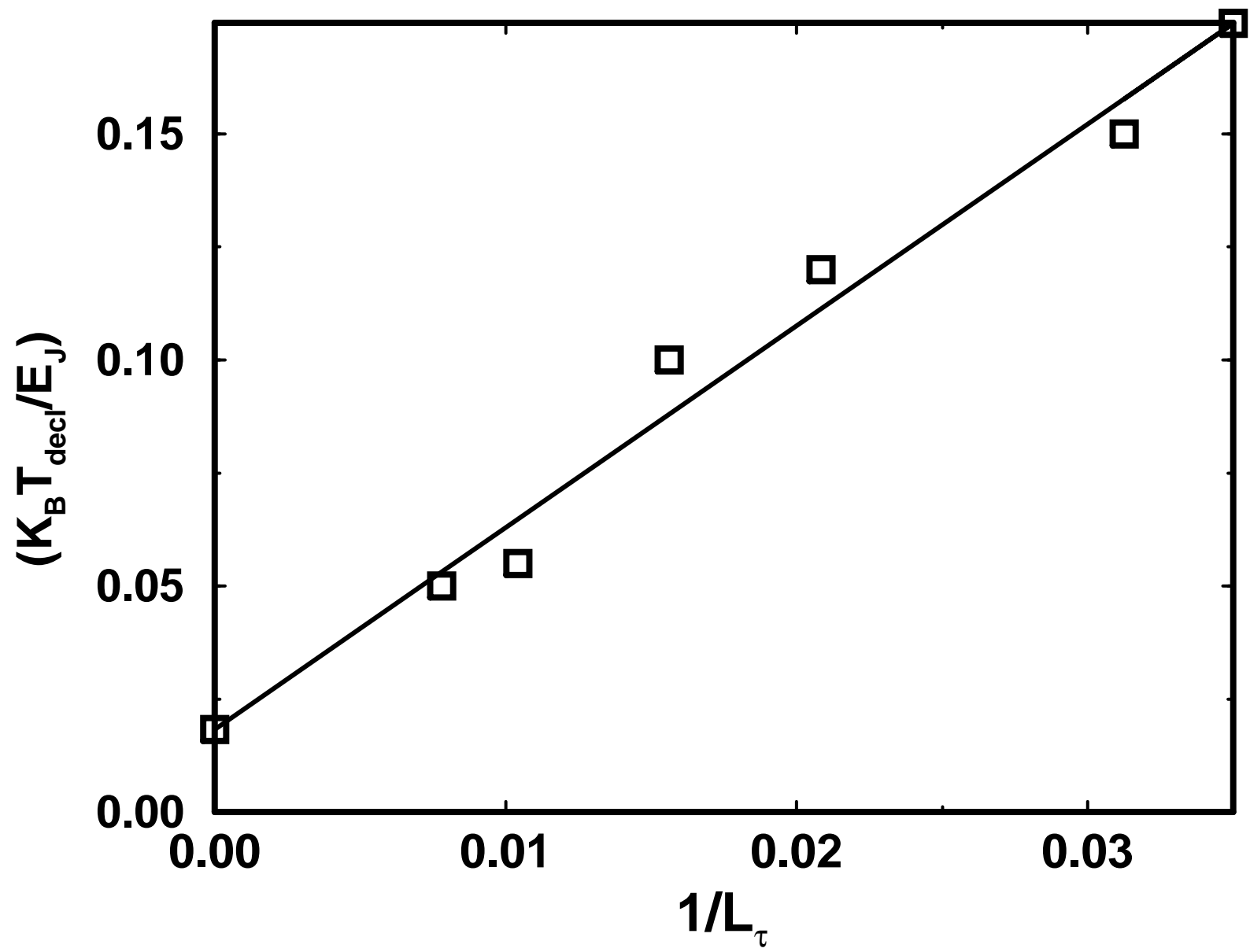
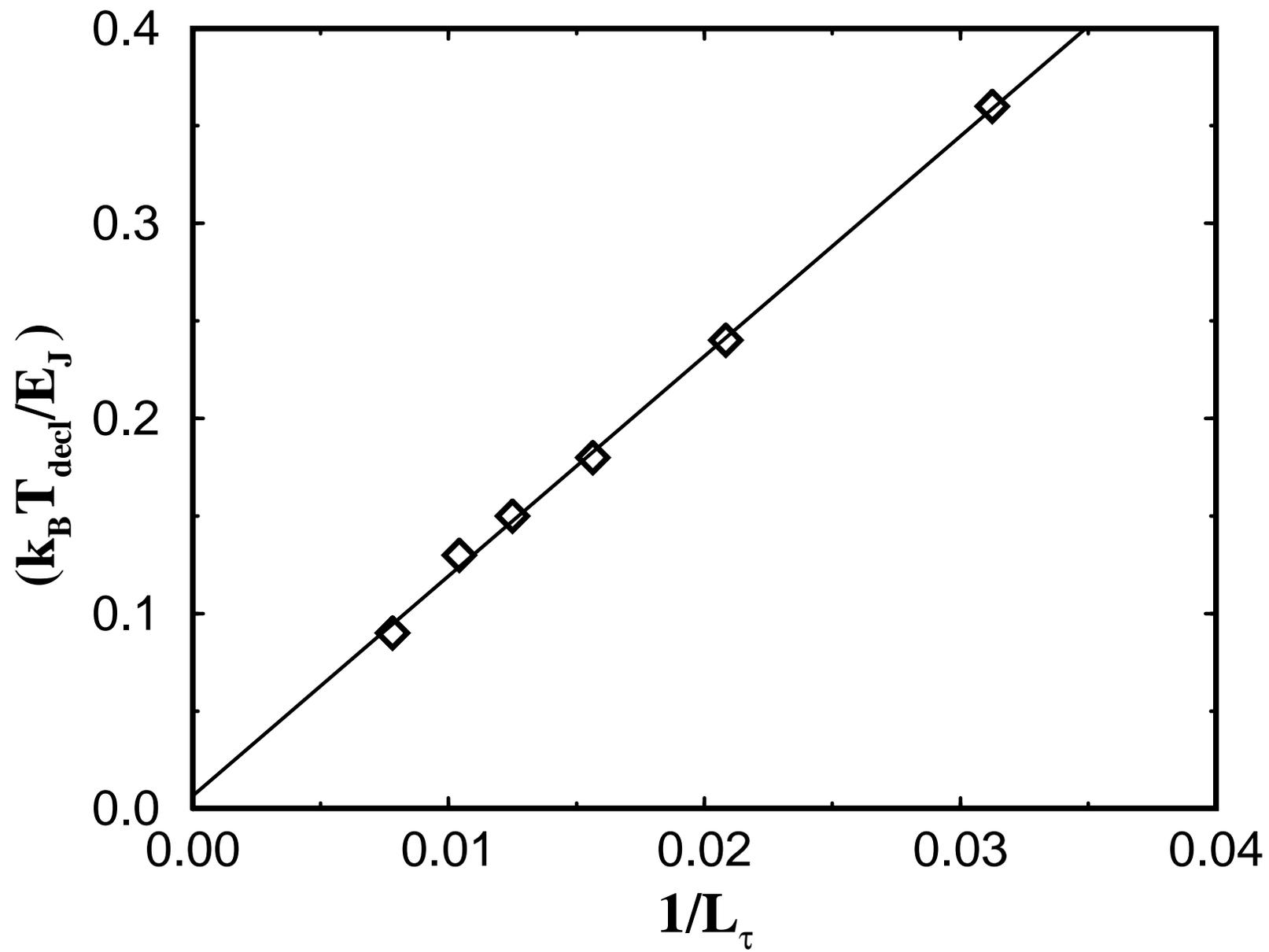
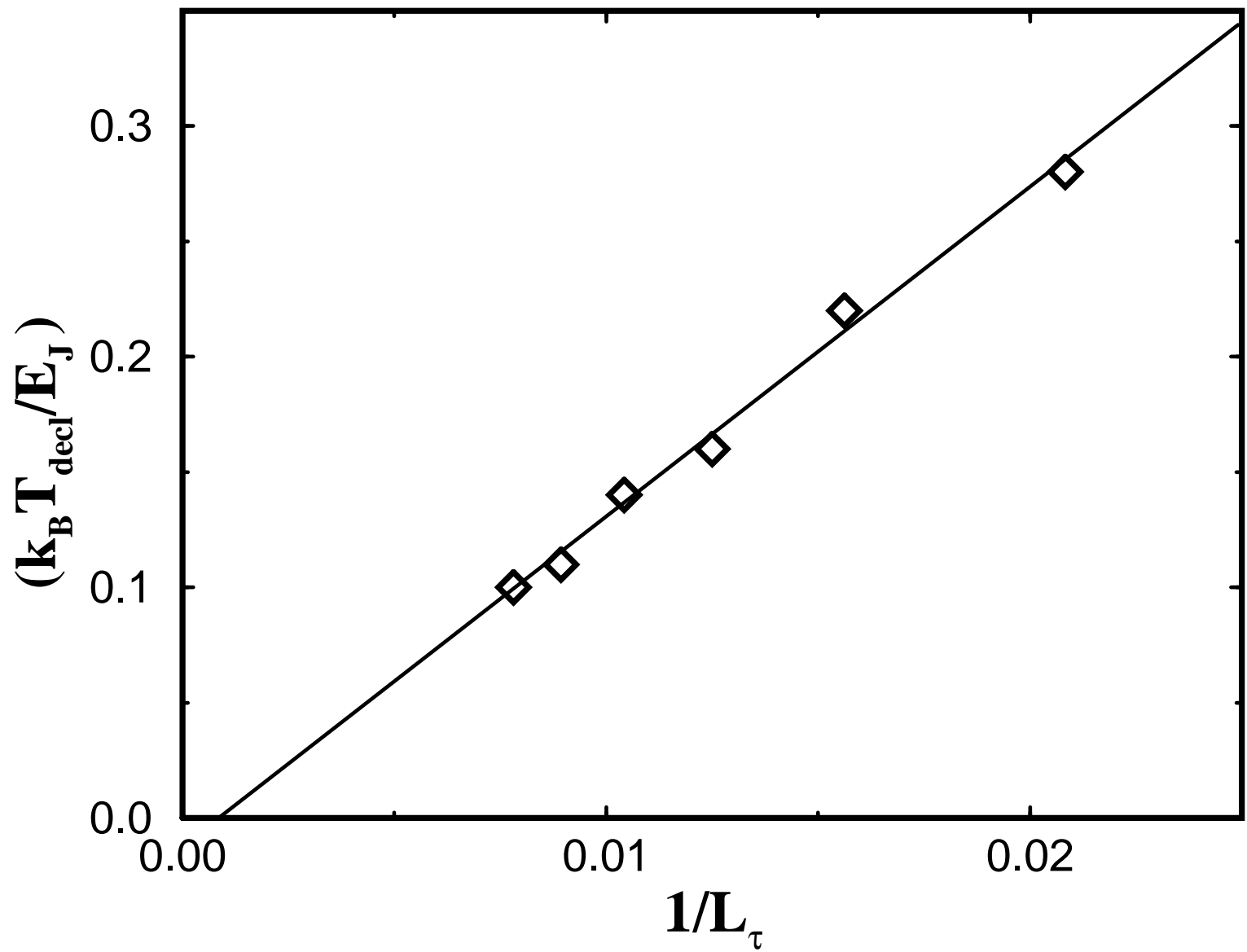
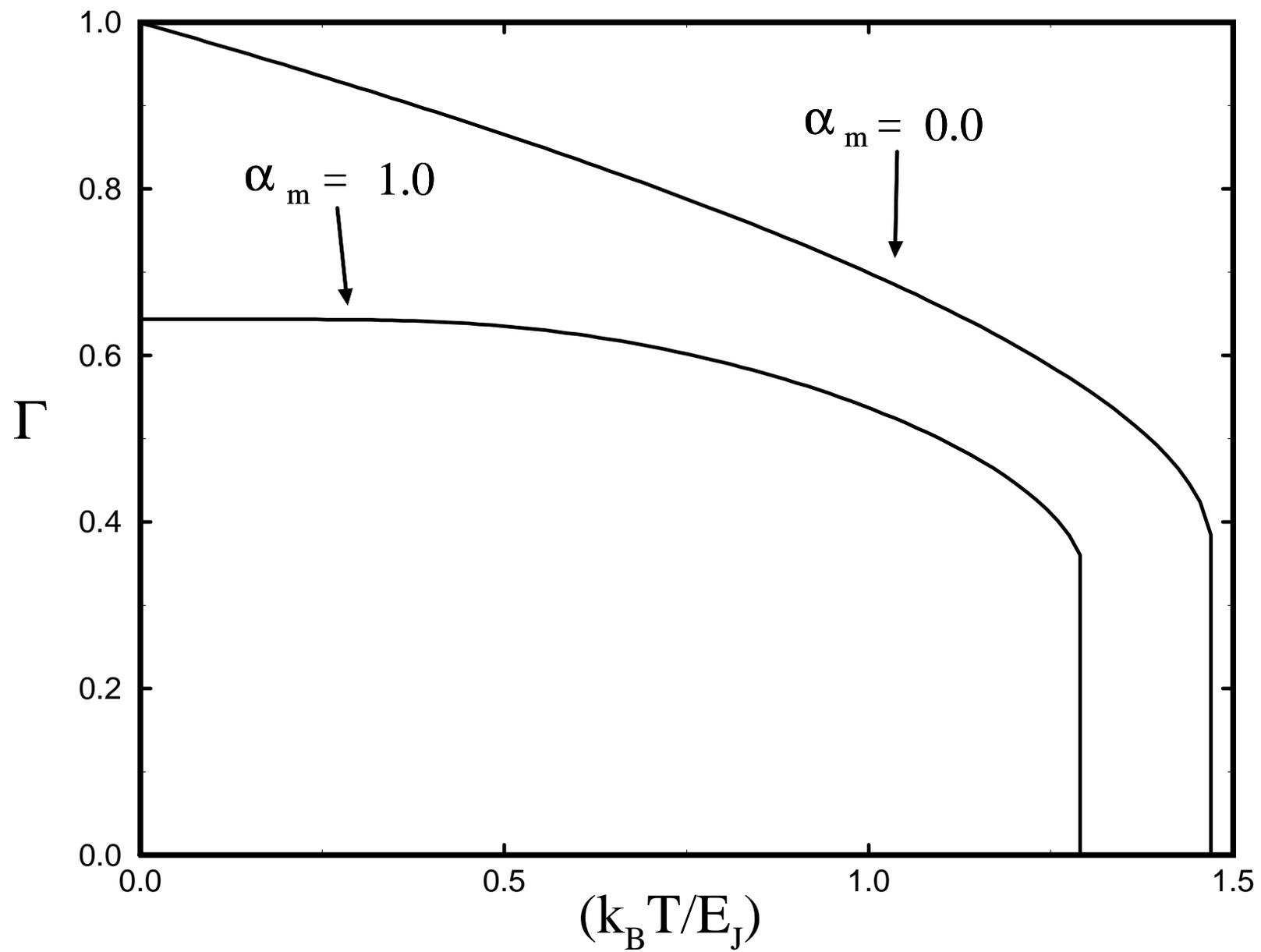
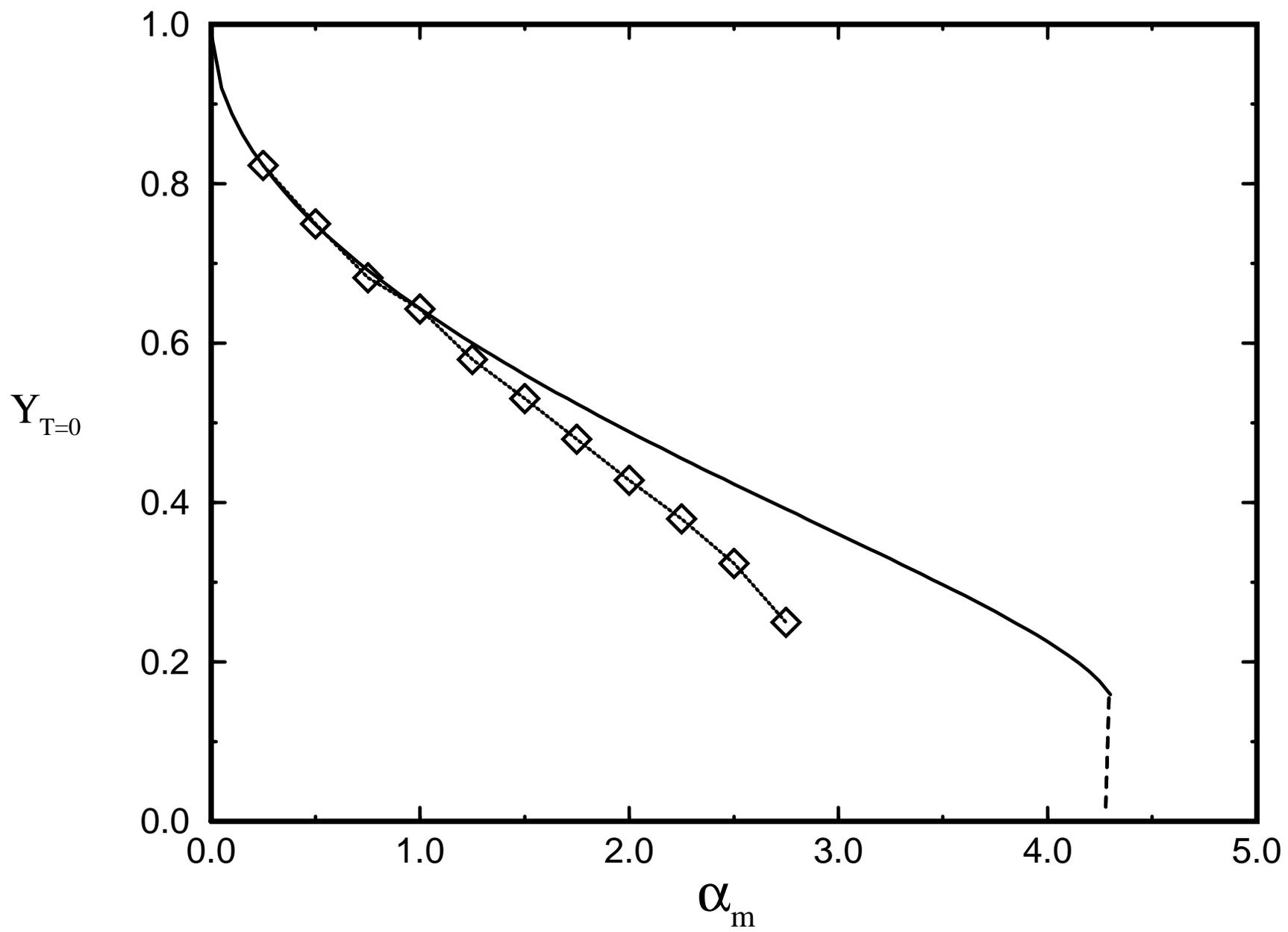
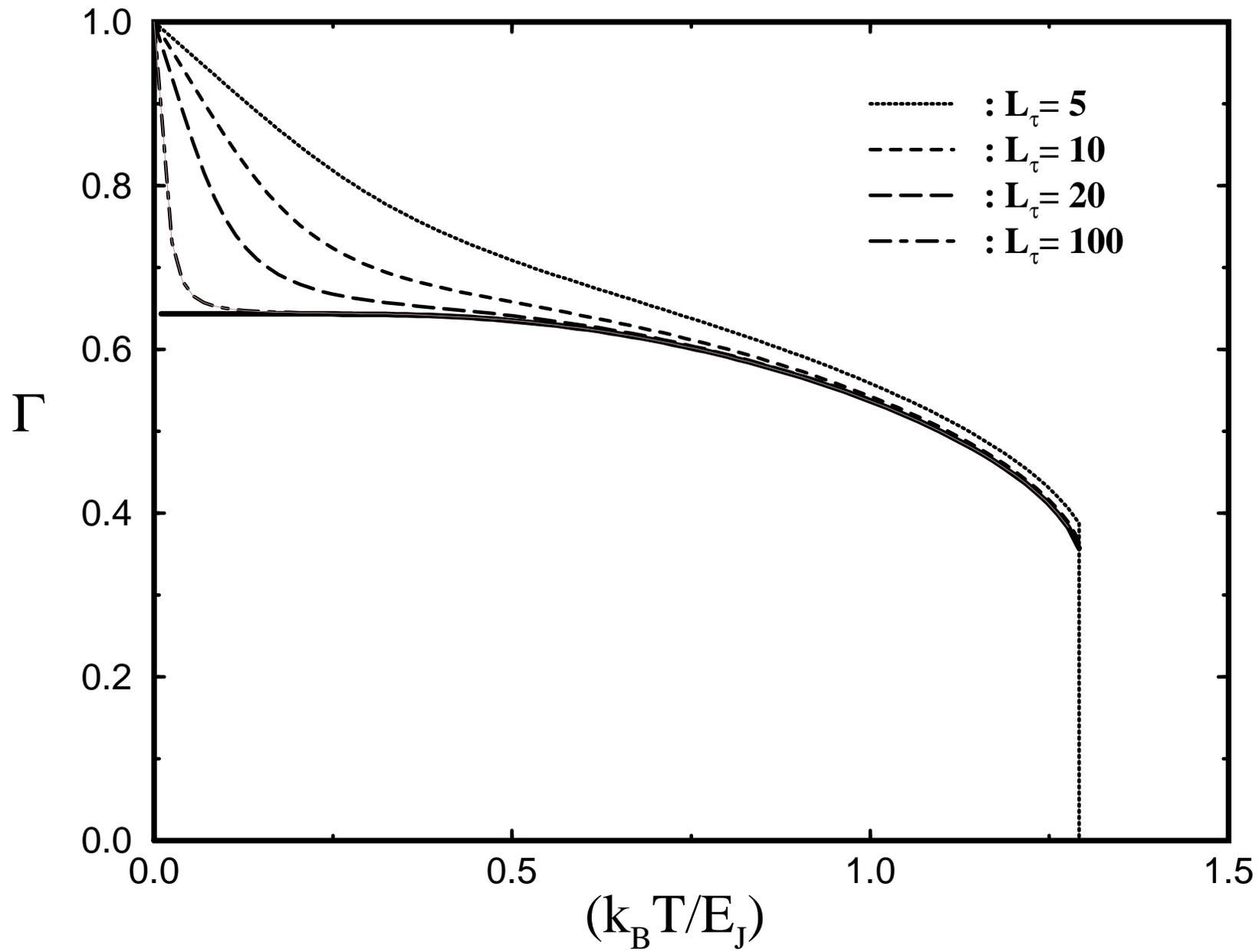
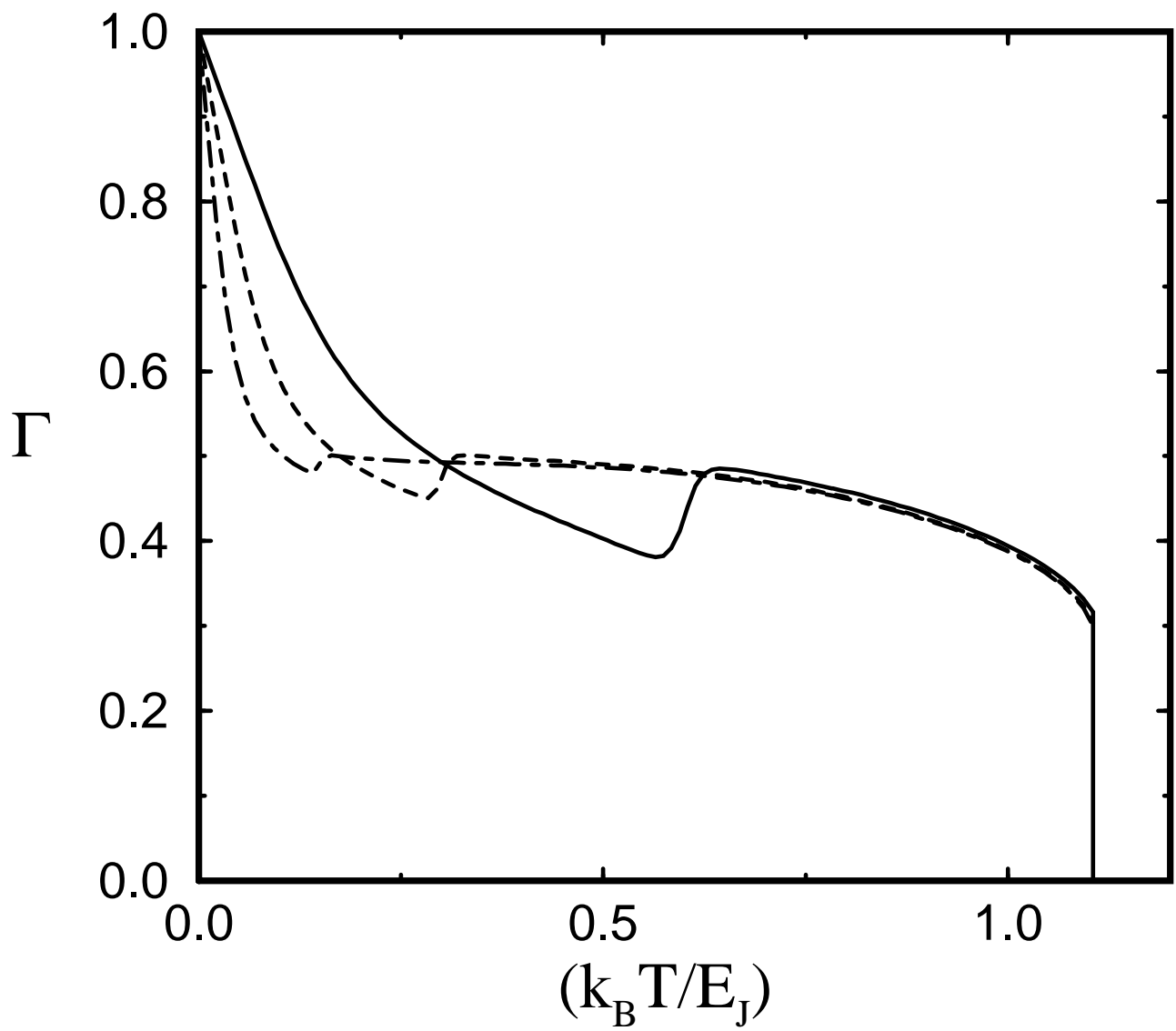
\end{document}